\documentclass{aa}
\usepackage{txfonts}
\usepackage{graphicx}
\usepackage{caption}
\providecommand{\tabularnewline}{\\}
\newcommand{\lyxdot}{.}

\RequirePackage[normalem]{ulem} 
\RequirePackage{color}\definecolor{RED}{rgb}{1,0,0}\definecolor{BLUE}{rgb}{0,0,1} 
\providecommand{\DIFadd}[1]{{\protect{#1}}} 
\providecommand{\DIFdel}[1]{}  
\providecommand{\DIFaddbegin}{} 
\providecommand{\DIFaddend}{} 
\providecommand{\DIFdelbegin}{} 
\providecommand{\DIFdelend}{} 
\providecommand{\DIFdelFL}[1]{\DIFdel{#1}} 
\providecommand{\DIFaddbeginFL}{} 
\providecommand{\DIFaddendFL}{} 
\providecommand{\DIFdelbeginFL}{} 
\providecommand{\DIFdelendFL}{} 
\newcommand{\DIFscaledelfig}{0.5}
\RequirePackage{settobox} 
\RequirePackage{letltxmacro} 
\newsavebox{\DIFdelgraphicsbox} 
\newlength{\DIFdelgraphicswidth} 
\newlength{\DIFdelgraphicsheight} 
\LetLtxMacro{\DIFOincludegraphics}{\includegraphics} 
\newcommand{\DIFaddincludegraphics}[2][]{{\color{blue}\fbox{\DIFOincludegraphics[#1]{#2}}}} 
\newcommand{\DIFdelincludegraphics}[2][]{
\sbox{\DIFdelgraphicsbox}{\DIFOincludegraphics[#1]{#2}}
\settoboxwidth{\DIFdelgraphicswidth}{\DIFdelgraphicsbox} 
\settoboxtotalheight{\DIFdelgraphicsheight}{\DIFdelgraphicsbox} 
\scalebox{\DIFscaledelfig}{
\parbox[b]{\DIFdelgraphicswidth}{\usebox{\DIFdelgraphicsbox}\\[-\baselineskip] \rule{\DIFdelgraphicswidth}{0em}}\llap{\resizebox{\DIFdelgraphicswidth}{\DIFdelgraphicsheight}{
\setlength{\unitlength}{\DIFdelgraphicswidth}
\begin{picture}(1,1)
\thicklines\linethickness{2pt} 
{\color[rgb]{1,0,0}\put(0,0){\framebox(1,1){}}}
{\color[rgb]{1,0,0}\put(0,0){\line( 1,1){1}}}
{\color[rgb]{1,0,0}\put(0,1){\line(1,-1){1}}}
\end{picture}
}\hspace*{3pt}}} 
} 
\LetLtxMacro{\DIFOaddbegin}{\DIFaddbegin} 
\LetLtxMacro{\DIFOaddend}{\DIFaddend} 
\LetLtxMacro{\DIFOdelbegin}{\DIFdelbegin} 
\LetLtxMacro{\DIFOdelend}{\DIFdelend} 
\DeclareRobustCommand{\DIFaddbegin}{\DIFOaddbegin \let\includegraphics\DIFaddincludegraphics} 
\DeclareRobustCommand{\DIFaddend}{\DIFOaddend \let\includegraphics\DIFOincludegraphics} 
\DeclareRobustCommand{\DIFdelbegin}{\DIFOdelbegin \let\includegraphics\DIFdelincludegraphics} 
\DeclareRobustCommand{\DIFdelend}{\DIFOaddend \let\includegraphics\DIFOincludegraphics} 
\LetLtxMacro{\DIFOaddbeginFL}{\DIFaddbeginFL} 
\LetLtxMacro{\DIFOaddendFL}{\DIFaddendFL} 
\LetLtxMacro{\DIFOdelbeginFL}{\DIFdelbeginFL} 
\LetLtxMacro{\DIFOdelendFL}{\DIFdelendFL} 
\DeclareRobustCommand{\DIFaddbeginFL}{\DIFOaddbeginFL \let\includegraphics\DIFaddincludegraphics} 
\DeclareRobustCommand{\DIFaddendFL}{\DIFOaddendFL \let\includegraphics\DIFOincludegraphics} 
\DeclareRobustCommand{\DIFdelbeginFL}{\DIFOdelbeginFL \let\includegraphics\DIFdelincludegraphics} 
\DeclareRobustCommand{\DIFdelendFL}{\DIFOaddendFL \let\includegraphics\DIFOincludegraphics} 

\begin{document}

\title{N-body simulations of planet formation via pebble accretion II. How various giant planets form}

\titlerunning{N-body simulations of planet formation via pebble accretion II}

\author{Soko Matsumura\inst{1} \and Ramon Brasser\inst{2} \and Shigeru
Ida\inst{2} }

\institute{School of Engineering, Physics, and Mathematics, University of Dundee,
DD1 4HN, UK\\
 \email{s.matsumura@dundee.ac.uk} \and Earth-Life Science Institute,
Tokyo Institute of Technology, Meguro-ku, Tokyo, 152-8550, Japan}

\DIFdelbegin 

\DIFdelend 
\abstract{}
{\DIFdelbegin \DIFdel{We }\DIFdelend  
The connection between initial disc conditions and final orbital and physical properties of planets is not well-understood. In this paper, we numerically study the formation of planetary systems
via pebble accretion and investigate the effects of disc properties
such as masses, dissipation timescales, and metallicities  on planet formation outcomes.} {We improved the N-body code SyMBA that
was modified for our paper I 
by taking account of new planet-disc interaction models and type II migration. We adopted the `two-$\alpha$' disc model to \DIFdelbegin \DIFdel{approximate
}\DIFdelend  mimic the
effects of both the standard disc turbulence and the mass accretion
driven by the magnetic disc wind.} {We successfully reproduced
\DIFadd{the overall distribution trends}  of semi-major axes, eccentricities, 
and planetary masses of extrasolar \DIFadd{giant planets}. 
There are two types of giant planet formation trends, depending
on whether or not the disc's dissipation timescales are comparable to
the planet formation timescales. When planet formation happens fast enough,
giant planets are fully grown (Jupiter mass or higher) and are distributed
widely across the disc. On the other hand, when planet formation is
limited by the disc's dissipation, discs generally form low-mass
cold Jupiters (CJs). Our simulations also naturally explain why hot
Jupiters (HJs) tend to be alone and how the observed eccentricity-metallicity
trends arise. The low-metallicity discs tend to form nearly circular
and coplanar HJs in situ, because planet formation is slower than
high-metallicity discs, and thus protoplanetary cores migrate significantly
before gas accretion.
The high-metallicity discs, on the other hand, generate HJs in situ or via tidal circularisation of eccentric orbits.
Both pathways usually involve dynamical instabilities, and thus HJs
tend to have broader eccentricity and inclination distributions.
When giant planets with very wide orbits ('super-cold Jupiters')
are formed via pebble accretion followed by scattering, we predict
that they belong to metal-rich stars, have eccentric orbits, and tend
to have ($\sim80\%$) companions interior to their orbits.}

\maketitle

\section{Introduction \label{sec:Introduction}}

 Accretion of both pebbles and planetesimals is likely to contribute
to planet formation \citep[e.g.][]{Ida16a,Johansen17}. 
The pebbles are cm- to m-sized particles, which are very sensitive to gas drag
\citep{Adachi76,Weidenschilling77}. Since their migration timescale
is short compared to the growth timescale \citep{Birnstiel12a,Lambrechts14b},
they are expected to rapidly migrate towards the central star without
much growth. \DIFdelbegin \DIFdel{Such }\DIFdelend 
These migrating dust particles could be concentrated
to form $100-1000\,$km-sized planetesimals directly (which correspond
to $\sim10^{-6}-10^{-4}\,M_{\oplus}$) \citep[e.g.][]{Johansen15,Simon16},
either via the trapping of dust grains in pressure bumps of the protoplanetary
disc or via streaming instability \citep{Youdin05,Johansen07b,Johansen09b}.
These planetesimals are initially likely to grow via planetesimal-planetesimal
collisions \citep{Ida16a,Johansen17}, until the protoplanetary cores
become massive enough to have a 
large capturing radius of pebbles
comparable to the Hill radius \citep{Ormel10,Kretke14}. The stage
of a protoplanetary core growth via pebbles is called `pebble accretion'.
 The pebble accretion slows down when the capturing radius becomes
larger than the pebble-disc scale height, and thus the accretion becomes
two-dimensional. At such a stage, both pebble and planetesimal accretion
may contribute to mass growth \mbox{
\citep[e.g.][]{Ida16a}}\hspace{0pt}
.

\citet{Ormel10} first developed the analytical model of particle-protoplanet
interactions in a gas disc, and pointed out the efficiency of pebble
accretion in the settling regime where gas drag effects are significant.
The pebble accretion was further studied by considering the growth
of a single core \citep{Bitsch15} and by using sophisticated N-body
simulations \citep{Levison15a,Chambers16}. All of these studies confirmed
the efficiency of pebble accretion and showed that a variety of planetary
systems could be formed within a typical gas disc lifetime. However,
the model by \citet{Bitsch15} was not designed to assess the planet-planet
interaction effects, which are important to determine the final orbital
architecture of planetary systems. The models by \citet{Levison15a}
and \citet{Chambers16} focused on the detailed physics of growth
and destruction of particles ranging from pebbles to planets, and
did not take account of planet migration effects, which are also important
to determine the orbital distribution of planets.

\DIFdelbegin \DIFdel{The goal of the current work, and that of our previous work \mbox{
\citep{Matsumura17}}\hspace{0pt}
,
is to compare planetary systems formed via pebble accretion with observed
extrasolar planetary systems, and to investigate how initial disc
conditions are related to the physical and orbital properties of planets.
From this point of view, giant planets are very useful in constraining
formation and evolution scenarios of planetary systems, because giant
planets are observed over a wide range of orbital radii, from less
than $0.01\,$au to over $100\,$au. On the other hand, observed Super
Earths and lower-mass planets, though they are more abundant, are
limited to the inner disc region ($\lesssim1\,$au). In this work,
we will generate both types of planets but pay a particular attention
to disc conditions which lead to different types of giant planets
ranging from close-in hot Jupiters to very distant super cold Jupiters.
}

\DIFdel{This kind of a study of connecting planetary systems with protoplanetary
discs is becoming increasingly important. ALMA observations now regularly
find gaps, spirals, and asymmetries in protoplanetary discs, which
may be attributed to planet formation processes or forming planets \citep[e.g.][]{Andrews18,Huang18}.
There are also some potential planetary candidates discovered in protoplanetary discs including
PDS 70 \mbox{
\citep{Keppler18}}\hspace{0pt}
, HD 100546 \mbox{
\citep{Brittain19,Casassus19}}\hspace{0pt}
,
and TW Hya \citep{Tsukagoshi19}. 
The future observations would allow
us to further constrain planet formation processes occurring in protoplanetary
discs.
}

\DIFdelend \citet[][Paper I]{Matsumura17} implemented the analytical pebble accretion model
by \citet{Ida16a} into an N-body code SyMBA \citep{Duncan98}, and
performed global N-body simulations of pebble accretion over a range
of disc parameters. Although the work has confirmed that pebble accretion
leads to a variety of planetary systems, it failed to reproduce the
overall distribution \DIFadd{trends}  of the physical and orbital parameters of extrasolar \DIFadd{giant} planets. 
In particular, few giant planets were formed, because most
planetary cores were lost to the disc's inner edge due to migration,
in a similar manner to planetesimal accretion simulations \citep[e.g. ][]{Coleman16a}.
The problem was that a range of planetary masses leading to outward
migration \citep{Paardekooper11} is very narrow, so growing planets
eventually start migrating inward \citep[also see][]{Brasser17}.
Therefore, although pebble accretion may be more efficient compared
to planetesimal accretion \citep[e.g.][]{Lambrechts12,Kretke14},
the giant planet formation timescale still appeared to be too long compared
to the migration timescale, and thus the migration problem remained.

There has been a significant development regarding the formation of `cold'
giant planets (i.e. Jupiter-like planets beyond $\sim1\,$au) in the
past few years. The formation of such planets has been particularly difficult
\DIFdelbegin \DIFdel{partly }\DIFdelend 
largely because of the efficient migration of protoplanetary cores
discussed above \citep[e.g.][]{Coleman16a,Matsumura17}. \citet{Bitsch15}
mitigated the issue and successfully formed cold Jupiters (CJs) by
placing cores far from the central star (beyond $\sim15\,$au) in
an evolved disc (up to $\sim2\,$Myr) with a relatively short disc
lifetime of $3\,$Myr. \citet{Coleman16b} resolved the migration
issue by considering the temporal planet `traps' due to variations
in the effective viscous stresses. More recently, \citet{Ida18} proposed
an elegant solution by combining two recent key developments in the
field \textemdash{} the new type II migration formula \DIFdelbegin \DIFdel{\mbox{
\citep{Kanagawa18}
}\hspace{0pt}
}\DIFdelend  \mbox{
\citep{Kanagawa18a}
}\hspace{0pt}
and the wind-driven accretion disc with the nearly laminar interior
\citep[e.g. ][]{Bai13,Bai17}, where both contribute to slowing down
type II migration and make the formation of CJs possible.

One of \DIFdelbegin \DIFdel{the }\DIFdelend  these key developments comes from the realisation that the
migration of a gap-opening planet is not tied to the disc evolution.
This is because, differently from the assumption of the classical type
II migration model \citep{Lin86b,Lin93}, the disc gas proved
to cross the gap opened by a planet. The new type II migration formula
by \DIFdelbegin \DIFdel{\mbox{
\citet{Kanagawa18}
}\hspace{0pt}
}\DIFdelend  \mbox{
\citet{Kanagawa18a} }\hspace{0pt}
reflects the results of recent hydrodynamic
simulations \citep{Duffell13,Fung14} as well as their own, and shows
that the migration is slower for a planet with the deeper gap (i.e.
for a smaller viscosity disc, a lower disc temperature, or a larger
mass planet, also see Equation \ref{eq:taua2}). Compared to the classical
type II formula, the new formula typically gives $\sim1$ order of
magnitude longer migration timescales \DIFdelbegin \DIFdel{\mbox{
\citep[e.g.][]{Ida18}}\hspace{0pt}
}\DIFdelend  \mbox{
\citep{Ida18}}\hspace{0pt}
.

\DIFdelbegin \DIFdel{Another }\DIFdelend  The other key development is the accretion mechanism of a protoplanetary
disc. In the classical picture, the disc's viscosity drives the angular
momentum transfer in the protoplanetary disc, and the chief explanation for the viscosity is the magnetorotational instability turbulence
\citep[e.g.][]{Balbus98}. Recent non-ideal magnetohydrodynamic simulations,
however, have shown that the protoplanetary discs are likely to be
largely laminar due to the combined effects of ambipolar diffusion,
Hall effects and Ohmic dissipation, and the fact that the disc's angular momentum
is mostly removed by the magnetic disc wind \citep[e.g.][]{Bai13,Bai17}.
Following this idea, \citet{Ida18} adopted two different alphas:
one representing the disc accretion driven by the mass loss due to
the magnetic disc wind $\alpha_{{\rm acc}}$, and the other representing
the local disc turbulence $\alpha_{{\rm turb}}$. Since the gap opening
is controlled by the disc turbulence and since the turbulence is expected
to be very weak $\alpha_{{\rm turb}}\lesssim10^{-4}$ \citep[e.g.][]{Bai17},
the estimated type II migration is slower than it is using the $\alpha$
required to explain the observed stellar mass accretion rate ($\alpha\sim10^{-2}$
in \citet{Hartmann98}, see Section \ref{subsec:PF_DiscEvol} as well).
 A similar study was conducted by \mbox{
\citet{Wimarsson20} }\hspace{0pt}
using N-body simulations
with pebble accretion. Besides the new type II mechanism described
above, they show that the transition zone between viscously
and radiatively heated disc regions could work as a temporal planet
trap and promote more efficient planetary growth via collisions.

Recently, a trilogy of papers on N-body pebble accretion simulations
was published, with two of them exploring the formation of super-Earths
(SEs) or lower mass planets \citep{Lambrechts19,Izidoro19ap}, and
the other focusing on the formation of giant planets \citep{Bitsch19}.
For SEs and lower mass planets, they proposed that Earth-like
planets were formed in the low pebble flux disc with little migration,
while SE-like planets were formed in the higher pebble flux disc.
In the latter case,  their growing cores migrated to the disc's inner
edge in a similar manner to \citet{Matsumura17}, but \DIFdelbegin \DIFdel{the cores
are }\DIFdelend  these cores
were more efficiently trapped in mean-motion resonances (MMRs) because
they chose the disc dissipation timescale comparable to the migration
timescale so that the migration would be halted completely near the inner
edge of the disc. As a result, their protoplanets \DIFdelbegin \DIFdel{become }\DIFdelend  became dynamically
unstable as the gas disc \DIFdelbegin \DIFdel{dissipates, collide }\DIFdelend  dissipated and the protoplanets collided with one another and
\DIFdelbegin \DIFdel{form }\DIFdelend  formed non-resonant multiple SEs \citep[cf.][]{Ogihara10}. However,
the disc-planet interactions near the disc edge is not well understood,
and the planet trapping efficiency depends on the sharpness of the
disc edge \citep{Ogihara10} and the details of the torque balance
\citep{Brasser18}. For giant planets, on the other hand, \citet{Bitsch19}
compared the classical $\alpha$ disc model with $\alpha=5.3\times10^{-3}$
with a `two-$\alpha$' disc model similar to that of \citet{Ida18} with
$\alpha_{{\rm acc}}=5.3\times10^{-3}$ and $\alpha_{{\rm turb}}=5.3\times10^{-4}-10^{-4}$,
and successfully produced CJs. However, they failed to form highly
eccentric giant planets (see discussion in Section \ref{subsec:Comp_Bitsch19}).

\DIFdelbegin \DIFdel{In this paper, we will explore the effects of disc properties such
as masses, dissipation timescales, and metallicities on formation of planetary
systems,
and show that such simulations reproduce the
observed distributions of planetary systems well. }\DIFdelend 
The goal of this work, like that of our previous work \mbox{
\citep{Matsumura17}}\hspace{0pt}
,
is to investigate how initial disc conditions are related to physical
and orbital properties of planets by comparing planetary systems formed
via numerical simulations with observed extrasolar planetary systems.
This kind of study of connecting planetary systems with protoplanetary
discs is becoming increasingly important. ALMA observations now regularly
find gaps, spirals, and asymmetries in protoplanetary discs, which
may be attributed to planet formation processes or forming planets
\mbox{
\citep[e.g.][]{Andrews18,Huang18}}\hspace{0pt}
. There are also potential planetary
candidates discovered in protoplanetary discs including PDS 70 \mbox{
\citep{Keppler18}}\hspace{0pt}
,
HD 100546 \mbox{
\citep{Brittain19,Casassus19}}\hspace{0pt}
, and TW Hya \citep{Tsukagoshi19}.
The future observations would allow us to further constrain planet
formation processes occurring in protoplanetary discs.

As described in Section \ref{sec:Methods}, we started 
with disc conditions motivated by observations, follow planet formation and
orbital evolution by using an N-body code SyMBA \mbox{
\citep{Duncan98}}\hspace{0pt}
,
and compare distributions of orbital and physical properties of simulated
planets with those of observed ones. From this point of view, giant
planets are particularly useful, because they are observed over a
wide range of orbital radii, from less than $0.01\,$au to over $100\,$au.
Super-Earths and lower-mass planets, on the other hand, are more abundant
but are limited to the inner disc region ($\lesssim1\,$au). Therefore,
in this paper, we pay particular attention to disc conditions
that lead to different types of giant planets, ranging from close-in
hot Jupiters to very distant super-cold Jupiters, though numerical
simulations will generate all kinds of planets.

In Section 2, we introduce the numerical methods we used
by highlighting the changes made since \citet{Matsumura17}. We improve
the N-body code SyMBA  \mbox{
\citep{Duncan98} }\hspace{0pt}
by adopting a more self-consistent
disc model and by taking account of recent developments of the field.
In Section 3, we show that our numerical simulations reproduce
\DIFdelbegin \DIFdel{the }\DIFdelend  
overall \DIFadd{trends of} distributions of semi-major axis  $a$, eccentricity $e$, 
\DIFdelbegin \DIFdel{eccentricity, }\DIFdelend and
mass $M_{p}$ of \DIFadd{extrasolar giant planets. }
We describe the effects of different parameters
on planet formation and also show how different types of giant planets
(such as HJs and CJs) are formed. We    discuss the results further
in Section 4, and summarise our findings in Section 5.

\section{Methods \label{sec:Methods}}

 We ran all of the simulations with the N-body integrator SyMBA
\citep{Duncan98}, which was modified to include
 models of planet formation,
disc-planet interactions, and disc evolution. In this \DIFdelbegin \DIFdel{work, we
will also adopt this ``two-$\alpha$'' disc model for simplicity,
and represent the wind-driven disc accretion with $\alpha_{{\rm acc}}$
and the disc turbulence with $\alpha_{{\rm turb}}$}\DIFdelend  section, we
highlight the updates made since \mbox{
\citet{Matsumura17}}\hspace{0pt}
.

In \DIFdelbegin \DIFdel{the following, we discuss the improved equation of motion as proposed
by \mbox{
\citet{Ida20} }\hspace{0pt}
(Section \ref{subsec:OrbEvol}), which formulates
}\DIFdelend  Section \ref{subsec:OrbEvol}, we introduce the planet-disc \DIFdelbegin \DIFdel{interactions based on the dynamical friction and provides
the behaviour consistent with hydrodynamic simulations both in the
subsonic and supersonic regimes}\DIFdelend  interaction
model adopted from \mbox{
\citet{Ida20},}\hspace{0pt}
 and we show how planet migration as
well as eccentricity and inclination damping of protoplanetary orbits
are modelled in our code. We also updated the type II migration
prescription \DIFdelbegin \DIFdel{as proposed by \mbox{
\citet{Kanagawa18}}\hspace{0pt}
,
and scaled
the eccentricity and inclination damping prescriptions in the type
II regime in a similar manner to planet migration (i.e. semimajor
axis damping). Planet }\DIFdelend  by following \mbox{
\citet{Kanagawa18a}}\hspace{0pt}
. In Section \ref{subsec:PF_DiscEvol},
we introduce planet formation and disc evolution \DIFdelbegin \DIFdel{prescriptions are
introduced in Section \ref{subsec:PF_DiscEvol}, which uses a more
self-consistent disc model compared to \mbox{
\citet{Matsumura17}}\hspace{0pt}
. We also
discuss our choice of the pebble accretion efficiency and the pebble
isolation mass}\DIFdelend  models. We adopted
the $\alpha$ disc model \citep{Shakura73}, 
but we modified it to take
account of a recent development in the field. Finally, in Section
\ref{subsec:Initial-Conditions}, we present the initial conditions
of our simulations\DIFdelbegin \DIFdel{. We also }\DIFdelend  , and show a simple example of planet formation
based on the prescriptions introduced in this section\DIFdelbegin \DIFdel{, before presenting the results of the bulk numerical
simulations in the following section}\DIFdelend .

%
%
\subsection{Orbital evolution of protoplanets }\label{subsec:OrbEvol}

The orbital evolution of protoplanets is largely determined by their
interactions with the gas disc. However, there are a few different
definitions of the equation of motion used in the planetary literature.
 The difference is not only confusing but could lead to very different
dynamical outcomes \citep[e.g.][]{Ida20,Matsumura17}. 
Recently, \mbox{
\citet{Ida20}
}\hspace{0pt}
compared a few such equations in the literature and highlighted which
equations are appropriate under which conditions. Based on this work,
we will use a new, more appropriate equation of motion.

In \citet{Matsumura17}, we adopted the equation proposed by
\citet{Coleman14}:

\begin{equation}
\frac{d{\bf v}}{dt}=-\frac{{\bf v}}{\tau_{m}}-\frac{v_{r}}{\tau_{e}}{\bf e}_{r}-\frac{0.5\left(v_{\theta}-v_{K,a}\right)}{\tau_{e}}{\bf e}_{\theta}-\frac{\left(2.176v_{z}+0.871z\Omega_{K,a}\right)}{\tau_{i}}{\bf e}_{z}\,,\label{eq:CN14}
\end{equation}
but we replaced the `migration timescale' $\tau_{m}=-\dot{L}/L$
in the first term with the semi-major axis evolution timescale $\tau_{a}=-\dot{a}/a$.
The motivation behind replacing $\tau_{m}$ with $\tau_{a}$ in \citet{Matsumura17}
was to avoid an artificial outward migration at the high eccentricity,
supersonic regime. Here, $v_{K,a}$ \DIFdelbegin \DIFdel{is }\DIFdelend  and $\Omega_{K,a}$ are the Keplerian
orbital speed  and the corresponding angular speed evaluated at the
semi-major axis $a$\DIFdelbegin \DIFdel{and $\Omega_{K,a}$ is the Keplerian
angular speed at $a$, while }\DIFdelend  , while ${\bf v}$ is the planetary velocity evaluated
at the instantaneous orbital radius $r$, with $v_{r}$, $v_{\theta}$,
and $v_{z}$ \DIFdelbegin \DIFdel{are the }\DIFdelend  being its polar coordinate components \DIFdelbegin \DIFdel{of the planetary velocity that
are }\DIFdelend defined with the
unit vectors ${\bf e}_{r}$, ${\bf e}_{\theta}$, and ${\bf e}_{z}$,
respectively. 
%
\citet{Ida20} showed that the azimuthal component of
eccentricity damping is implicitly included in the first term of Equation
\ref{eq:CN14}, $-\frac{{\bf v}}{\tau_{m}}$, and its explicit addition
as the second term is not necessary.
\DIFdelbegin \DIFdel{Furthermore, \mbox{
\citet{Ida20} }\hspace{0pt}
}\DIFdelend  

In this work, we adopted the equation of motion proposed by \mbox{
\citet{Ida20}}\hspace{0pt}
,
since the equation describes planet-disc interactions well both in
subsonic and supersonic regimes compared to the other equations proposed
so far. They introduced an intuitive model of planet-disc interactions
based on dynamical friction by combining the work of \citet{Muto11}
in the supersonic regime with that of \citet{Tanaka04} in the subsonic
regime. \DIFdelbegin \DIFdel{In this work, we will apply
the equation of motion proposed by \mbox{
\citet{Ida20} }\hspace{0pt}
since the equation
best describes planet-disc interactions both in subsonic and supersonic
regimes among various equations proposed so far. }\DIFdelend The equation reads as follows:

\begin{eqnarray}
\frac{d{\bf v}}{dt} & = & -\frac{v_{K}}{2\tau_{a}}{\bf e}_{\theta}-\frac{v_{r}}{\tau_{e}}{\bf e}_{r}-\frac{v_{\theta}-v_{K}}{\tau_{e}}{\bf e}_{\theta}-\frac{v_{z}}{\tau_{i}}{\bf e}_{z}\,.\label{eq:EOM}
\end{eqnarray}
We note that $v_{K}$  is the Keplerian orbital speed evaluated at
the instantaneous orbital radius $r $. In this model, the evolution
timescales for the semi-major axis, eccentricity, and inclination are expressed
as follows:
\begin{align}
\tau_{a} & =  -\frac{t_{{\rm wave}}}{2h_{g}^{2}}\left[\frac{\Gamma_{L}}{\Gamma_{0}}\left(1-\frac{1}{\pi}\frac{\Gamma_{L}}{\Gamma_{0}}\sqrt{\hat{e}^{2}+\hat{i}^{2}}\right)^{-1} \right. \nonumber \\
&\qquad\qquad\left.+\frac{\Gamma_{C}}{\Gamma_{0}}\exp\left(-\frac{\sqrt{e^{2}+i^{2}}}{e_{f}}\right)\right]^{-1}\label{eq_taua},\\
\tau_{e} & =  \frac{t_{{\rm wave}}}{0.780}\,\left(1+\frac{1}{15}\left(\hat{e}^{2}+\hat{i}^{2}\right)^{3/2}\right)\label{eq_taue},\\
\tau_{i} & =  \frac{t_{{\rm wave}}}{0.544}\,\left(1+\frac{1}{21.5}\left(\hat{e}^{2}+\hat{i}^{2}\right)^{3/2}\right)\ ,\label{eq_taui}
\end{align}
where $h_{g}=H_{g}/a$ is the gas disc's aspect ratio, and eccentricities
and inclinations scaled with $h_{g}$ are defined as $\hat{e}=e/h_{g}$
and $\hat{i}=i/h_{g}$, respectively. The exponential factor in $\tau_{a}$
is introduced with $e_{f}=0.01+h_{g}/2$ following \citet{Fendyke14},
so that the contribution from the corotation torque disappears in
the supersonic regime. Also, $\Gamma/\Gamma_{0}$ represents a normalised
torque, and the subscripts $L$ and $C$ respectively correspond to Lindblad and
corotation torques from \citet{Paardekooper11}, with
the normalisation torque being

\begin{equation}
\Gamma_{0}=\left(\frac{M_{p}}{M_{*}}\right)^{2}\Sigma_{g}a^{4}\,h_{g}^{-2}\Omega_{K,a}^{2}.
\end{equation}
Finally, $t_{{\rm wave}}$ is the characteristic time of the orbital
evolution  by \citet{Tanaka04}: 
\begin{equation}
t_{{\rm wave}}=\left(\frac{M_{*}}{M_{p}}\right)\left(\frac{M_{*}}{\Sigma_{g}a^{2}}\right)h_{g}^{4}\Omega_{K,a}^{-1}\ ,
\end{equation}
where $M_{p}$ and $M_{*}$ are planetary and stellar masses, respectively,
and $\Sigma_{g}$ is the gas disc's surface mass density. We note that
$t_{{\rm wave}}$ is defined by using the semi-major axis $a$, not
the instantaneous orbital radius $r$, so the timescales defined in
Equations \ref{eq_taua}-\ref{eq_taui} are orbit-averaged timescales.

The above equations only hold for fully embedded type I migrators
\DIFdelbegin \DIFdel{.
Recently, \mbox{
\citet{Kanagawa18} }\hspace{0pt}
}\DIFdelend  and need to be modified for gap-opening type II migrators. For planet
migration, \mbox{
\citet{Kanagawa18a} }\hspace{0pt}
have that type II migration
is merely type I migration with the reduced surface mass density in
the gap\DIFdelbegin \DIFdel{and showed }\DIFdelend  , and showed that such migration is well described by the following equation:

\begin{equation}
\tau_{a}^{\prime}\simeq\frac{\Sigma_{g}}{\Sigma_{{\rm gap}}}\tau_{a}\simeq\left(1+0.04K\right)\tau_{a}\,,\label{eq:taua2}
\end{equation}
where $K=\left(\frac{M_{p}}{M_{*}}\right)^{2}\left(\frac{H_{g}}{a}\right)^{-5}\alpha_{{\rm turb}}^{-1}$.
\DIFdelbegin \DIFdel{Following this approach, we also scale }\DIFdelend  Since there is no consensus on eccentricity and
inclination damping in the type II regime, we adjusted eccentricity
and inclination damping timescales in a similar manner ($\tau_{e}^{\prime}=(1+0.04K)\tau_{e}$
and $\tau_{i}^{\prime}=(1+0.04K)\tau_{i}$) and \DIFdelbegin \DIFdel{adopt }\DIFdelend  implemented the equation
of motion \ref{eq:EOM} with timescales replaced by $\tau_{a}^{\prime}$,
$\tau_{e}^{\prime}$, and $\tau_{i}^{\prime}$ \DIFdelbegin \DIFdel{. Thus, in our model}\DIFdelend  in our code. In this
prescription, the eccentricity and inclination damping \DIFdelbegin \DIFdel{timescale }\DIFdelend  timescales
are always short compared to the migration timescale. This is an important
condition for the `eccentricity trap' \citep{Ogihara10}, where
planets can be resonantly trapped near the disc's inner edge. The
condition also avoids spurious outward migration at the disc's edge.

\begin{figure*}
\noindent %
\noindent\begin{minipage}[t]{1\columnwidth}%
\begin{center}
\includegraphics[width=0.9\paperwidth]{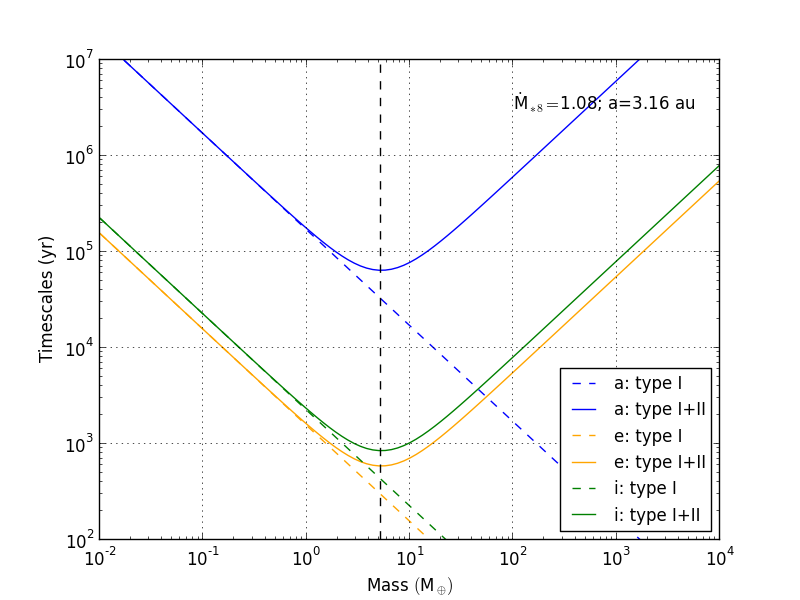} 
\par\end{center}%
\end{minipage}\caption{Comparison of evolution timescales of semi-major axis $\tau_{a}$ (blue),
eccentricity $\tau_{e}$ (orange), and inclination $\tau_{i}$ (green)
for a planet at 3.16 au with the circular and coplanar orbit in Disc
6. Solid and dashed blue lines are from Equations~\ref{eq:taua2}
and \ref{eq_taua}, respectively. The vertical black dashed line indicates
where migration timescale takes the minimum value (Eq.~\ref{eq_Mpcrit}).
\label{fig_typeItypeII} }
\end{figure*}

\DIFdelbegin \DIFdel{These timescales are plotted in }\DIFdelend 
Figure~\ref{fig_typeItypeII}  shows these timescales as a function
of a protoplanetary mass for a circular and coplanar case. The vertical
line indicates a critical mass $M_{{\rm crit}}$ above which \DIFdelbegin \DIFdel{a planet
}\DIFdelend  planet
migration switches from type I to type II regimes, and  it is defined
as the \DIFdelbegin \DIFdel{value giving }\DIFdelend  mass having the minimum timescale of Eq.~\ref{eq:taua2}:

\begin{eqnarray}
M_{{\rm crit}} & = & \left[\frac{1}{0.04}\alpha_{{\rm turb}}\left(\frac{H_{g}}{a}\right)^{5}\right]^{1/2}M_{*} \\
&\simeq& 9.3\left(\frac{\alpha_{{\rm turb}}}{10^{-4}}\right)^{1/2}\left(\frac{H/a}{0.05}\right)^{5/2}\left(\frac{M_{*}}{M_{\odot}}\right)\;M_{\oplus}.\label{eq_Mpcrit}
\end{eqnarray}
Since this critical mass gives $K=\frac{1}{0.04}$ and thus $\frac{\Sigma_{{\rm gap}}}{\Sigma_{g}}\simeq\frac{1}{1+0.04K}=\frac{1}{2}$,
the transition occurs when the gap depth is 50\% of the nominal surface
mass density \citep[also see][]{Johansen19}. The transition mass
is sensitive to the disc's aspect ratio, and we have $M_{{\rm crit}}=2.6\,M_{\oplus}$
for $\frac{H}{a}=0.03$, $\alpha_{{\rm turb}}=10^{-4}$, and $M_{*}=M_{\odot}$.

\subsection{Planet formation and disc evolution \label{subsec:PF_DiscEvol}}

The mass growth rate of a protoplanet $\dot{M}_{p}$ is written as follows:

\begin{equation}
\dot{M}_{p}=\dot{M}_{{\rm gas}}+\dot{M}_{{\rm core}},
\end{equation}
where \DIFdelbegin \DIFdel{$\dot{M}_{gas}$ }\DIFdelend  $\dot{M}_{{\rm gas}}$ and $\dot{M}_{{\rm core}}$ represent
gas and core accretion rates  of protoplanets, respectively. In this
study, we assume that a protoplanetary core grows via pebble accretion
and mutual collisions among \DIFdelbegin \DIFdel{cores}\DIFdelend  themselves, and we do not take the effect of planetesimal accretion into account. This choice is justified
when the pebble accretion is 3D (i.e. when the impact
radius of a core is small compared to the scale height of the pebble
disc $b<h_{p}$ ), because the timescale of 3D pebble accretion is
much shorter than that of the planetesimal accretion of the oligarchic
growth stage \citep{Ida16a}. The pebble and planetesimal accretion
timescales may become comparable to each other once the impact parameter
of the core becomes larger than the pebble disc's scale height $b\gg h_{p}$
\citep{Ida16a}.

The core growth rate  via pebble accretion is written as

\begin{equation}
\dot{M}_{{\rm core}}=\epsilon\dot{M}_{F},
\end{equation}
by using the \DIFdelbegin \DIFdel{pebble }\DIFdelend accretion efficiency $\epsilon$ and the pebble mass
flux $\dot{M}_{F}$. In the following, we describe the model of the
stellar mass accretion rate $\dot{M}_{*}$ and how $\dot{M}_{F}$
is related to it. We also discuss the choice of pebble accretion
efficiencies, and the pebble isolation masses. Finally, we present
the model of the gas accretion rate onto a protoplanetary core $\dot{M}_{{\rm gas}}$.

\subsubsection{Stellar mass flux $\dot{M}_{*}$ }

\DIFdelbegin \DIFdel{Following \mbox{
\citet{Ida16a}}\hspace{0pt}
, our disc
model assumes the disc}\DIFdelend  The evolution of a gas disc is represented by the stellar mass flux
$\dot{M}_{*}$, which is also related to the evolution of a dust disc
$\dot{M}_{F}$ as it is discussed in the following sub-section. In
this work, we used the same disc model as \mbox{
\citet{Matsumura17}}\hspace{0pt}
, which
is adopted from \mbox{
\citet{Ida16a}}\hspace{0pt}
. The disc's midplane temperature $T$
and the gas surface mass density $\Sigma_{g}$  are modelled as the
power-law functions of the orbital radius $r$ \DIFdelbegin \DIFdel{: }\DIFdelend  as $T\propto r^{-q}$
and $\Sigma_{g}\propto r^{-p}$. Since the inner disc is heated by
viscous dissipation while the outer disc is heated by the irradiation
from the central star \citep[e.g.][]{Hueso05,Oka11}, \DIFdelbegin \DIFdel{our }\DIFdelend  both $T$ and
$\Sigma_{g}$ are \DIFdelbegin \DIFdel{both }\DIFdelend dual power-law functions  in our disc model.

The disc evolution \DIFdelbegin \DIFdel{can be }\DIFdelend  is described by the diffusion equation for the
disc's surface mass density $\Sigma_{g}$:

\begin{equation}
\frac{\partial\Sigma_{g}}{\partial t}=-\frac{1}{r}\frac{\partial}{\partial r}\left[3r^{1/2}\frac{\partial}{\partial r}\left(\Sigma_{g}\nu r^{1/2}\right)\right]\ ,
\end{equation}
where $\nu=\alpha_{{\rm acc}}c_{s}H$ is the disc's `viscosity'
representing the accretion. \DIFdelbegin \DIFdel{Compared to the sophisticated wind-driven
accretion models by \citet{Bai16} and \citet{Suzuki16}, 
this corresponds
to approximating the combined effects of the wind-driven and turbulence-driven
accretions with a single effective $\alpha_{{\rm acc}}$, and ignoring
the effect of the wind mass loss. The equation has the following similarity
solution when the effective }\DIFdelend  Assuming that the accretion is steady
and that the effective viscosity is written as a power-law \DIFdelbegin \DIFdel{in
radius }\DIFdelend  function
in radius as $\nu\propto r^{p}$ \citep{LyndenBell74,Hartmann98}\DIFdelbegin \DIFdel{.
}\DIFdelend  ,
the mass accretion rate is related to the surface mass density as
follows:

\begin{equation}
\dot{M}_{*}\simeq3\pi\Sigma_{g}\nu=\frac{M_{D,0}}{2(2-p)t_{{\rm diff}}}\tilde{t}^{-(5/2-p)/(2-p)}\ .\label{eq:dotM}
\end{equation}

Here, $M_{D,0}$ is the initial disc mass, $\tilde{t}\equiv t/t_{{\rm diff}}+1$
\DIFdelbegin \DIFdel{so that the time is normalised
by the diffusion timescale:
}\DIFdelend  with the diffusion timescale of

\begin{equation}
t_{{\rm diff}}\equiv\frac{1}{3(2-p)^{2}}\frac{r_{D}^{2}}{\nu_{D}}\ ,
\end{equation}
where $r_{D}$ is the characteristic disc size \DIFdelbegin \DIFdel{which we take as the
disc's outer radius }\DIFdelend and $\nu_{D}$ is the
corresponding viscosity.
%
%
%

 As can be seen from the equation, there are several parameters
we could use to specify the disc evolution, \DIFdelbegin \DIFdel{namely, }\DIFdelend 
 such as initial values
of $\dot{M}_{*}$, $M_{D}$, and $\Sigma_{g,D;}$ as well as $r_{D}$,
$\alpha_{{\rm acc}}$, and $t_{{\rm diff}}$. However, they are dependent
on one another, and we need to choose three parameters out of these
to specify a disc model. For example, \DIFdelbegin \DIFdel{\mbox{
\citet{Ida04a} }\hspace{0pt}
}\DIFdelend  \mbox{
\citet{Ida08a} }\hspace{0pt}
chose $M_{D,0}$,
$t_{{\rm diff}}$, and $\alpha_{{\rm acc}}$ as control parameters
since they are relevant to planetesimal accretion as well as gap-opening
criteria. For pebble accretion, however, the disc radius $r_{D}$
is important because the pebble mass flux decreases sharply once the
pebble formation front reaches the outer disc radius \citep[e.g. ][]{Sato16}.
In \DIFdelbegin \DIFdel{our }\DIFdelend  this 
work, we considered a few different sets of $M_{D,0}$ and
$t_{{\rm diff}}$, along with $r_{D}=100\,{\rm au}$. The choice of
the characteristic disc size is rather arbitrary, but it is motivated
by a typical size of observed protoplanetary discs \citep[e.g.][]{Andrews10,Andrews07,Vicente05}.
The $\alpha_{{\rm acc}}$ \DIFdelbegin \DIFdel{representing the angular momentum transport
rate chiefly by the disc wind }\DIFdelend is calculated from these values as follows:

\begin{equation}
\alpha_{{\rm acc}}=\frac{h_{D}^{-2}}{6\pi\left(2-p\right)^{2}}\frac{t_{{\rm orb,D}}}{t_{{\rm diff}}},\label{eq:alpha_acc}
\end{equation}
where $h_{D}=H_{D}/r_{D}$ is the disc's aspect ratio at \DIFdelbegin \DIFdel{the outer
edge of the disc}\DIFdelend  $r_{D}$,
and $t_{{\rm orb,D}}$ is the corresponding orbital period. \DIFdelbegin \DIFdel{Note that,
as seen in Table \ref{tab:discs}, $\alpha_{{\rm acc}}$
defined with our choices of parameters ($t_{{\rm diff}}=0.1-10\,$Myr
and $r_{D}=100\,{\rm AU}$)is higher than
our default turbulent viscosity
alpha value of $\alpha_{{\rm turb}}=10^{-4}$. In Section \ref{subsec:singlePF}, 
we discuss the effects of other values of }\DIFdelend 
 In \citet{Matsumura17}, 
we defined the stellar mass accretion rate and the alpha parameter
independently of each other and calculated the surface mass density
based on these parameters and the disc temperature models in \mbox{
\citet{Ida16a}}. 
In this work, we instead calculated the stellar mass accretion rate
for given disc masses and diffusion timescales (and thus $\alpha_{{\rm acc}}$),
and we used these with the temperature model to determine the surface
mass density profile. In this sense, our current disc model is more
self-consistent compared to the previous one.

As discussed in Section \ref{sec:Introduction}, recent studies showed
that disc accretion is driven by the magnetic disc wind rather than
by the MRI turbulence \mbox{
\citep[e.g.][]{Bai13,Bai17}.}\hspace{0pt}
 The follow-up
studies proposed to replace the classical $\alpha$ disc model with
the one incorporating the disc wind \citep{Bai16,Suzuki16}, 
and the
model has been adopted by N-body studies such as \mbox{
\citet{Ogihara17}}\hspace{0pt}
.
Some other works, on the other hand, opt to use two-$\alpha$
disc models by representing the mass accretion driven mainly by the
disc wind, and the local disc evolution driven by turbulence with two
different alphas \citep[e.g.][]{Ida18,Johansen19,Bitsch19}.
In this work, we also adopted this two-$\alpha$ disc model for simplicity,
and we represent the wind-driven disc accretion with $\alpha_{{\rm acc}}$
and the disc turbulence with $\alpha_{{\rm turb}}$ . Compared to the
sophisticated wind-driven accretion models by \mbox{
\citet{Bai16} }\hspace{0pt}
and \mbox{
\citet{Suzuki16}}\hspace{0pt}
,
this corresponds to approximating the combined effects of the wind-driven
and turbulence-driven accretions with a single effective $\alpha_{{\rm acc}}$,
and ignoring the effect of the wind mass loss. 

\begin{figure*}
\begin{minipage}[t]{0.45\textwidth}%
\begin{center}
\includegraphics[width=0.45\paperwidth]{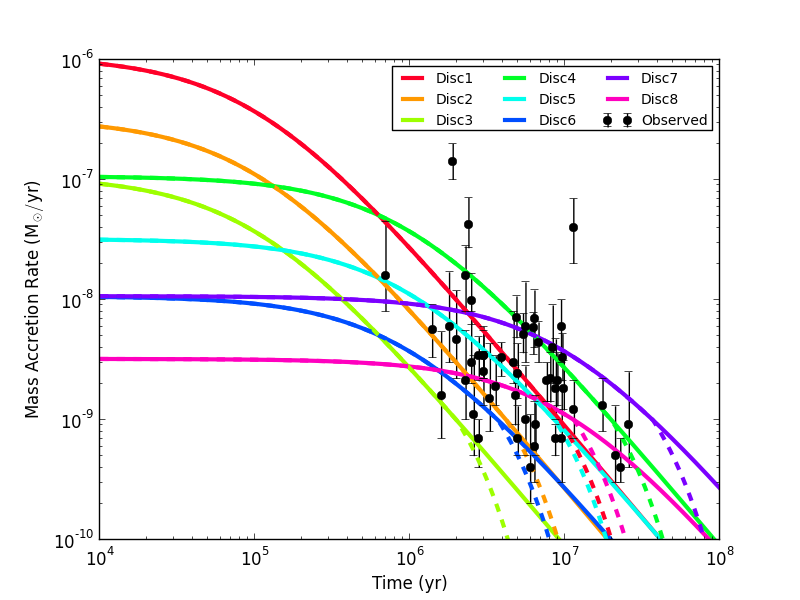} 
\par\end{center}%
\end{minipage}\hfill{}%
\begin{minipage}[t]{0.45\textwidth}%
\begin{center}
\includegraphics[width=0.45\paperwidth]{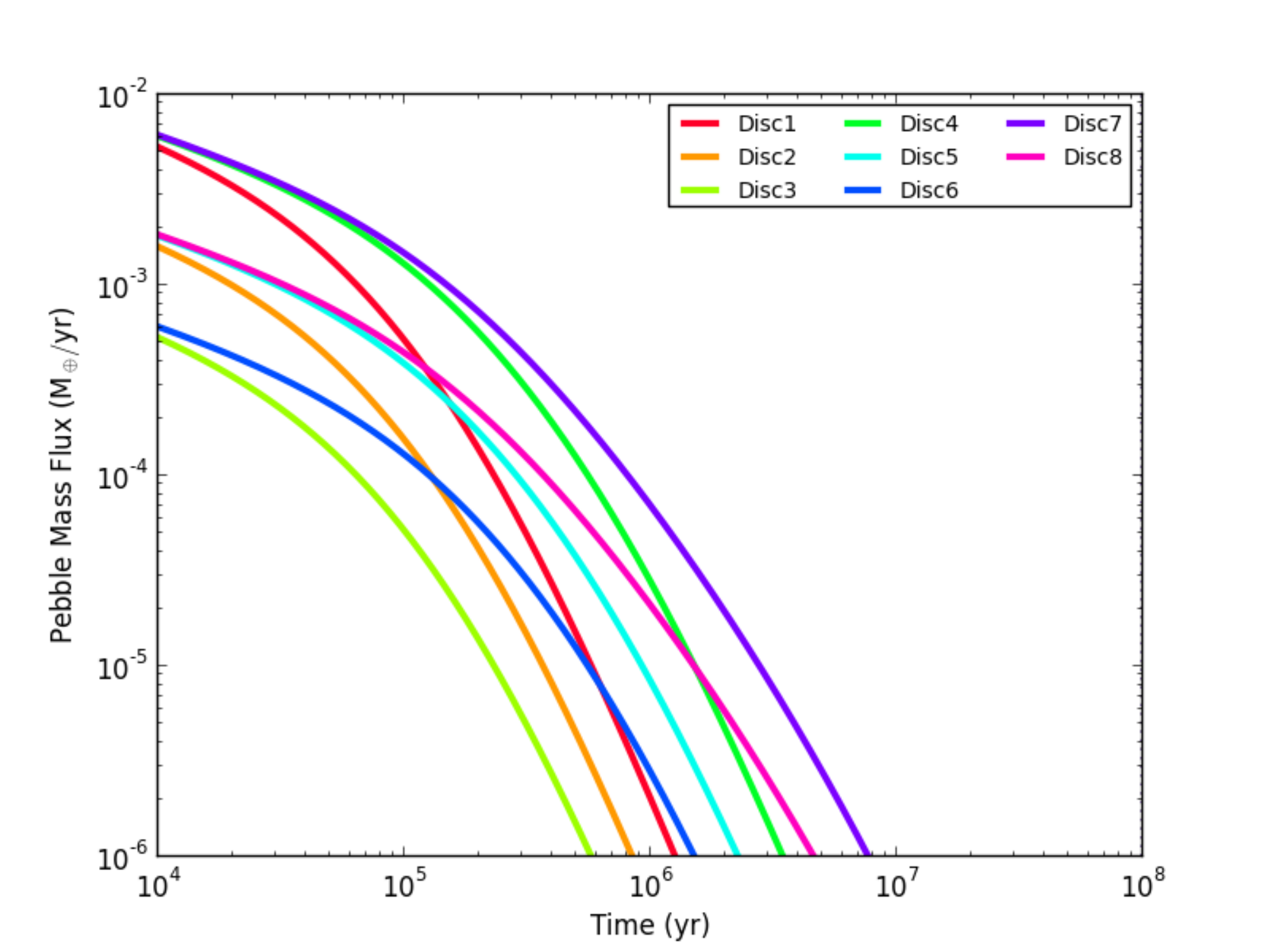} 
\par\end{center}%
\end{minipage}\caption{Left: Stellar mass accretion rates $\dot{M}_{*}$ for disc models
1-8 from Table \ref{tab:discs} (Eq \ref{eq:dotM}). The accretion
is exponentially reduced after the critical mass accretion rate of
$10^{-9}\,M_{\odot}/{\rm yr}$ is reached (dashed lines). The black
circles with error bars are observed stellar accretion rates from
\citep{SiciliaAguilar10} (data courtesy of Sicilia-Aguilar). Right:
Corresponding pebble mass fluxes $\dot{M}_{F}$ (Eq \ref{eq:dotMf_final}).
\label{fig:dotM}}
\end{figure*}

\DIFdelbegin \DIFdel{We }\DIFdelend  Finally, to choose our disc models, we constrained the combination
of $\left(M_{D,0},\,t_{{\rm diff}}\right)$ by using the observed
stellar mass accretion rates. Figure \ref{fig:dotM} shows eight disc
models using Equation \ref{eq:dotM}, along with the observed stellar
mass accretion rates from \citet{SiciliaAguilar10} (data courtesy
of Sicilia-Aguilar)\footnote{We note that we assume the solar mass for all of our simulations, while
observed data include GKM stars.}. Here, we did not attempt to find the best fit to all the data, as
this was done in \citet{Hartmann98}, but instead we tried to find a
set of reasonable $\left(M_{D,0},\,t_{{\rm diff}}\right)$ that \DIFdelbegin \DIFdel{goes
}\DIFdelend  makes
$\dot{M}_{*}$ go through at least some part of the distribution of
observed accretion rates. This is because, as \citet{Hartmann98}
pointed out, the best fit to all data does not necessarily represent
a typical evolution of the stellar mass accretion rates.

As can be seen in the figure, our chosen disc models tend to have
long lifetimes. Towards the oldest ages (about a few to several tens
of Myr), the mass accretion rate data are likely dominated by long-lived,
potentially less common protoplanetary discs. However, a recent study
shows that $\sim30\,\%$ of stars may have disc lifetimes longer than
10 Myr \citep{Pfalzner14}, making it possible that these long-lived
discs may not be so uncommon. Therefore, in this work, we consider
a relatively wide range of disc lifetimes. The combination of $\left(M_{D,0},\,t_{{\rm diff}}\right)$
and the corresponding $\alpha_{{\rm acc}}$ for eight disc models
are summarised in Table \ref{tab:discs}. To mimic the photoevaporation
effect, we reduced the mass accretion rate exponentially once it became
lower than the critical value of $10^{-9}\,M_{\odot}/{\rm yr}$. The
choice of this critical value is arbitrary, but we chose it to be close
to the minimum observed mass accretion rate seen in the figure.  We note
that, as seen in the table, $\alpha_{{\rm acc}}$ defined with our
choices of parameters ($t_{{\rm diff}}=0.1-10\,$Myr and $r_{D}=100\,{\rm AU}$)
is higher than our default turbulent viscosity alpha value of $\alpha_{{\rm turb}}=10^{-4}$,
which is consistent with our assumption that the mass accretion is
dominated by the disc's wind effects. In Section \ref{subsec:singlePF},
we discuss the effects of other values of $\alpha_{{\rm turb}}$.

\subsubsection{Pebble mass flux $\dot{M}_{F}$}

To determine the pebble accretion rate by a protoplanet, we need to
estimate the pebble mass flux. \DIFdelbegin \DIFdel{Pebbles are formed as dust particles
grow from micron sizes, potentially via collisional agglomeration
\mbox{
\citep[e.g.][]{Blum08}}\hspace{0pt}
. As dust particles grow, they become decoupled
from the gas disc. Since the gas disc revolves around the star slower
than dust particles when the pressure gradient is negative (i.e. $\Omega_{g}\simeq\Omega_{K}\left(1-\eta\right)<\Omega_{K}\sim\Omega_{d}$,
where $\eta=-\frac{h_{g}^{2}}{2}\mid\frac{\partial\ln P}{\partial\ln r}\mid$
and subscripts $g$ and $d$ correspond to gas and dust, respectively),
the dust particle experiences headwinds, loses angular momentum, and
migrates toward the central star \mbox{
\citep[e.g.][]{Adachi76,Weidenschilling77}}\hspace{0pt}
.
Since the pebbles start migrating when the migration timescale $t_{{\rm mig}}$
becomes short compared to the growth timescale $t_{{\rm grow}}$,
the pebble mass flux is mostly generated at the pebble formation front
where $t_{{\rm mig}}\sim t_{{\rm grow}}$.
}
%
\DIFdelend \citet{Lambrechts14b} wrote the pebble
mass flux as follows:

\begin{equation}
\dot{M}_{F}=2\pi r_{{\rm pf}}\Sigma_{p}\frac{dr_{{\rm pf}}}{dt},
\end{equation}
which describes the pebble mass swept up by the pebble formation front
per unit time. Here, $r_{{\rm pf}}$ is the radius of the pebble formation
front, and $\Sigma_{p}$ is the pebble surface mass density. Since
the dust particles are likely to convert to pebbles over a long period
of time, the dust surface mass density $\Sigma_{d}$ should
be different from that of pebbles $\Sigma_{p}$. At the pebble formation
front, however, we assume $\Sigma_{d}=\Sigma_{p}$ so that all
dust particles are converted to pebbles there.

Another effect we need to take into account is that the pebble mass
flux decreases drastically once the pebble formation front reaches
the outer disc radius \citep[e.g.][]{Sato16}. \citet{Ida19} fitted
the numerical simulations by \citet{Sato16} and proposed the following
relation:

\begin{equation}
\Sigma_{pg}=\Sigma_{pg,0}\left(1+\DIFdelbegin \DIFdel{\frac{t}{t_{{\rm pff}}}}\DIFdelend  \frac{t}{t_{{\rm pf}}}\right)^{-\gamma},\label{eq:Sigma_pg}
\end{equation}
where $\Sigma_{pg}=\Sigma_{p}/\Sigma_{g}$, the subscript 0 indicates
the initial value, $\gamma=1+\gamma_{a}\left(\frac{300\,{\rm au}}{r_{D}}\right)$
with $\gamma_{a}\sim0.15$, $r_{D}$ is the disc radius, and \DIFdelbegin \DIFdel{$t_{{\rm pff}}$
}\DIFdelend  $t_{{\rm pf}}$
is the time it takes for the pebble front to reach $r_{D}$.

In the Epstein regime (i.e. when a dust particle is small enough compared
to the mean free path: $R\lesssim\frac{9}{4}\lambda_{{\rm mfp}}$),
the growth timescale of the particle can be approximated as follows
by assuming that the relative velocities between particles are dominated
by turbulence \citep{Ida19}:

\begin{eqnarray}
t_{{\rm grow}} & \simeq & \frac{4}{\sqrt{3\pi}}\Sigma_{pg,0}^{-1}\Omega^{-1}\nonumber \\
 & = & \frac{200}{\sqrt{3\pi^{3}}}\left(\frac{\Sigma_{pg,0}}{0.01}\right)^{-1}\left(\frac{M_{*}}{M_{\odot}}\right)^{-1/2}\left(\frac{r}{{\rm au}}\right)^{3/2}\:{\rm yr}.
\end{eqnarray}
The growth timescale of pebbles from ${\rm \mu m}$ to cm sizes can
be estimated as $t_{{\rm grow,peb}}=\ln\frac{R_{{\rm peb}}}{R_{0}}\cdot t_{{\rm grow}}$
\citep{Lambrechts14b,Ida16a}, where $R_{0}$ and $R_{{\rm peb}}$
are the initial size of dust particles and the `final' size of
pebbles where they start to migrate, respectively. We adopted the nominal
values of $R_{{\rm peb}}\sim10\,{\rm cm}$ and $R_{0}\sim1\,{\rm \mu m}$
in the above equation.

From these, we can define the timescale for the pebble formation front
to reach the outer disc radius as follows:

\begin{equation}
t\DIFdelbegin \DIFdel{_{{\rm pff}}}\DIFdelend  _{{\rm pf}}\simeq\ln\frac{R_{{\rm peb}}}{R_{0}}\cdot\frac{200}{\sqrt{3\pi^{3}}}\left(\frac{\Sigma_{pg,0}}{0.01}\right)^{-1}\left(\frac{M_{*}}{M_{\odot}}\right)^{-1/2}\left(\frac{r_{D}}{{\rm au}}\right)^{3/2}\:{\rm yr}\label{eq:tpff}
.\end{equation}
Thus, the pebble front location is in turn given by

\begin{eqnarray}
\left(\DIFdelbegin \DIFdel{\frac{r_{{\rm pff}}}{{\rm au}}}\DIFdelend  \frac{r_{{\rm pf}}}{{\rm au}}\right) & \simeq & \left(\frac{1}{\ln\frac{R_{{\rm peb}}}{R_{0}}}\frac{\sqrt{3\pi^{3}}}{200}\right)^{2/3}\left(\frac{\Sigma_{pg,0}}{0.01}\right)^{2/3}\left(\frac{M_{*}}{M_{\odot}}\right)^{1/3}\left(\frac{t}{{\rm yr}}\right)^{2/3},
\end{eqnarray}
and we also have

\begin{eqnarray}
\frac{dr_{{\rm pf}}}{dt} & = & \left(\frac{1}{\ln\frac{R_{{\rm peb}}}{R_{0}}}\frac{\sqrt{3\pi^{3}}}{200}\right)^{2/3}\left(\frac{\Sigma_{pg,0}}{0.01}\right)^{2/3}\left(\frac{M_{*}}{M_{\odot}}\right)^{1/3} \nonumber \\
&&\cdot\frac{2}{3}\left(\frac{t}{{\rm yr}}\right)^{-1/3}\,{\rm {\rm au\,}yr^{-1}}.\label{eq:drpffdt}
\end{eqnarray}
By putting these together, the pebble mass flux is written as follows:

\begin{eqnarray}
\dot{M}_{F} & = & 2\pi r_{{\rm pf}}\Sigma_{g}\DIFdelbegin \DIFdel{\frac{\Sigma_{p}}{\Sigma_{g}}}\DIFdelend \frac{dr_{{\rm pf}}}{dt}\DIFdelbegin 
\DIFdel{= }
\DIFdel{2\pi r_{{\rm pf}}\Sigma_{g}\frac{dr_{{\rm pf}}}{dt}}\DIFdelend \cdot\Sigma_{pg,0}\left(1+\frac{t}{t_{{\rm pff}}}\right)^{-\gamma}.\label{eq:dotMf}
\end{eqnarray}

Finally, we calculated the pebble mass flux in our disc model. Since
the pebble formation front is in the outer disc, we used the gas surface
mass density in the irradiation region from \citet{Ida16a}:

\begin{eqnarray}
\Sigma_{g,{\rm irr}}&=&2514\left(\frac{T_{2}}{150\,{\rm K}}\right)^{-1}L_{*0}^{-2/7}M_{*0}^{9/14} \\ \nonumber
&&\cdot\alpha_{3}^{-1}\dot{M}_{*8}\left(\frac{r}{{\rm au}}\right)^{(q_{2}-3/2)}\;{\rm g/cm^{2}},\label{eq:Sigma_irr}
\end{eqnarray}
where $T_{2}$ is a characteristic disc temperature in the irradiation
region and $T_{2}=150\,$K in our default model, $q_{2}=3/7$ is the
power in $T\propto r^{-q}$, the $L_{*0}=L_{*}/L_{\odot}$ and $M_{*0}=M_{*}/M_{\odot}$
are the stellar luminosity and mass scaled to the solar values, $\alpha_{3}=\alpha_{{\rm acc}}/10^{-3}$
is the disc accretion $\alpha_{{\rm acc}}$ scaled to a typical value,
and $\dot{M}_{*8}=\dot{M}_{*}/\left(10^{-8}\,{\rm M_{\odot}/yr}\right)$.

Substituting Equations \ref{eq:Sigma_pg}, \ref{eq:tpff}-\ref{eq:drpffdt},
and \ref{eq:dotMf} into Equation \ref{eq:dotMf}, we obtain the following:

\begin{eqnarray}
\dot{M}_{F} & = & 2.066\times10^{-2}\left(\frac{\ln\frac{R_{{\rm peb}}}{R_{0}}}{\ln10^{4}}\right)^{-1}\left(\frac{T_{2}}{150\,{\rm K}}\right)^{-1} \nonumber \\
 & & L_{*0}^{-2/7}M_{*0}^{8/7}\alpha_{3}^{-1}\dot{M}_{*8} \left(\frac{\Sigma_{pg,0}}{0.01}\right)^{2}\left(\frac{r_{D}}{{\rm au}}\right)^{q_{2}-1}\nonumber \\
 & & \left(\frac{t}{t_{{\rm pff}}}\right)^{\frac{2}{3}\left(q_{2}-1\right)}\left(1+\frac{t}{t_{{\rm pff}}}\right)^{-\gamma}\,{\rm M_{\oplus\,}yr^{-1}}\,.\label{eq:dotMf_final}
\end{eqnarray}
The evolution of this equation is shown in the right panel of Figure~\ref{fig:dotM}.

In our simulations, we considered five different stellar metallicities:
${\rm [Fe/H]}=\left(-0.5,\,-0.3,\,0.0,\,0.3,\,0.5\right).$ Thus, we covered
a range of metallicities of planet-hosting stars that are related
to the dust-to-gas surface mass densities $\Sigma_{dg}=\Sigma_{d}/\Sigma_{g}$
as follows:

\begin{equation}
\frac{\Sigma_{dg}}{\Sigma_{dg,\odot}}\simeq10^{{\rm [Fe/H]}},
\end{equation}
where $\Sigma_{dg,\odot}=0.01$ is the value for the Solar System.
Thus, in our model, the pebble mass flux is proportional to the square
of the dust-to-gas surface mass density ratio.  The dependence is not
exactly the same, but similar to that of \mbox{
\citet{Lambrechts14b}, }\hspace{0pt}
where
$\dot{M}_{F}\propto\Sigma_{dg,0}^{5/3}$.

 The last column of Table \ref{tab:discs} shows the total masses of
pebbles accreting towards the star during the simulations for the solar
metallicity case ${\rm [Fe/H]}=0.0$. The total masses vary over $19.4-597\,M_{E}$
for ${\rm [Fe/H]}=0.0$ in our disc models, and they vary by a factor
of 3 above and below these values for the entire range of metallicities
we considered: $\sim6.5-1791\,M_{E}$. In comparison, \mbox{
\citet{Bitsch19}
}\hspace{0pt}
obtained total pebble masses of $70-700\,M_{E}$. 
\citet{Tychoniec18} estimated that dust disc masses of Class 0 objects vary over $10-6000\,M_{E,}$
where a typical value is $248\,M_{E}$ and less than 10\% of systems
have masses above $1000\,M_{E}$. Therefore, the range of total pebble
masses we considered are consistent with observations.
%

\subsubsection{Pebble accretion efficiency $\epsilon$}

The protoplanetary cores are exposed to the flux of pebbles, but they
only accrete a fraction of the incoming flux. As stated before, we
write the pebble accretion rate \DIFadd{as $\dot{M}_{{\rm core}}=\epsilon\dot{M}_{F}$}\DIFaddend. 
%
\citet{Matsumura17} adopted the accretion efficiency defined by \citet{Ida16a},
which is valid in the settling regime:

\begin{equation}
\epsilon_{{\rm IGM16}}={\rm min}\left(\frac{C_{\epsilon}\zeta^{-1}\chi b^{2}}{4\sqrt{2\pi}\tau_{s}h_{p}}\left(1+\frac{3b}{2\chi\eta}\right),\,1\right)\ \label{eq:epsIda16}
.\end{equation}
The $1$ in parentheses ensures that the accretion efficiency does
not become larger than unity. In this equation, $\zeta$ and $\chi$
are functions of the Stokes number $\tau_{s}$ and defined as

\begin{eqnarray}
\chi & = & \frac{\sqrt{1+4\tau_{s}^{2}}}{1+\tau_{s}^{2}},\\
\zeta & = & \frac{1}{1+\tau_{s}^{2}}\ .
\end{eqnarray}
The pebble aspect ratio is defined as $h_{p}=H_{p}/r,$ where $H_{p}$
is the pebble scale height \citep{Ida16a}:

\begin{equation}
H_{p}\simeq\left(1+\frac{\tau_{s}}{\alpha_{{\rm turb}}}\right)^{-1/2}H_{g}.\label{eq:Hp}
\end{equation}
Similarly, $b=B/r$ and $B$ is an impact parameter for a pebble passing
by the protoplanet with mass $M_{p}$:

\[
B\simeq\min\left(\sqrt{\frac{3\tau_{s}^{1/3}R_{H}}{\chi\eta r}},\,1\right)\cdot2\kappa _{{\rm OK12}}\tau_{s}^{1/3}R_{H}\,,
\]
where the left and the right terms in parentheses correspond to the Bondi
and Hill regimes, respectively. The $R_{H}=\left(\frac{M_{p}}{3M_{*}}\right)^{1/3}r$
is the Hill radius of the protoplanet, and the \DIFdelbegin \DIFdel{$\kappa$ }\DIFdelend  $\kappa_{{\rm OK12}}$
is a reduction factor of the accretion proposed by \citet{Ormel12}
for $\tau_{s}\gg1$ as:

\[
\kappa _{{\rm OK12}}=\exp\left(-\left(\frac{\tau_{s}}{\min\left(2,\,\tau_{s}^{*}\right)}\right)^{0.65}\right),
\DIFdelbegin \DIFdel{,
}\DIFdelend \]
\DIFdelbegin \DIFdel{where }\DIFdelend  with $\tau_{s}^{*}=4\left(M_{p}/M_{*}\right)/\eta^{3}$. Also, $C_{\epsilon}$
is defined as

\begin{equation}
C_{\epsilon}=\min\left(\sqrt{\frac{8}{\pi}}\frac{h_{p}}{b},\,1\right)\ ,
\end{equation}
where the left and right terms represent 2D and 3D accretion, respectively.
For the simulations, we scaled $C_{\epsilon}$ by a factor $C_{\epsilon,i}$
to take account of the orbital inclination effect. By assuming that
the pebble volume density scales as $\exp\left(-z^{2}/\left(2\,H_{p}^{2}\right)\right)$,
the surface mass density between $z\pm B$ scales with:

\begin{eqnarray}
&&\int_{z-B}^{z+B}\exp\left(-\frac{z^{\prime2}}{2\,H_{p}^{2}}\right)dz^{\prime}= \nonumber \\
&&\frac{\sqrt{2\pi}H_{p}}{2}\left({\rm erf\left(\frac{z+B}{\sqrt{2}H_{p}}\right)-{\rm erf\left(\frac{z-B}{\sqrt{2}H_{p}}\right)}}\right).
\end{eqnarray}
By normalising this factor with the value at $z=0$, we obtain the following:

\begin{equation}
C_{\epsilon,i}=\frac{1}{2}\frac{\left({\rm erf\left(\frac{z+B}{\sqrt{2}H_{p}}\right)-{\rm erf\left(\frac{z-B}{\sqrt{2}H_{p}}\right)}}\right)}{{\rm erf}\left(\frac{B}{\sqrt{2}H_{p}}\right)}.
\end{equation}
More recently, \citet{Ormel18} studied the 3D pebble accretion efficiency
by considering the effects of eccentricity, inclination, and
disc turbulence:

\begin{equation}
\epsilon_{{\rm OL18}}=\frac{A_{3}}{\eta h_{{\rm p,eff}}}\left(\frac{M_{p}}{M_{*}}\right)f_{{\rm set}}^{2}\ ,\label{eq:epsOL18}
\end{equation}
where $A_{3}=0.39$ is a fitting constant to their pebble accretion
simulations. The effective pebble aspect ratio is defined as

\begin{equation}
h_{{\rm p,eff}}\sim\sqrt{h_{p}^{2}+\frac{\pi\,i_{p}^{2}}{2}\left(1-\exp\left[-\frac{i_{p}}{2h_{p}}\right]\right)}\ ,
\end{equation}
where $i_{p}$ is the inclination of a planetary orbit. The settling
fraction is determined by integrating the accretion probability over
the velocity distribution and is written as follows for turbulence
operating only in the vertical direction:

\begin{equation}
f_{{\rm set}}=\exp\left[-a_{{\rm set}}\left(\frac{\Delta v_{y}^{2}}{v_{*}^{2}}+\frac{\Delta v_{z}^{2}}{v_{*}^{2}+a_{{\rm turb}}\sigma_{P,z}^{2}}\right)\right]\frac{v_{*}}{\sqrt{v_{*}^{2}+a_{{\rm turb}}\sigma_{P,z}^{2}}}\ ,
\end{equation}
where $a_{{\rm set}}=0.5$ and $a_{{\rm turb}}=0.33$ are other fit
constants, $\Delta v_{y}$ and $\Delta v_{z}$ are azimuthal and vertical
approach velocities, respectively, and $\sigma_{P,z}$ is the vertical
component of pebble root mean square velocity.

\begin{figure*}
\begin{minipage}[t]{0.45\textwidth}%
\begin{center}
\includegraphics[width=0.45\paperwidth]{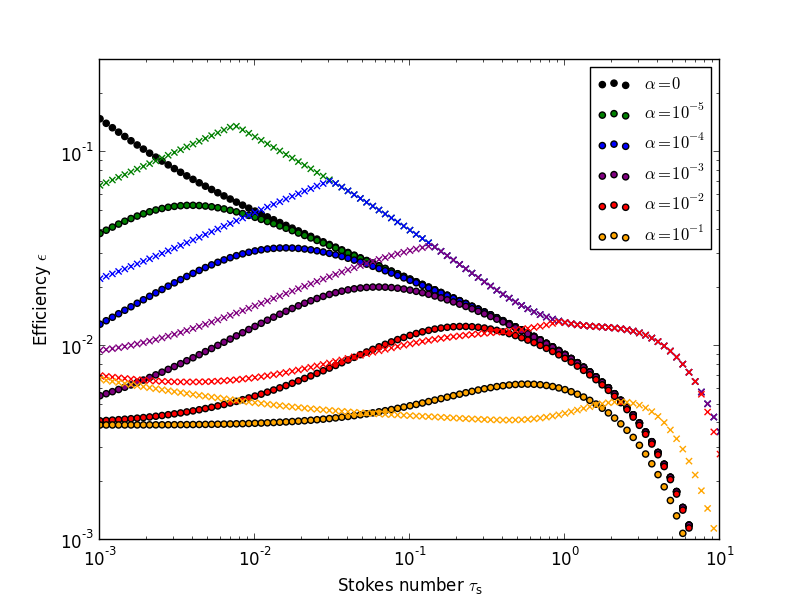} 
\par\end{center}%
\end{minipage}\hfill{}%
\begin{minipage}[t]{0.45\textwidth}%
\begin{center}
\includegraphics[width=0.45\paperwidth]{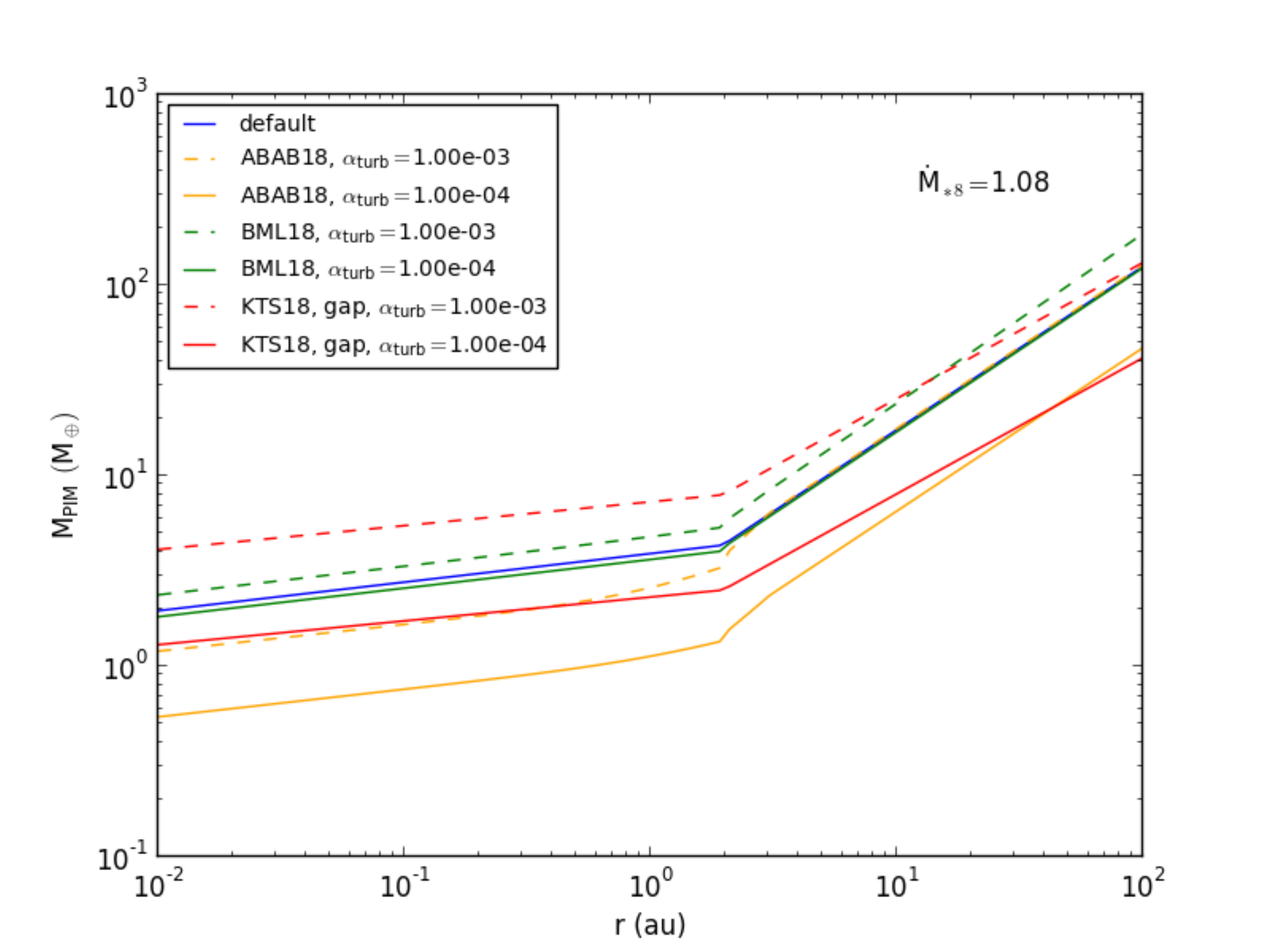} 
\par\end{center}%
\end{minipage}

\caption{Left: Pebble accretion efficiency $\epsilon$ and the Stokes number
for different values of the vertical turbulence strength $\alpha_{{\rm turb}}$.
The figure uses the planet-to-star mass ratio of $3\times10^{-7}$
(i.e. $\sim0.1\,M_{E}$ for a Sun-like star), the gas disc aspect
ratio of \DIFdelbeginFL \DIFdelFL{$\hat{h}=0.03$}\DIFdelendFL $h_{g}=0.03$, and the disc radial pressure gradient of $\eta=1.0\times10^{-3}$
as in Figure 4 of \citet{Ormel18}. Circles show $\epsilon$ from
\citet{Ormel18} and crosses show corresponding $\epsilon$ from \citet{Ida16a}.
Right: Pebble isolation masses estimated by \citet{Ataiee18} (orange)
and \citet{Bitsch18} (green), compared with our default case from \citet[blue]{Ida16a}.
Also plotted are the critical migration transition
masses (Equation \ref{eq_Mpcrit}, red). \label{fig:eps}}
\end{figure*}

The left panel of Figure \ref{fig:eps} compares the accretion efficiencies
for a $\sim0.1M_{\oplus}$ planet from \citet{Ida16a} (crosses) and
from \citet{Ormel18} (circles) as a function of $\tau_{s}$. The estimated
accretion efficiencies are similar for \citet{Ida16a} and \citet{Ormel18}
over a wide range of $\tau_{s}$, though the values from \citet{Ida16a}
are generally slightly higher than those from \citet{Ormel18}. In
this work, we tested both types of the accretion efficiencies,
and confirmed that the general trends are similar in both cases except
that the mass growth is less efficient in the cases of \citet{Ormel18}.
In this paper, we focus on the simulations with the efficiency
by \citet{Ida16a} $\epsilon_{{\rm IGM16}}$, but we discuss the effects
of that by \citet{Ormel18} $\epsilon_{{\rm OL18}}$ briefly in Section
\ref{subsec:epsOL18}.

\subsubsection{Pebble isolation mass}

The pebble isolation mass (PIM) is the protoplanetary
mass at which the embryo becomes large enough to perturb the gas disc,
modify the pressure gradient so that pebbles locally feel a tailwind
rather than a headwind, and thus to halt the inward radial drift of pebbles
\citep{Morbidelli12}. This critical mass marks the end of the pebble
accretion stage and thus is one of the key parameters to determine
the outcome of planet formation. There are different formulations
for the PIM in the literature, which we briefly discuss below.

\citet{Lambrechts14a} estimated the PIM as $M_{{\rm iso}}\sim20\left(\frac{a}{5\,{\rm au}}\right)^{3/4}M_{\oplus}$
from hydrodynamic simulations, which we adopted for \citet{Matsumura17}.
However, this study did not take account of the effects of the disc's
turbulent viscosity. More recently, \citet{Ataiee18} and \citet{Bitsch18}
extended this study and investigated the dependence of the PIM on
the disc aspect ratio, the pressure gradient of the disc as well as
the viscosity. \citet{Ataiee18} used 2D gas and dust hydrodynamic simulations,
while \citet{Bitsch18} used 3D gas-only hydrodynamic simulations
with 2D integrations of particle trajectories. As discussed in \citet{Ataiee18}
and also shown in Figure \ref{fig:eps}, the two studies \DIFdelbegin \DIFdel{show similar
}\DIFdelend  have similar
PIM trends, but details are different. Moreover, the differences become
larger for the lower values of the viscosity $\alpha_{{\rm turb}}$.

\citet{Bitsch18} performed 3D hydro simulations over $\alpha_{{\rm turb}}=\left[2\times10^{-4},\:6\times10^{-3}\right]$
to determine the PIM as follows:

\begin{align}
M_{{\rm PIM,B18}} & \simeq  \left(25+10.5\frac{\alpha_{{\rm turb}}}{\tau_{s}}\right)f_{{\rm fit}}M_{\oplus},\\
f_{{\rm fit}} & =  \left(\frac{h_{g}}{0.05}\right)^{3}\left(0.34\left(\frac{\log10^{-3}}{\log\alpha_{{\rm turb}}}\right)^{4}+0.66\right)\nonumber \\
&\quad \left(1-\frac{\frac{\partial\ln P}{\partial\ln r}+2.5}{6}\right).
\end{align}
A potential problem with this fit formula is that the saturation
might not quite have been achieved for the lower end of $\alpha_{{\rm turb}}$,
as indicated in Figure 3 of \citet{Johansen19}.

On the other hand, \citet{Ataiee18} performed 2D hydro simulations
over $\alpha_{{\rm turb}}=\left[5\times10^{-4},\:1\times10^{-2}\right]$
and proposed the following PIM:

\begin{align}
M_{{\rm PIM,A18}} & \simeq  h_{g}^{3}\sqrt{37.3\alpha_{{\rm turb}}+0.01}\nonumber \\
&\quad \cdot\left(1+0.2\left(\frac{\sqrt{\alpha_{{\rm turb}}}}{h_{g}}\sqrt{\frac{1}{\tau_{s}^{2}}+4}\right)^{0.7}\right)M_{*.}\label{eq:PIM_A18}
\end{align}
As seen in their Figure 8, the expression agrees well with the 2D hydro simulations in the low-viscosity limit ($\alpha_{{\rm turb}}\sim10^{-3}$),
while the difference is about a factor of 3 for $\alpha_{{\rm turb}}\sim10^{-2}$.

The right panel of Figure \ref{fig:eps} compares the pebble isolation
mass model by \citet{Ataiee18} with that of \citet{Bitsch18}. The
figure also shows the transition mass from type I to type II migration
as shown in the previous sub-section. Since the PIM measures the first
appearance of the zero pressure gradient outside the planetary gap,
the required depth is about $15\%$ \citep{Lambrechts14a} and is shallower
than that for the migration transition of $\sim50\%$ \citep{Johansen19}.
In other words, the planets are expected to first obtain the PIM and
then reach the migration transition mass via gas accretion (or planetesimal
accretion or core-core collisions). This implies that, unless the
gas accretion is fast enough, the protoplanets are likely to migrate
toward the central star on type I migration timescale \citep{Johansen19}.
However, since the PIM decreases towards the central star, it is likely
for such \DIFdelbegin \DIFdel{cores undergoing type I migrating }\DIFdelend  type I migrators to eventually start gas accretion, and for migration
to slow down as a result.

\DIFdelbegin \DIFdel{For the standard disc model, we have compared the PIMs by \mbox{
\citet{Ataiee18}
}\hspace{0pt}
and \mbox{
\citet{Bitsch18} }\hspace{0pt}
with the migration transition mass. }\DIFdelend The PIM by \citet{Bitsch18} is lower than the migration transition
mass within $\sim10\,$au for $\alpha_{{\rm turb}}=1\times10^{-3}$
(see dashed green and dashed red lines, respectively), but it is higher
for $\alpha_{{\rm turb}}=1\times10^{-4}$ (see solid green and solid
red lines, respectively). The latter is counter-intuitive, since it
indicates that \DIFdelbegin \DIFdel{the gap depth deeper than
that required for the migration transition is necessary to reach }\DIFdelend  a deeper gap is required for migration transition than
for reaching the PIM. On the other hand, for the PIM model by \citet{Ataiee18},
the PIM is generally lower than the migration transition mass. Thus,
for this work, we adopted the PIM by \citet{Ataiee18}\footnote{As seen in the figure, there is a possibility that the required gap
depth becomes deeper than $50\%$ for the PIM by \citet{Ataiee18}
as well. However, we confirm that within our simulation parameters,
the effects are not important.}.

\subsubsection{Gas accretion onto a protoplanet \label{subsec:GasAccretion}}

The rapid gas accretion onto a protoplanet starts when a protoplanetary
core becomes massive enough so that the (quasi-)\DIFdelbegin \DIFdel{hyrdostatic gas pressure
cannot }\DIFdelend  hydrostatic gas pressure
can no longer support the envelope against the gravity. Following
\citet{Ikoma00}, we assume this critical core mass for gas accretion
is

\begin{equation}
M_{{\rm core,crit}}\simeq10\left[\left(\DIFdelbegin \DIFdel{\frac{\dot{M}_{{\rm core}}}{10^{-6}{\rm M_{\oplus}/yr}}}\DIFdelend  \frac{\dot{M}_{{\rm core}}}{10^{-6}{\rm M_{E}/yr}}\right)\left(\frac{\kappa}{1\,{\rm cm^{2}/g}}\right)\right]^{s}M\DIFdelbegin \DIFdel{_{\oplus}}\DIFdelend  _{E}, \label{eq:crit_core}
 \end{equation}
where $\dot{M}_{{\rm core}}$ is the core accretion rate, $s=0.2-0.3$,
and $\kappa$ is the grain opacity. For our simulations, we adopted
$s=0.25$ and $\kappa=1\,{\rm cm^{2}/g}$, respectively. The equation
assumes \DIFdelbegin \DIFdel{the planetesimal accretion heating up the envelope, }\DIFdelend  that the envelope is heated by the planetesimal accretion,
and it is \DIFdelbegin \DIFdel{unclear whether the }\DIFdelend  not immediately clear whether pebble accretion would provide
a comparable heating. \DIFdelbegin \DIFdel{However, we will adopt this critical core mass for the lack
of a better model . }\DIFdelend  Recently, \mbox{
\citet{Ogihara20} }\hspace{0pt}
explored this topic
by using the 1D hydrostatic model with pebble accretion rates over
$10^{-12}-10^{-2}\,M_{E}/{\rm yr}$. They obtained the fitting function:
$M_{{\rm crit}}=13\left(\frac{\dot{M}_{F}}{10^{-6}{\rm M_{E}/yr}}\right)^{0.23}\,M_{E}$,
which is consistent with the equation adopted here. 
We only became aware of their work while this manuscript was under revision, and thus 
we adopted Equation \ref{eq:crit_core} instead of theirs. 
The model
implies that the gas accretion starts once the core accretion stops
either because the PIM is reached or because the pebble flux runs
out.

The gas accretion onto a protoplanet is limited both by how quickly
a protoplanet can accrete gas and by how quickly a disc can supply
gas to the protoplanet \citep[e.g.][]{Ida18}. The former  condition
is given by the Kelvin-Helmholtz (KH) timescale $\tau_{{\rm KH}}$
as follows:

\begin{equation}
\frac{dM_{{\rm KH}}}{dt}\simeq\frac{M_{p}}{\tau_{{\rm KH}}},
\end{equation}
where

\begin{equation}
\tau_{{\rm KH}}=10^{b}\left(\frac{M_{p}}{M_{\oplus}}\right)^{-c}\left(\frac{\kappa}{1\,{\rm cm^{2}/g}}\right)\,{\rm yrs}.\label{eq:tauKH}
\end{equation}
Parameters $b$ and $c$ are not well constrained. \citet{Ikoma00}
proposed $\left(b,\,c\right)=\left(8,\,2.5\right)$ based on their
numerical simulations, while \citet{Ida04a} suggested that the parameters
could take values up to $\left(b,\,c\right)=\left(10,\,3.5\right)$
depending on different opacity tables, and they adopted $\left(b,\,c\right)=\left(9,\,3.0\right)$
for their work. For this work, we adopted $\left(b,\,c\right)=\left(8,\,3\right),$
though we also tested other combinations. The KH timescale becomes
shorter for smaller $b$ and larger $c$, and shorter timescales lead
to more massive final giant planet masses. However, beyond a certain
point, the fast KH timescale does not help the growth any more because
the gas accretion is limited by the efficiency of gas supply by the
disc.

\DIFdelbegin \DIFdel{Nominally}\DIFdelend  Regarding the latter condition, the disc  nominally supplies gas at
a rate $\dot{M}_{*}$ (see Equation \ref{eq:dotM}). For an actively
accreting protoplanet, however, we need to take account of the reduced
surface mass density effect as proposed by \citet{Tanigawa16}:

\begin{equation}
\dot{M}_{{\rm TT16}}=\frac{\Sigma_{{\rm gap}}}{\Sigma_{g}}\dot{M}_{*}\simeq\frac{1}{\left(1+0.04K\right)}\dot{M}_{*}.\label{eq:M_TT16}
\end{equation}
Putting these together, we adopted the following gas accretion rate
as in \citet{Ida18}:

\begin{equation}
\dot{M}_{{\rm gas}}\simeq{\rm min}\left[\dot{M}_{{\rm KH}},\;\dot{M}_{*},\;\dot{M}_{{\rm TT16}}\right].\label{eq:Mgas}
\end{equation}

\subsection{Initial conditions \label{subsec:Initial-Conditions}}

\begin{table*}[h!]
\centering
\begin{tabular}{|c|c|c|c|c|} \hline 
 & $M_{d,0}\,(M_{\odot})$  & $t_{{\rm diff}}\,({\rm Myr})$  & $\alpha_{{\rm acc}}$  & $M_{{\rm peb,tot}}\,(M_{\oplus})$ \tabularnewline
\hline 
\hline 
Disc 1  & $0.2$  & $0.1$  & $\sim7.4\times10^{-2}$ & $194$ \tabularnewline
\hline 
Disc 2  & $0.06$  & $0.1$  & $\sim7.4\times10^{-2}$ & $58.2$ \tabularnewline
\hline 
Disc 3  & $0.02$  & $0.1$  & $\sim7.4\times10^{-2}$ & $19.4$ \tabularnewline
\hline 
Disc 4  & $0.2$  & $1$  & $\sim7.4\times10^{-3}$ & $446$ \tabularnewline
\hline 
Disc 5  & $0.06$  & $1$  & $\sim7.4\times10^{-3}$ & $134$ \tabularnewline
\hline 
Disc 6  & $0.02$  & $1$  & $\sim7.4\times10^{-3}$ & $44.5$ \tabularnewline
\hline 
Disc 7  & $0.2$  & $10$  & $\sim7.4\times10^{-4}$ & $597$ \tabularnewline
\hline 
Disc 8  & $0.06$  & $10$  & $\sim7.4\times10^{-4}$ & $179$ \tabularnewline
\hline 
\end{tabular}

\caption{Disc models 1-8. Columns 2, 3, and 4 show the initial disc mass,
the disc diffusion timescale, and the corresponding viscosity $\alpha_{{\rm acc}}$,
respectively. The last column is the total pebble mass throughout
the simulations for ${\rm [Fe/H]}=0.0$.  For all the simulations,
the turbulent viscosity $\alpha_{{\rm turb}}=10^{-4}$ is used, which
ensures $\alpha_{{\rm turb}}\ll\alpha_{{\rm acc}}$ (however, see Section
\ref{subsec:singlePF} for the cases using other values). }
\label{tab:discs} 
\end{table*}

In this work, we considered eight different disc models  around
a Sun-like star as shown in Table \ref{tab:discs}\DIFdelbegin \DIFdel{around a Sun-like star}\DIFdelend , which lead to
the stellar mass accretion rates and the corresponding pebble accretion
rates as in Figure \ref{fig:dotM}. For each of these discs, we
assumed five different stellar metallicities of ${\rm [Fe/H]}=-0.5,\,-0.3,\,0.0,\,0.3,\,{\rm and\,}0.5$.
As stated earlier, we assume that the global angular momentum
transfer is carried out mainly by the disc wind with $\alpha_{{\rm acc}}$
shown in the table, while the local disc-planet interactions \DIFdelbegin \DIFdel{is }\DIFdelend  are controlled 
by the disc's turbulent viscosity constant $\alpha_{{\rm turb}}=10^{-4}$ 
(see Section \ref{subsec:singlePF} for other values).

\begin{figure*}
\noindent %
\noindent\begin{minipage}[t]{1\columnwidth}%
\begin{center}
\includegraphics[width=0.9\paperwidth]{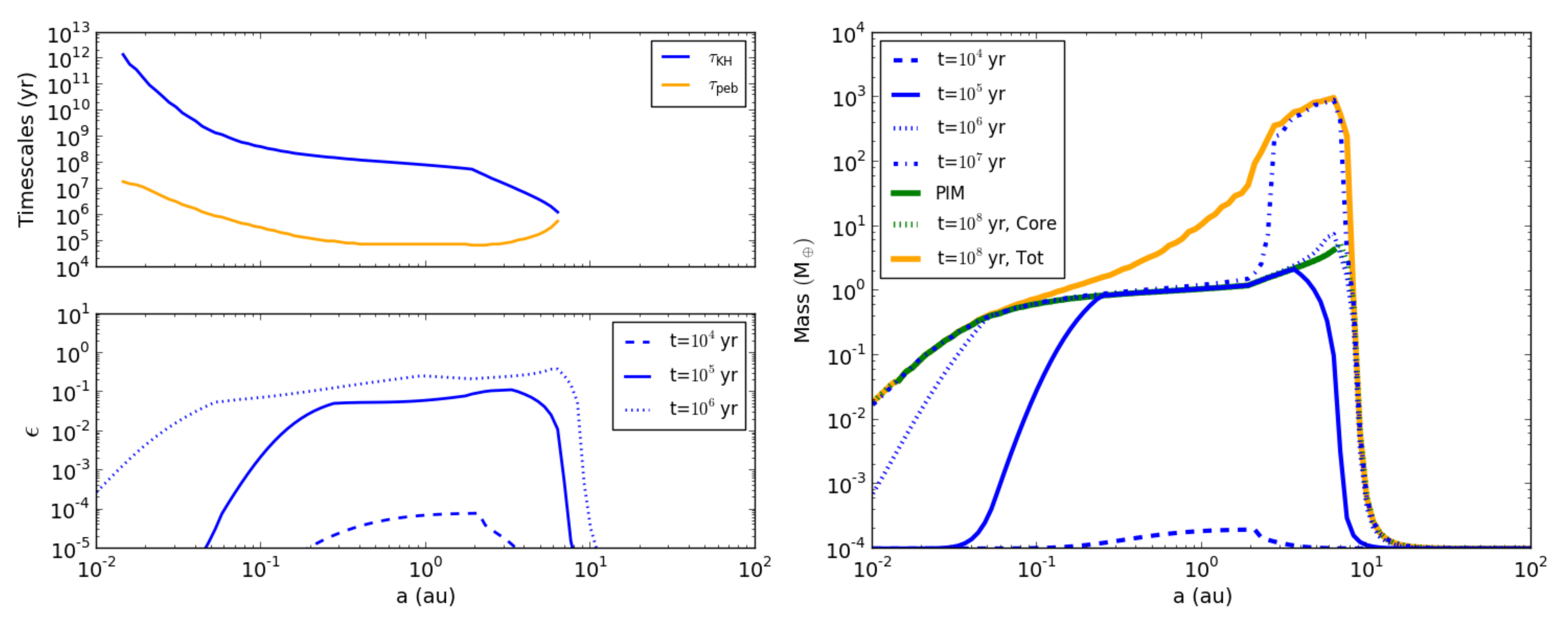} 
\par\end{center}%
\end{minipage}

\caption{Top left panel compares the total time to form the protoplanetary
core with the pebble isolation mass (orange, $\tau_{{\rm peb}}$)
with the Kelvin-Helmholtz gas accretion timescale for the PIM core
(blue, $\tau_{{\rm KH}}$). The right panel shows the planetary mass
growth at different times starting from $10^{-4}\,M_{E}$ in Disc
5. The solid green line shows the PIM, while the dotted green line
represents the core masses at 100 Myr (i.e. at the end of the simulations)
for the cores not reaching PIMs. The orange line shows the final total
mass. Here, the effects of the snow line are ignored. The bottom left
panel shows the corresponding pebble accretion efficiencies by \citet{Ida16a} 
at different times as a core grows from $10^{-4}\,M_{E}$.\label{fig:insituPF}}
\end{figure*}

Figure \ref{fig:insituPF} shows the outcome of \emph{\emph{in situ}} planet
formation (i.e. no migration) in Disc 5 with the solar metallicity,
starting with a protoplanetary core mass of $10^{-4}M_{\oplus}$.
This initial mass is comparable to a Ceres mass, which is a characteristic
mass for a planetesimal formed via streaming instability \citep{Johansen15,Simon16}.
Differently from simulations presented in Section \ref{sec:Results},
the effect of the snowline is ignored here for simplicity. The top
left panel compares the Kelvin-Helmholtz gas accretion timescale for
the PIM core (orange) with the core formation timescale (i.e. the
total time required for a core to reach the PIM, blue). The total
in situ formation timescale of a gas giant planet will be comparable
to the sum of these two timescales. For example, although the core
accretion time is the shortest around 1 au, the PIM core there is
too small to accrete gas efficiently. Thus, in our model, it is likely
difficult to form a gas giant in situ at around 1 au, and the gas
giant planet formation timescale becomes shortest around several
au.

The right panel shows the total mass  of a protoplanet reached after
100 Myr across the disc (orange), along with the total mass \DIFdelbegin \DIFdel{of a protoplanet }\DIFdelend at different
times  of the growth calculations (blue curves). The solid green curve
shows the PIM, while the dotted green curve shows the final core masses
in the regions where the planetary masses never reach the PIMs. The
figure indicates that protoplanets achieve the PIMs in most parts of the disc 
within a few au by $\sim10^{5}\,$yr and within $\sim10\,$au by $\sim10^{6}\,$yr.
In this disc model, planetary growth is inefficient beyond $\sim10\,$au,
because the accretion cross-section and thus the pebble accretion
efficiency sharply decreases for a low-mass protoplanet in the outer
disc. The corresponding pebble accretion efficiencies are shown in
the bottom left panel. Although this figure is made based on the accretion
efficiency by \citet{Ida16a}, the trend is similar with the efficiency
by \citet{Ormel18}.

 We compiled similar figures for other discs and found that gas giant
planets do not form in situ in our disc models beyond $\sim20\,$au,
if the initial core mass is comparable to that of Ceres. If typical planetesimals
formed are about the size of Ceres \citep[e.g.][]{Johansen15,Simon16}
 throughout the disc, planetesimals have to grow via planetesimal-planetesimal
collisions or via some other manner to reach $\sim10^{-2}\,M_{\oplus}$
to form gas giants via pebble accretion beyond $\sim20\,$au. The
trend shown here is largely consistent with \citet{Johansen17}, where
growing a core from $10^{-5}\,M_{\oplus}$ to $10^{-1}\,M_{\oplus}$
via pebble accretion takes a few Myr at \DIFdelbegin \DIFdel{30 }\DIFdelend  $30\,$au.

However, typical planetesimal sizes may be different depending on
the disc environment. We can roughly estimate the planetesimal mass
formed via streaming instability by considering the gravitational
instability in the dust layer:

\begin{eqnarray}
M_{{\rm plsml,\,init}} & \sim & \rho_{{\rm Roche}}\cdot\frac{4\pi}{3}H_{p}^{3}\sim3M_{*}\left(\frac{H_{p}}{H_{g}}\right)^{3}\left(\frac{H_{g}}{a}\right)^{3}\label{eq:Mcoreinit}\\
 & \simeq & 13.8\,M_{\oplus}\left(1+\frac{\tau_{s}}{\alpha_{{\rm turb}}}\right)^{-3/2}L_{*0}^{3/7}M_{*0}^{-5/7}\left(\frac{a}{{\rm au}}\right)^{6/7}.\nonumber 
\end{eqnarray}
To get the final expression, we used  the Roche density $\rho_{{\rm Roche}}\sim\frac{9}{4\pi}\frac{M_{*}}{a^{3}}$,
Equation \ref{eq:Hp} for $H_{p}/H_{g}$, and the disc aspect ratio
of the irradiation region  \mbox{
\citep{Ida16a}}\hspace{0pt}
:

\begin{equation}
\left(\frac{H_{g}}{a}\right)_{{\rm irr}}=0.024\left(\frac{T_{2}}{150\,{\rm K}}\right)^{1/2}L_{*0}^{1/7}M_{*0}^{-4/7}\left(\frac{a}{{\rm au}}\right)^{2/7}.
\end{equation}
With Equation \ref{eq:Mcoreinit}, the initial planetesimal mass becomes
$10^{-4}\,M_{\oplus}$ at $\sim2.8\,$au and increases to $10^{-2}\,M_{\oplus}$
at $\sim83\,$au in Disc 5 (i.e. the same disc as in Figure \ref{fig:insituPF}),
which makes the giant planet formation possible out to $\sim40\,$au.
For longer-lived, more massive, or more metal-rich discs, giant planets
can form out to or beyond $100\,$au. In this paper, we choose
conservative initial conditions and do not place the planetary cores
beyond 20 au.

Exercises similar to that of Figure \ref{fig:insituPF} also suggest that
the maximum giant planet mass achievable in our simulations is a few
to several thousand $M_{\oplus}$. The upper mass limit is determined
partly by the planet's efficiency of gas accretion and partly by the
disc's capability of providing gas (see Section \ref{subsec:GasAccretion}).
It may be possible to achieve $\sim10^{4}\,M_{\oplus}$ with an extreme
choice of parameters such as $(b,c)=(7,3)$ in a highly turbulent
disc with $\alpha_{{\rm turb}}=10^{-2}$ or in a massive, long-lived
disc (e.g. $0.2\,M_{\odot}$ with $t_{{\rm diff}}\sim10\,$\DIFdelbegin 
\DIFdelend  Myr). However,
the former tends to lose planetary cores to the star due to efficient
migration, while the latter-type of discs may be relatively rare.
Taking a more extreme set of parameters would not help the further
mass growth because the gas accretion is limited by the disc's gas
supply as seen in Equation \ref{eq:Mgas}. By assuming the initial
masses of protoplanets with Equation \ref{eq:Mcoreinit}, the maximum
mass becomes \DIFdelbegin \DIFdel{$\sim50\,M_{J}$ }\DIFdelend  $\sim50\,M_{J}\sim1.6\times10^{4}\,M_{\oplus}$ for massive,
metal-rich, long-lived discs. We discuss this issue further in Section
\ref{subsec:Max_Mp_a}.

We ran \DIFdelbegin \DIFdel{240 }\DIFdelend  $240$ multiple-planet-core simulations as well as about
a hundred more single-planet core simulations. In all of these \DIFdelbegin \DIFdel{simulations}\DIFdelend  runs,
we assumed a Sun-like star. For the main simulations, we have
ten cores per system with the initial mass of $10^{-2}\,M_{\oplus}\sim100\,M_{{\rm Ceres}}$
over $0.5-15\,$au  \footnote{The number of cores is chosen arbitrarily. However, \mbox{
\citet{Bitsch20}
}\hspace{0pt}
recently studied pebble accretion with 15, 30, and 60 cores across
$3-17\,$au and found that the number of cores generally has little
effect on the final outcome of planetary distributions. The exception
is the case when the eccentricity and inclination damping are very
slow so that the higher number of cores leads to more random orbits.
Since the number of our cores is lower and they are distributed over
a wider orbital range compared to their work, our initial conditions
are less prone to dynamical instabilities compared to theirs.}. The inner edge of the disc is set \DIFdelbegin \DIFdel{to be }\DIFdelend at $0.1\,$au for all of our simulations.
This is rather arbitrary, but is adopted partly because the timesteps
required to resolve an orbit become too small within this radius to
run a simulation for \DIFdelbegin \DIFdel{100 }\DIFdelend  $100\,$Myr, and partly because tidal interactions
with the star also become important there, and we do not explicitly
include this effect in our code (however,  see Section \ref{subsec:HJs}).
Each core has a randomly assigned initial eccentricity of $e=\left[0,\,0.01\right]$,
the corresponding inclination of $i\,{\rm (rad)}=0.5e$, and randomly
chosen phase angles. Planets are also considered `removed' beyond
$1000\,$au in our simulations. In summary, we simulated the
growth and orbital evolution of the ten cores in the eight different disc
models shown in Table \ref{tab:discs} with five different stellar
metallicities. Six simulations are run for each combination of parameters,
which makes the total number of main simulations \DIFdelbegin \DIFdel{240.
}\DIFdelend  $240$.

Besides the effects of disc models and stellar metallicities, we have
also explored the effects of the Kelvin-Helmholtz timescales, the
turbulent viscosity $\alpha_{{\rm turb}}$, and the pebble accretion
efficiency. We discuss some of these in Section \ref{subsec:singlePF}
for single-planet simulations. For main simulations, these parameters
are set as $(b,\,c)=(8,\,3)$, $\alpha_{{\rm turb}}=10^{-4}$,
and $\epsilon=\epsilon_{{\rm IGM16}}$, respectively.

\section{Results \label{sec:Results}}

In Section \ref{subsec:multiplePF}, we \DIFdelbegin \DIFdel{discuss }\DIFdelend  compare the results of
\DIFdelbegin \DIFdel{the }\DIFdelend main numerical simulations of multiple-planet cores  with observed
systems, by focusing on the effects of disc parameters such as mass,
dissipation timescale, and metallicity. In Section \ref{subsec:singlePF},
we discuss the effects of the turbulent viscosity alpha $\alpha_{{\rm turb}}$ as
well as the pebble accretion efficiency $\epsilon$ through single-planet
simulations. 
Throughout this paper, we use the radial-velocity
detected, confirmed extrasolar planets obtained from the NASA Exoplanet
Archive\footnote{https://exoplanetarchive.ipac.caltech.edu}
as `observed planets'. We do not include transit planets for comparison with
simulations  unless we state otherwise, because the transit surveys
have a large bias towards close-in planets due to the detectability
dependence of $\propto a^{-1}$.
%
 \subsection{The effects of disc mass, dissipation timescale, and stellar metallicity \label{subsec:multiplePF}}

First, we present the overall results of simulations in Section \ref{subsec:overall}
and show that \DIFadd{the overall trends of} distributions of giant planets 
are well-reproduced.
{
%
Then we will focus on formation of different types of giant planets
in Sections \ref{subsec:HJs} and \ref{subsec:SCJs}. 

%
\subsubsection{Overall results of multiple protoplanetary-core simulations }\label{subsec:overall}}

\begin{figure*}
\noindent %
\noindent\begin{minipage}[t]{1\columnwidth}%
\begin{center}
\includegraphics[width=0.9\paperwidth]{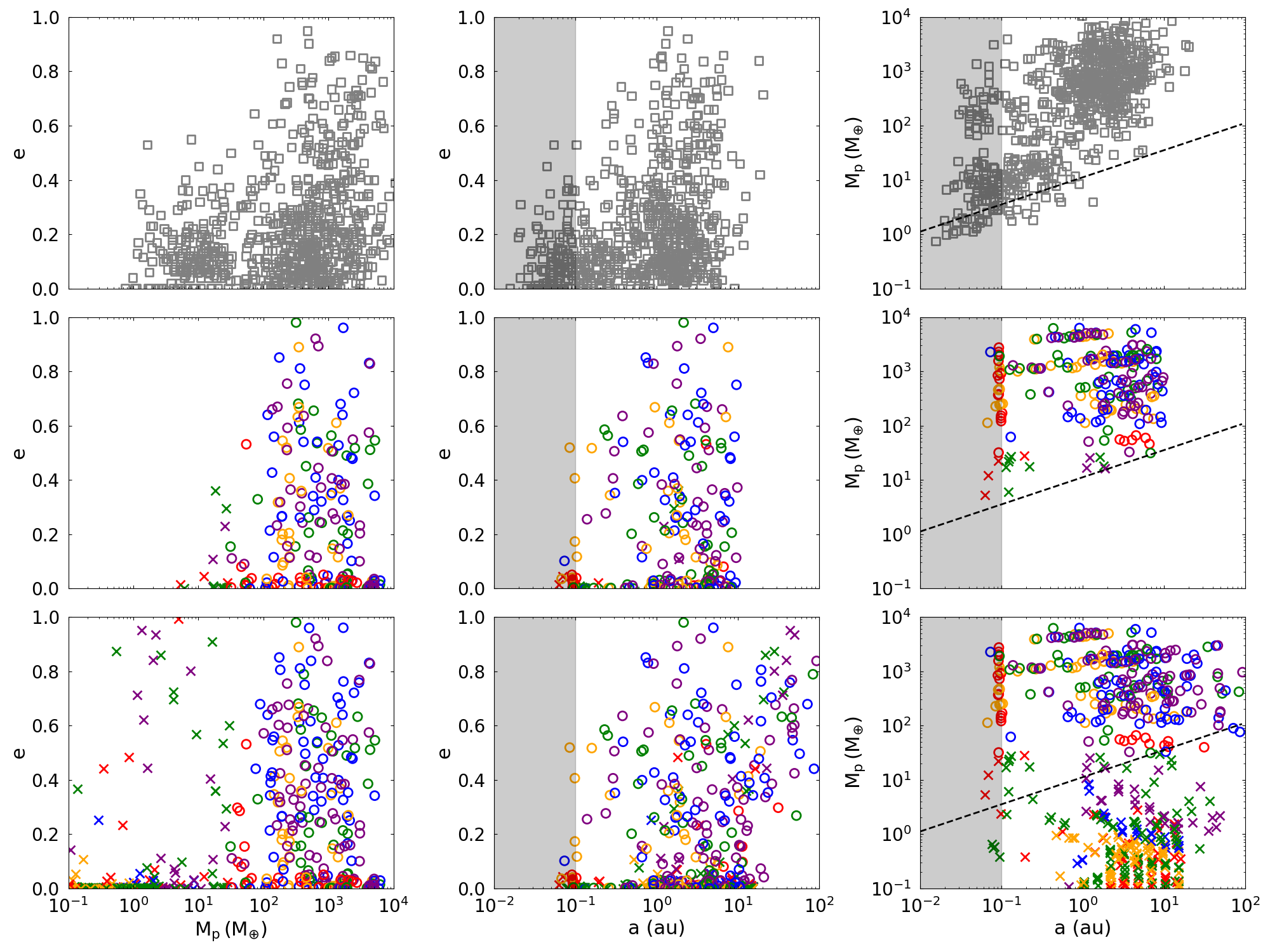} 
\par\end{center}%
\end{minipage}

\caption{Comparison of distributions of parameters for the observed exoplanets
and left, middle, and right panels \DIFdelbeginFL \DIFdelFL{correspond to
}\DIFdelendFL are $M_{p}-e$,
$a-e$, and $a-M_{p}$ distributions, respectively. The top panels
show observed data from the RV-detected planets. The bottom panels
show all the simulated planets at the end of the simulations (100
Myr), while the middle panels show observable planets with the
RV detection limit of $1\,$m/s and $a\leq10$ au. The black dashed
line shown on $a-M_{p}$ panels corresponds to $1\,$m/s limit. 
\DIFadd{The shaded areas in $a-e$ and $a-M_{p}$ distributions indicate that 
the inner disc edge of our model is at $0.1\,$au, and thus we do not intend to 
reproduce planet distributions there.}
Circles and crosses represent giant ($\gtrsim0.1\,M_{J}$) and low-mass ($<0.1\,M_{J}$)
planets, respectively, and the red, orange, green, blue, and purple
colours correspond to stellar metallicities of -0.5, -0.3, 0.0, 0.3,
and 0.5, respectively. Some planets are clustered around 0.1 au because
the disc's inner edge is set there. \label{fig:aM_Me_2panels}}
\end{figure*}

\DIFdelbegin \DIFdel{Therefore, it might be tempting to define that planets with masses
above $M_{p}\sim4\,M_{\oplus}$ as ``giant planets''. However, as
seen in Figure \ref{fig:aM_Me_2panels}, planets with $M_{p}\sim4\,M_{\oplus}$
clearly have different dynamical behaviours compared to more massive
planets. Therefore, in this paper, we take $M_{p}=0.1\,M_{J}\sim30\,M_{\oplus}$
as a critical mass to divide giant planets from }\DIFdelend  In this paper, we define planetary bodies as having a mass $\geq0.1\,M_{\oplus}$,
because this is about the Mars mass and also roughly the smallest mass
of observed exoplanets. We also divided planets into two groups
for simplicity: giant planets ($\geq0.1\,M_{J}\sim30\,M_{E}$)
and low-mass \DIFdelbegin \DIFdel{planets.
This mass corresponds to a planet with the core mass fraction of $\sim10\,\%$
or lower,
or equivalently, the gas envelope mass fraction of $\sim90\,\%$
or higher. The mass also roughly coincides with }\DIFdelend  planets ($<0.1\,M_{J}$). This division is rather arbitrary,
but it is motivated by a mass above which the semi-major axis mass distribution
of observed exoplanets appears to change. 

\DIFdelbegin 
{
\DIFdelFL{The mass fractions of the rocky core (blue) $\frac{M_{{\rm core}}}{M_{p}}$
and the gaseous envelope (orange) $\frac{M_{{\rm env}}}{M_{p}}$ for
all the planets formed in our simulations. The core and the envelope
have comparable masses $M_{{\rm core}}\sim M_{{\rm env}}$ when a
planetary mass is $\sim4\,M_{\oplus}$. }
}

\DIFdelend Figure \ref{fig:aM_Me_2panels} shows the overall results of our simulations
compared to the observed planetary systems. The left, middle, and
right panels correspond to $M_{p}-e$, $a-e$, and $a-M_{p}$ distributions,
respectively, for all the simulated planets (bottom panels), observable
simulated planets (middle panels), and observed planets (top panels).
For the `observable' planets, we \DIFdelbegin \DIFdel{adopted }\DIFdelend  applied the RV detection
limit of $1\,$m/s and $a\leq10$ au. As seen in the figure, the simulated
planets \DIFdelbegin \DIFdel{successfully }\DIFdelend reproduce 
\DIFadd{the overall trends of} $M_{p}-e$, $a-e$, and $a-M_{p}$ distributions
of \DIFadd{extrasolar giant planets well. } 
We note that the disc
inner edge of our simulations is set at 0.1 au and the code does not
include the tidal evolution effects. Therefore, the cluster of simulated
planets near 0.1 au and their slightly elevated eccentricities compared
to observed systems are not surprising.

As seen in the $a-M_{p}$ distribution, our simulations successfully
\DIFdelbegin \DIFdel{produce }\DIFdelend  generate all kinds of giant planets including hot Jupiters (HJs, $a\lesssim0.1\,$au),
warm Jupiters (WJs, $0.1\,{\rm au}\,\lesssim a\lesssim1\,{\rm au}$),
cold Jupiters (CJs, $1\,{\rm au}\,\lesssim a\lesssim20\,{\rm au}$),
and super-cold Jupiters (SCJs, $a\gtrsim20\,$au). \DIFdelbegin \DIFdel{The success of
forming CJs is due to }\DIFdelend  As already shown
by \mbox{
\citet{Ida18}}\hspace{0pt}
, CJs are formed successfully since the new type
II migration \DIFdelbegin \DIFdel{formula and the }\DIFdelend  is slower than the classical one and since the low disc
turbulence $\alpha_{{\rm turb}}=10^{-4}$ is assumed in the two-$\alpha$
disc model\DIFdelbegin \DIFdel{assumption as already shown by \mbox{
\citet{Ida18}}\hspace{0pt}
}\DIFdelend .  We discuss the formation of HJs and SCJs further
in Sections \ref{subsec:HJs} and \ref{fig:SCJs}, respectively.

As in \citet{Matsumura17}, we still have trouble reproducing the
orbital properties of low-mass planets within $\sim1\,$au, because
many protoplanetary cores are lost to the star since type I migration
is too efficient and the disc edge is not sharp enough to retain them
efficiently. This may not bee too surprising because, although we
have improved our code significantly, as seen in Section \ref{sec:Methods},
none of those mechanisms help to slow type I migration. \citet{Lambrechts19}
and \citet{Izidoro19ap} successfully produced rocky super-Earth systems over $0.1-1\,$au, where we lost many such planets due
to rapid migration. 
 This is partly because their disc ages are much
shorter than ours and partly because their planets are trapped in
mean motion resonances more efficiently at the disc edge since they
terminate planet migration there. 
In our simulations, some gas discs'
lifetimes are as short as a few Myr like their simulations, but
others are 10 Myr or longer. The long lifetimes are motivated by
recent observations as stated before, but it also leads to the loss
of the majority of low-mass planets as expected.

Having this bias of low-mass planets in mind, our simulations suggest
that there may be a group of eccentric low-mass planets in the outer
disc ($\gtrsim10\,$au). In our simulations, they are coming from
Discs 2 and 3, which are not too massive ($0.06\,M_{\odot}$ and $0.02\,M_{\odot}$)
and have a short dissipation timescale of $0.1\,$Myr. 
As discussed further in the next sub-section, these are protoplanetary cores which
did not have enough time to become giant planets. 
Since we reproduce \DIFadd{the overall trends of distributions of giant planets well while those of low-mass planets
are poorer} due to the uncertainties in type I migration, 
we focus on the outcomes of giant planets for the rest
of this paper.

\begin{figure*}
\begin{minipage}[t]{0.45\textwidth}
\begin{center}
\includegraphics[width=0.43\paperwidth]{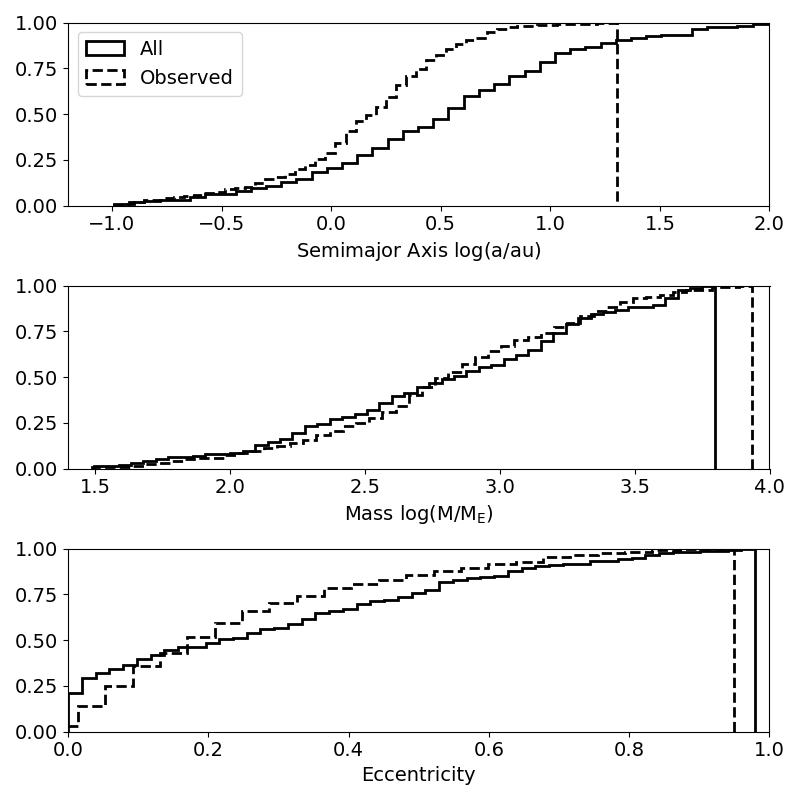}
\par\end{center}
\end{minipage}\hfill{}
\begin{minipage}[t]{0.45\textwidth}
\begin{center}
\includegraphics[width=0.43\paperwidth]{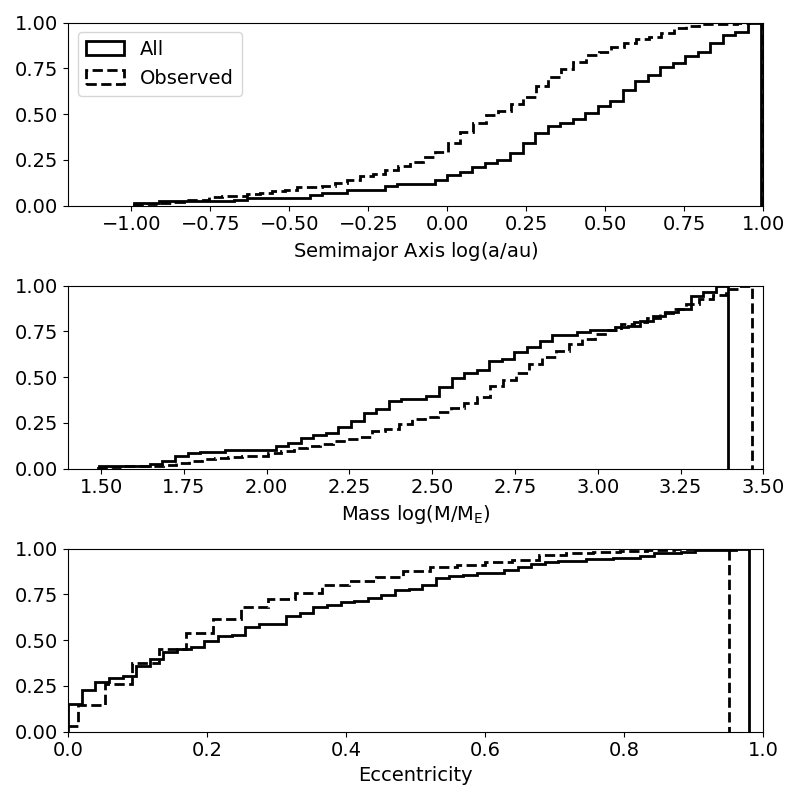}
\par\end{center}
\end{minipage}

\caption{Distributions of semi-major axis (top), planetary mass (middle), and
eccentricity (bottom). The solid and dashed lines correspond to simulated
and RV-detected planets, respectively. The left panels include all
the giant planets beyond $0.1\,$au, while the right ones are limited
to orbital radii of $0.1-10\,$au and masses up to $3000\,M_{E}$.
The K-S test cannot reject the null hypothesis for the mass distribution
on the left with the statistics of $0.092$ and the P value of $0.11$,
but rejects the null hypothesis for all other cases. Our simulations
clearly produce more giant planets in the outer region compared to
observed systems. However, the agreements are not bad by eye for
both masses and eccentricities. \label{fig:hist_aMe}}
\end{figure*}

Although we compare our simulations with observed giant planets, 
it is not our intention to exactly reproduce the distributions of
orbital and physical parameters of extrasolar planets. Such a comparison
does not make sense for this study since it would force us to make
assumptions, for example, that disc models 1-8 exist with an equal probability
and that the inner and outer disc radii are the same for all the discs,
which are unlikely to be true in reality. Having said that, Figure
\ref{fig:hist_aMe} presents the comparison of distributions of the semi-major
axis, planetary mass, and eccentricity for observed (dashed lines)
and simulated (solid lines) giant planets. The left panels include
all the giant planets ($\geq30\,M_{E}$) beyond $0.1\,$au, while
the right panels include only those with masses $30-3000\,M_{E}$
and semi-major axes $0.1-10\,$au. 
\DIFadd{Only the planets beyond $0.1\,$au are chosen because our inner disc edge 
sits there, and because we do not include tidal evolution in our simulations directly 
(however, see Section \ref{subsec:Form_HJs}).}\DIFaddend

By eye, the agreement between observed and simulated distributions
are good for both planetary masses and eccentricities. In fact, for
the mass distribution in the left panel, the Kolmogorov-Smirnov (K-S) 
test could not reject the null hypothesis with the statistics of $0.092$ and the P value
of $0.11$. The agreement is poor for semi-major axis distributions,
which indicates that we overproduced CJs compared to WJs. This may imply
that we need to understand type I migration better to produce cores of these giants 
and/or that the low disc turbulence $\alpha_{{\rm turb}}$ is not appropriate for all
the systems. Nevertheless, the overall good agreement with observed
properties is encouraging.

As seen in $M_{p}-e$ and $a-e$ distributions of Figure \ref{fig:aM_Me_2panels},
as well as the middle panels of Figure \ref{fig:hist_aMe}, our simulations
also reproduce the wide range of eccentricities observed for giant
planets. Compared to the lack of highly eccentric giant planets seen
in \citet{Matsumura17} and \citet{Bitsch19}, this is a great improvement.
We suspect that the success is partly due to a range of disc parameters
leading to a larger number of  giant planets per system than \citet{Matsumura17},
and partly due to \DIFdelbegin \DIFdel{a more self-consistent
disc model compared to \mbox{
\citet{Matsumura17} }\hspace{0pt}
and \mbox{
\citet{Bitsch19}}\hspace{0pt}
}\DIFdelend  low efficiencies of eccentricity and inclination
damping (see Section \ref{subsec:Comp_Bitsch19}). 

\begin{figure*}
\noindent %
\noindent\begin{minipage}[t]{1\columnwidth}%
\begin{center}
\includegraphics[width=0.9\paperwidth]{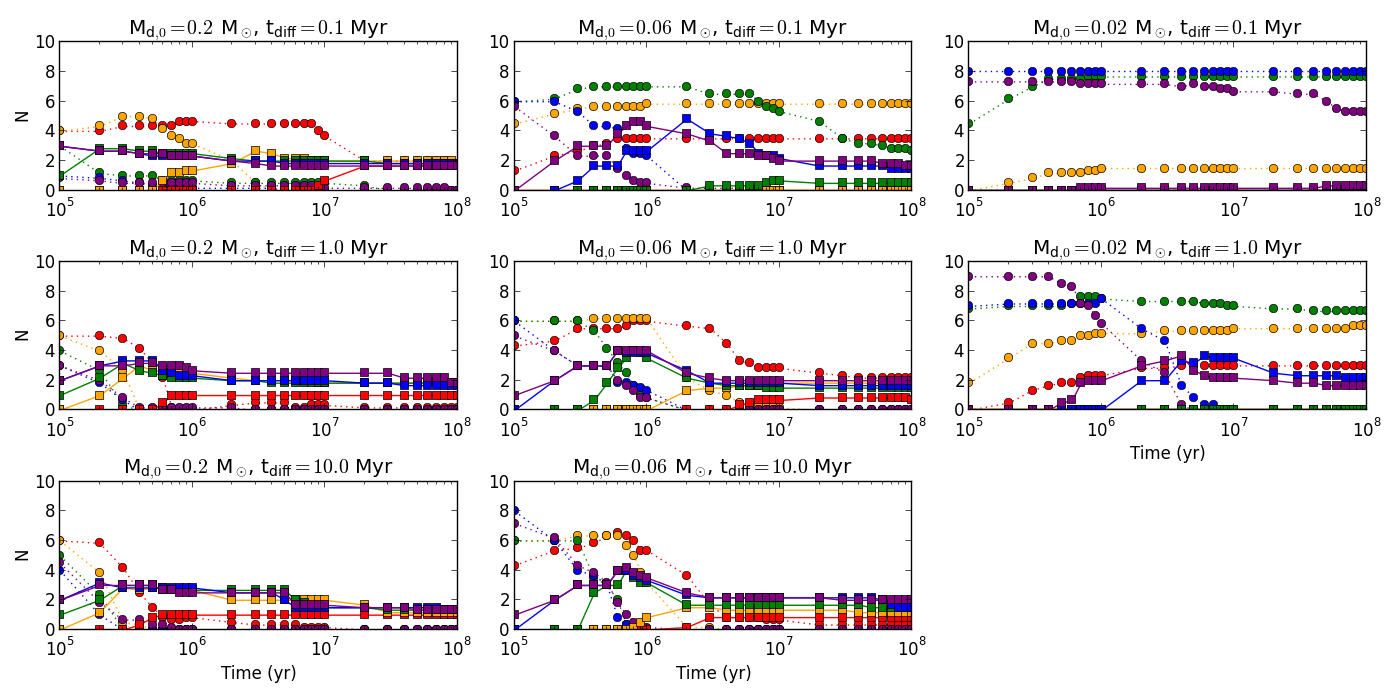} 
\par\end{center}%
\end{minipage}\caption{Evolution of the average number of giant planets (solid) and low-mass
planets (dotted) for different disc models.  Red, orange, green,
blue, and purple correspond to stellar metallicities of -0.5,
-0.3, 0.0, 0.3, and 0.5, respectively. \label{fig:evol_num}}
\end{figure*}

In \citet{Matsumura17}, the average number of giant planets formed
per system throughout the simulations was typically about two for
no-migration cases. In our new simulations, on the other hand, the
average number of giant planets can be much higher, as seen in Figure
\ref{fig:evol_num}. The number of giant planets decreases over time
due to dynamical instabilities, which lead to the population of giant
planets with high eccentricities. Overall, the number of giant planets
tends to be higher when a disc has higher mass, higher metallicity,
and longer dissipation timescale. For Disc 3, which has the lowest
mass and the shortest dissipation timescale, giant planet formation
is a rare event, and it happens only for the highest metallicity case.
The average number of giant planets formed in high-mass and/or high-metallicity
discs is typically a few or more and decreases over time to $\sim2$
in the end. However, we note that the number of giant planets per
system was also very high in \citet{Bitsch19} (typically about 5),
and we discuss this issue further in Section \ref{subsec:Comp_Bitsch19}.

\DIFdelbegin \DIFdel{Since we reproduce the distributions of giant planets well while those
of low-mass planets are poor, we will focus on the outcomes of giant
planets for the rest of this paper.
}

\DIFdelend \subsubsection{Formation of hot Jupiters \label{subsec:HJs}}

\DIFdelbegin \DIFdel{About $10\,\%$ of Sun-like stars are known to have giant planets,
while roughly $0.5-1\,\%$ of such stars have hot Jupiters \mbox{
\citep[HJs hereafter, ][]{Winn18}}\hspace{0pt}
.
This implies that $\sim5-10\,\%$ of giant-planet hosting stars have
at least one HJ. }\DIFdelend There are three main pathways to form \DIFdelbegin \DIFdel{these HJs }\DIFdelend  hot HJs:{} in situ formation \citep[e.g.][]{Batygin16}, disc migration
after the formation of a giant planet beyond the snow line \citep[e.g. ][]{Lin96},
and the tidal circularisation of a highly eccentric orbit of a giant planet
\citep[e.g.][]{Fabrycky07,Nagasawa08,Naoz11,Wu11}. Here, the in situ
formation includes both the cases where protoplanetary cores also
form in situ \citep[e.g.][]{Chiang13,Boley16} and the cases where
cores form further out, migrate to the inner disc, and then accrete
gas there \citep[e.g.][]{Coleman16a,Matsumura17}. In this section,
we evaluate these pathways by using our simulations and argue that
differences in stellar metallicities may naturally lead to different
pathways.

\begin{figure*}
\noindent %
\noindent\begin{minipage}[t]{1\columnwidth}%
\begin{center}
\includegraphics[width=0.9\paperwidth]{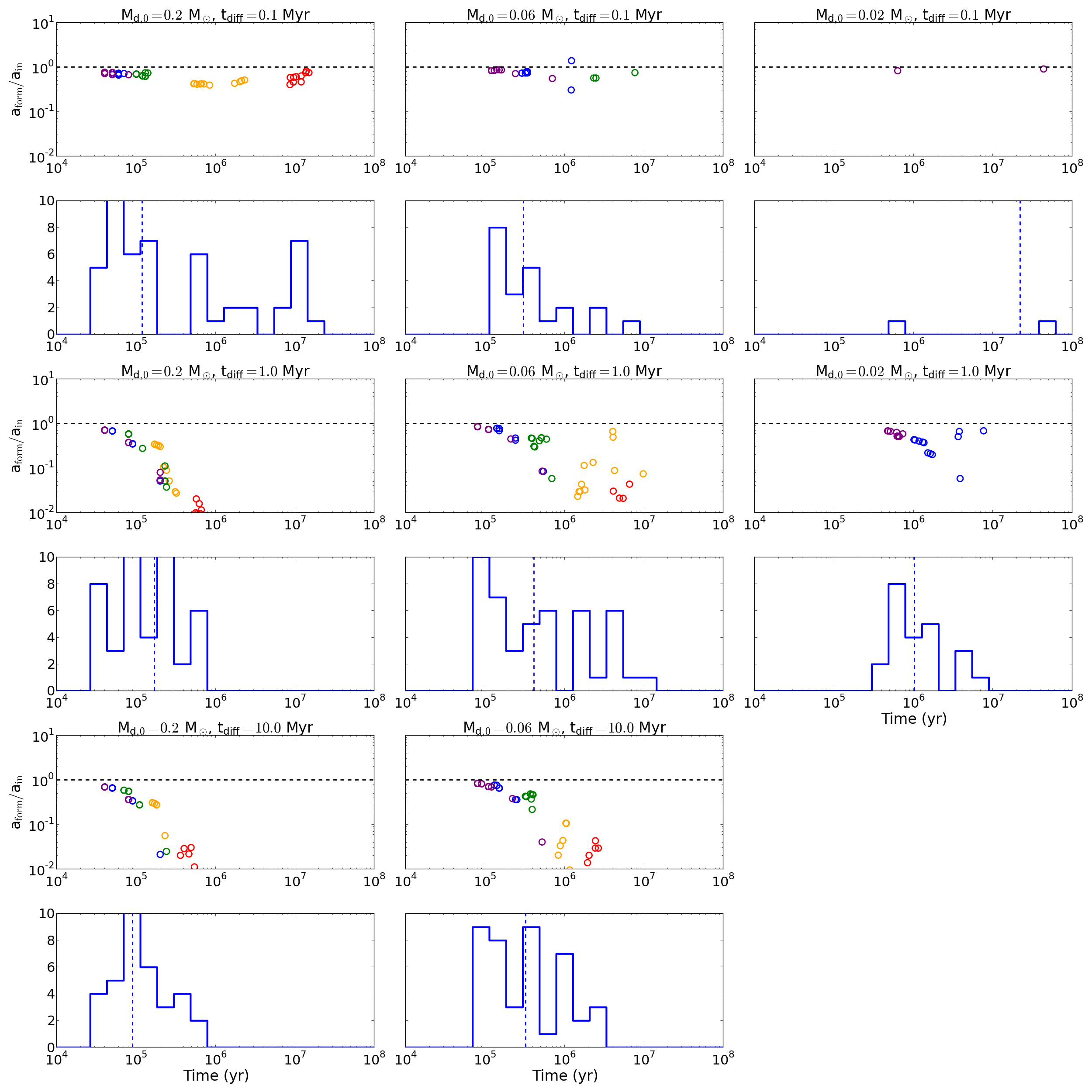} 
\par\end{center}%
\end{minipage}

\caption{Top: Ratio of the semi-major axis of a giant planet when it became
giant (i.e. mass reaches $0.1\,M_{J}$) with respect to the initial
semi-major axis for different disc models. The red, orange, green,
blue, and purple symbols correspond to stellar metallicities of -0.5,
-0.3, 0.0, 0.3, and 0.5, respectively. Bottom: Formation timescales
of giant planets. The vertical dashed line indicates the median value.
\label{fig:GPformtime}}
\end{figure*}

Figure \ref{fig:GPformtime} shows two-panel plots of eight different
disc models, where each disc is run with five different stellar metallicities.
For each disc model, the bottom panel shows a histogram of giant planet
formation timescales, while the top panel shows how far from the initial
orbital radii giant planets are formed. As discussed in the previous
section, we assume that a protoplanet becomes a giant planet once
its mass surpasses $0.1\,M_{J}$.

As seen in bottom panels of different disc models, almost all discs
form giant planets regularly except for Disc 3. For the disc models
we \DIFdelbegin \DIFdel{tested}\DIFdelend  considered, the median formation timescales are short ($\sim0.1-1\,$Myr)
compared to an often-quoted typical disc lifetime of $\sim3\,$Myr.
This confirms that pebble accretion is an efficient way \DIFdelbegin \DIFdel{to form }\DIFdelend  of forming
giant planets within a disc's lifetime.

The top panels of different disc models in Figure \ref{fig:GPformtime}
compare the ratio of the `formation' semi-major axis to the initial
semi-major axis: $a_{{\rm form}}/a_{{\rm in}}$ with the formation
timescales. Here, again, the formation semi-major axis is where
a planetary mass reaches $0.1\,M_{J}$. The red, orange, green, blue,
and purple symbols correspond to different metallicities of ${\rm [Fe/H]}=-0.5$,
$-0.3$, 0.0, 0.3, and 0.5, respectively. As expected, giant planets
form more quickly in higher metallicity environments within the same
disc model.

In discs with short dissipation timescales (Discs 1 to 3, $t_{{\rm diff}}=0.1\,$Myr),
we find that giant planets often form near their initial locations.
On the other hand, in discs with longer dissipation timescales (Discs
4 to 9, $t_{{\rm diff}}=1-10\,$Myr), there are two types of giant
planets: some are forming near their initial locations (often cores
beyond several au), and others are forming \emph{\emph{after}} some migration.
In the case of the latter, the formation radii tend to be about 1
to 2 orders of magnitude smaller than the initial orbital radii. In
Discs 4-9, when metallicities are low ${\rm [Fe/H]}\lesssim-0.3$,
cores tend to migrate and then accrete gas to become either WJs or
HJs. On the other hand, when metallicities are high ${\rm [Fe/H]}\gtrsim0.0$,
outer cores tend to become CJs, while inner cores migrate to form WJs
and/or HJs. However, in these cases, the inner cores are often removed
via planet-planet interactions as the gas disc dissipates, and the
final systems include either only CJs or HJs along with CJs. Thus,
in our simulations, when HJs form in situ after their cores migrated
to the inner disc, they tend to be in a low-metallicity disc, though
they can also be formed in a higher metallicity disc.

\begin{figure*}
\noindent %
\noindent\begin{minipage}[t]{1\columnwidth}%
\begin{center}
\includegraphics[width=0.9\paperwidth]{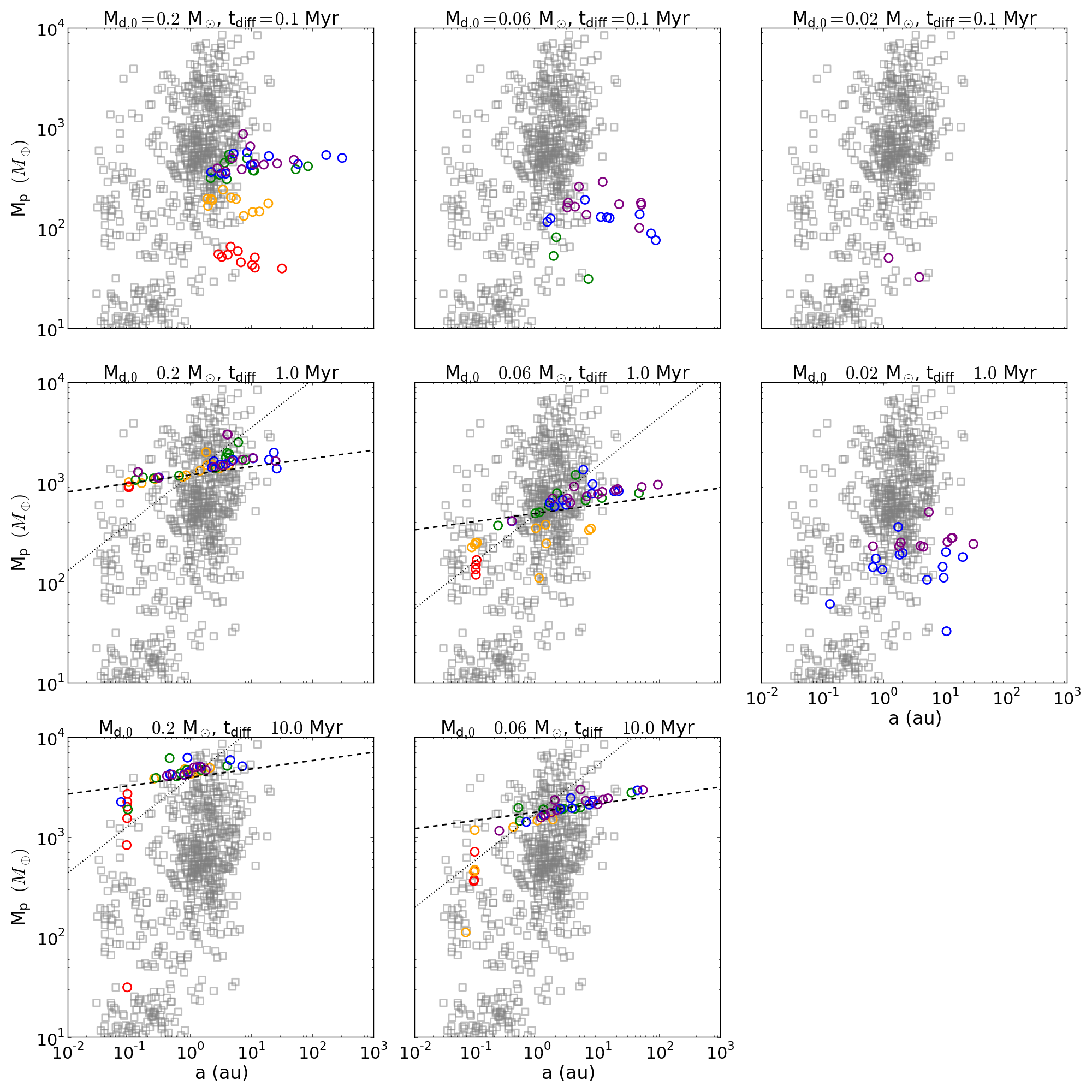} 
\par\end{center}%
\end{minipage}

\caption{Semi-major axis mass distribution of giant planets at the end of
all the simulations for different disc models. The red, orange, green,
blue, and purple colours correspond to stellar metallicities of -0.5,
-0.3, 0.0, 0.3, and 0.5, respectively. The dashed and dotted lines
in Discs 4, 5, 7, and 8 are expected mass trends in viscous and irradiation
regions, respectively (see text). \label{fig:aMgiants}}
\end{figure*}

Figure \ref{fig:aMgiants} shows the semi-major axis mass distribution
of giant planets for each disc model. Combined with the results of
Figure \ref{fig:GPformtime}, this figure shows two types of outcomes
of giant planet formation. In rapidly dissipating or low-mass discs
(Discs 1, 2, 3, and 6), low-mass (typically less than a few Jupiter
masses) and cold (orbiting beyond $\sim1\,$au) giant planets are
often produced. In these discs, lower metallicities lead to lower mass
giant planets as expected, and the range of masses across different
metallicities is about an order of magnitude. In more slowly dissipating,
moderate-to-high-mass discs (Discs 4, 5, 7, and 8), on the other hand,
giant planets have a much wider distribution of orbital radii (two
to three orders of magnitude), as expected from Figure \ref{fig:GPformtime}.
Planetary masses per disc model are largely comparable to one another
and \DIFdelbegin \DIFdel{depend }\DIFdelend weakly depend 
on metallicities (see Section \ref{subsec:plsml_vs_pebble}),
but cores migrated further tend to have lower masses.


This $a-M_{p}$ relation  of mass increasing with orbital radius is
likely to arise from the reduced gas accretion rate onto the gap-opening
core $\dot{M}_{{\rm TT16}}$ in Equation \ref{eq:Mgas}. Assuming
that the gas accretion timescale can be written as $\tau_{{\rm gas}}\sim\frac{M_{p}}{\dot{M}_{{\rm TT16}}}$,
the planetary mass has the dependence of: $M_{p}\propto h_{g}^{\frac{5}{3}}$.
Since the disc aspect ratios in viscous and irradiation regions in
our disc model have $h_{g,{\rm vis}}\propto\left(\frac{a}{{\rm au}}\right)^{\frac{1}{20}}$
and $h_{g,{\rm irr}}\propto\left(\frac{a}{{\rm au}}\right)^{\frac{2}{7}}$,
respectively \citep[see][]{Ida16a}, the expected mass dependences
on \DIFdelbegin \DIFdel{the semimajor axis }\DIFdelend  semi-major axes are $M_{p,{\rm vis}}\propto\left(\frac{a}{{\rm au}}\right)^{\frac{1}{12}}$
and $M_{p,{\rm irr}}\propto\left(\frac{a}{{\rm au}}\right)^{\frac{10}{21}}$,
respectively. For Discs 4, 5, 7, and 8 in Figure \ref{fig:aMgiants},
these trends are shown in dashed and dotted lines, respectively\DIFdelbegin \DIFdel{. The
}\DIFdelend  , and
the agreement is very good for these cases.

 \medskip{}

\begin{figure*}
\noindent %
\noindent\begin{minipage}[t]{1\columnwidth}%
\begin{center}
\includegraphics[width=0.9\paperwidth]{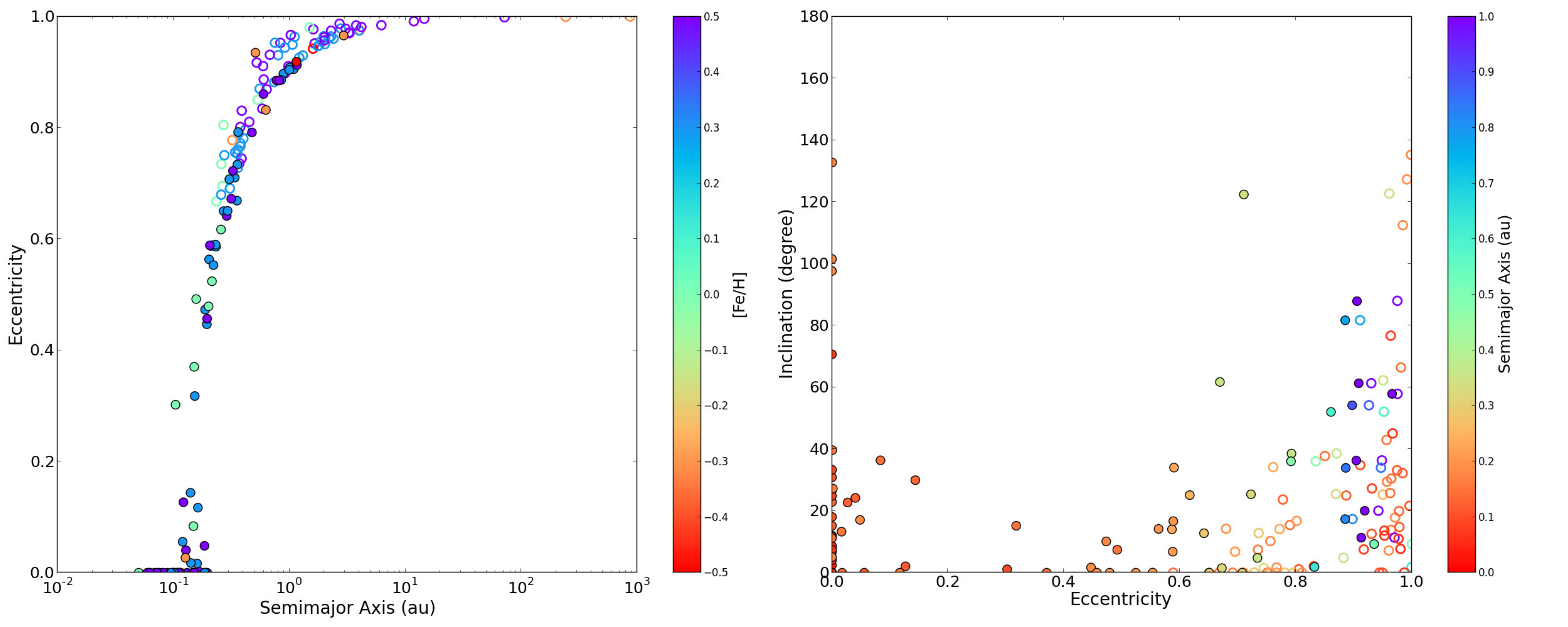} 
\par\end{center}%
\end{minipage}\caption{Left: Semi-major axis eccentricity distribution of planets before (open
circle) and after (filled circle) tidal evolution. Different colours
correspond to the stellar metallicity. Highly eccentric planets tend
to belong to higher metallicity discs. Right: the corresponding eccentricity-inclination
distribution. However, the colour does not represent the metallicity,
but the final semi-major axis. Closer in planets tend to have either
a low eccentricity and a range of inclinations, or a moderate eccentricity
with a low inclination. \label{fig:tide_aei}}
\end{figure*}

 Our simulations do not explicitly include tidal interaction effects
between the central star and planets, so we cannot directly measure
the production rate of a HJ via tidal circularisation. However, we selected planets that could have experienced significant tidal
evolution and applied the weak-friction tidal interaction model \citep[e.g. ][]{Hut81}.
For giant planets that have a pericentre radius of $a\left(1-e\right)\leq0.1\,$au
and a semi-major axis $a\geq0.2\,$au (i.e. an eccentricity of $e\geq0.5$)
at any time during the simulation, \DIFdelbegin \DIFdel{we calculate its tidal evolution for 3 Gyr}\DIFdelend  tidal evolution is calculated.
The results are plotted in Figure \ref{fig:tide_aei}. As in \citet{Matsumura10b},
we scaled a modified tidal quality factor of

\begin{equation}
Q^{\prime}=Q_{0}^{\prime}\frac{n_{0}}{n},
\end{equation}
where $Q^{\prime}\equiv1.5Q/k_{2}$ is a modified tidal quality factor,
$Q$ is a tidal quality factor, $k_{2}$ is the Love number of
degree 2, $n$ is the mean motion, and the subscript $0$ represents
their initial values. In our calculations, we assumed the initial
stellar and planetary modified tidal quality factors to be $Q_{*0}^{\prime}=10^{6}$
and $Q_{p0}^{\prime}=10^{5}$, respectively. Since a typical age of
a planet hosting star ranges from \DIFdelbegin \DIFdel{1 to 10 }\DIFdelend  $1$ to $10\,$Gyr, we have run
the tidal evolution simulations for \DIFdelbegin \DIFdel{3 }\DIFdelend  $3\,$Gyr.

The left panel of Figure \ref{fig:tide_aei} shows the semi-major axis eccentricity
distribution of these planets before (open circle) and after (filled
circle) tidal evolution, where different colours correspond to different
stellar metallicities. As expected from the higher planet formation
efficiency seen in high metallicity environments (see Figure \ref{fig:evol_num},
for example), most planets that achieved high enough eccentricities
in the first place \DIFdelbegin \DIFdel{tend to be }\DIFdelend  are in the high-metallicity discs with ${\rm [Fe/H]\gtrsim0.0}$. 
\DIFdelbegin \DIFdel{Out of 240 simulations
, 175 }\DIFdelend  

Since we do not know what kinds of discs or disc evolution paths dominate
in reality, the fraction of HJ systems with respect to all the simulations
cannot be compared with the observed fraction of HJ systems. However,
we can attempt to estimate the fractions of \emph{\emph{giant-planet}} systems
with HJs. About $10\,\%$ of Sun-like stars are known to have giant
planets, while roughly $0.5-1\,\%$ of such stars have HJs \citep{Winn18}. 
This implies that, roughly speaking, $\sim5-10\,\%$ of giant-planet
hosting stars have at least one HJ. In our simulations, out of $240$
runs, $175$ formed at least one giant planet at some point during
\DIFdelbegin \DIFdel{100 }\DIFdelend  $100\,$Myr simulations. Out of these \DIFdelbegin \DIFdel{175 systems, 10 }\DIFdelend  $175$ systems, $10$  formed HJs that have semi-major axes less than \DIFdelbegin \DIFdel{0.1 au after
3
}\DIFdelend  $0.1\,$au after
$3\,$Gyr. Thus $10/175\sim5.71\,\%$ of a giant-planet-forming system
led to the formation of at least one HJ. If we relax this condition to
the final semi-major axis within 0.2 au, the frequency becomes $38/175\sim21.7\,\%$.
These are reasonable values compared to the observed fraction of HJ
systems among systems with giant planets: $\sim5-10\,\%$. 
 We note that the occurrence rate depends on the choice of the initial distribution 
of disc parameters, and thus the high occurrence rate suggested by the simulations 
does not necessarily mean that our models overestimate the occurrence rate.
\DIFdelbegin \DIFdel{(see the beginning of this subsection)}\DIFdelend

The right panel of Figure \ref{fig:tide_aei} shows the corresponding
eccentricity-inclination distribution of these planets before (open
circles) and after (filled circles) applying tidal evolution. We note
that the colours here indicate the final semi-major axes rather than
the stellar metallicities. The overall trend is similar to what \citet{Nagasawa11}
pointed out for observed exoplanets \textemdash{} planets with low
eccentricities have a wide range of inclinations, while those with
moderate eccentricities have low inclinations. Our simulations indicate
that this trend breaks down at larger orbital radii, because the orbits
of some planets cannot be circularised quickly enough.

From the figure, it is also clear that high-inclination planets initially
had high eccentricities as well. This outcome of tidal evolution reflects
the fact that the orbital circularisation is largely controlled by
tidal dissipation inside the planet, while the spin alignment is controlled
by tidal dissipation inside the star (i.e. the stellar spin aligns
with the orbit normal rather than the other way around). Since the
former takes place faster than the latter, the orbital eccentricity
becomes small, while the orbital inclination does not change dramatically.

These outcomes lead to an expectation about  the effects of stellar
metallicities on HJ formation pathways \DIFdelbegin \DIFdel{and stellar metallicities \textemdash{} }\DIFdelend  \textemdash{} around low-metallicity
stars, HJs tend to form via the migration \DIFdelbegin \DIFdel{around low-metallicity stars, while HJs }\DIFdelend  of cores followed by in situ
gas accretion, while around high-metallicity stars, they can form
either via  migration or via the tidal circularisation of eccentric orbits\DIFdelbegin \DIFdel{or migration around high-metallicity
stars}\DIFdelend .
One way to distinguish these two cases is to check the relation between
the orbital inclination and the stellar metallicity. The \DIFdelbegin \DIFdel{migrated
planets are }\DIFdelend  HJs formed
in situ or via migration are more likely to have low inclinations,
while the tidally circularised planets are expected to have a wide
range of inclinations. \DIFdelbegin \DIFdel{Figure
\ref{fig:ecc-obl-met} shows observed eccentricities for close-in,
giant planets ($\leq0.1\,$au and $\geq0.1\,M_{J}$) against stellar
metallicities (left panel) and a similar figure for stellar obliquities
for giant planets ($\geq0.1\,M_{J},$ middle panel). Since both eccentricities
and inclinations of close-in planets are low for the low stellar metallicities
${\rm [Fe/H]\lesssim-0.3}$, this confirms our expectation. One potential
exception is HATS-11b seen on the eccentricity figure; its host star
has ${\rm [Fe/H]=-0.39\pm0.06}$, while $a=0.046\,$au and the $95\,\%$
confidence upper limit on the eccentricity is $e<0.340$ for fixed
circular orbit models \mbox{
\citep{Rabus16}}\hspace{0pt}
. The future observations could
confirm or disprove this trend.
}\DIFdelend  We discuss this matter further in Section
\ref{subsec:Form_HJs}. 

\DIFdelbegin 

{
\DIFdelFL{\textcolor{black}{Left: eccentricities of observed HJs (}$\leq0.1\,$au
and $\geq0.1\,M_{J}$\textcolor{black}{) compared to stellar metallicities.
One system with $e\sim0.4$ and ${\rm [Fe/H]}\sim-0.4$ is HAT-P-11b.
Colour bars show planetary masses in a log scale. Squares and crosses
correspond to RV-detected and transit-detected planets, respectively.
Middle: a similar figure for stellar obliquities of observed HJs.
Right: stellar metallicities and orbital distributions of giant planets
(}$\geq0.1\,M_{J}$\textcolor{black}{). Data is taken from The Extrasolar
Planets Encyclopaedia (http://exoplanet.eu) for this figure.}\textcolor{red}{{}
}}}

\DIFdelend \subsubsection{Formation of super-cold Jupiters \label{subsec:SCJs}}

The observations show that some giant planets have very wide orbital
radii, ranging from a few tens of au to a few thousand au. \DIFdelbegin \DIFdel{As
discussed in Section \ref{subsec:Initial-Conditions}, we do not place
protoplanetary cores beyond $20\,$au. Therefore, to form SCJs via
pebble accretion }\DIFdelend  In
the pebble accretion model, we expect either (1) an SCJ forms in situ,
\DIFdelbegin \DIFdel{either
directly }\DIFdelend  or (2) a gas giant forms within $\sim20\,$au and then gets scattered
into the outer disc. Again, the in situ formation here includes both
the case where an SCJ directly forms in the outer disc, \DIFdelbegin \DIFdel{or as }\DIFdelend  and the case
where a protoplanetary core \DIFdelbegin \DIFdel{is }\DIFdelend scattered into the outer disc \DIFdelbegin \DIFdel{and }\DIFdelend accretes
gas to become a giant planet there \citep{Kikuchi14}\DIFdelbegin \DIFdel{, or (2) a gas giant forms within $\sim20\,$auand then gets scattered into }\DIFdelend  . Since we did
not place cores beyond $20\,$au, SCJs did not form directly in the
outer disc \DIFdelbegin \DIFdel{. }\DIFdelend  in our simulations.

 As seen in Figure \ref{fig:GPformtime}, most giant planets in our
simulations `form' (i.e. mass  exceeds $0.1\,M_{J}$ in our
definition) near or inside the initial orbital radii. Therefore,
most SCJs in our simulations are generated via scenario (2) (i.e.
formed and then scattered), and not \DIFdelbegin \DIFdel{in-situ. Note that, since our disc only stretches out to 100 au,
we do not
take account of the effects of orbital circularisation of the scattered
planet by }\DIFdelend  via scenario (1). We note, however,
that our simulations preclude the scenario of \mbox{
\citet{Kikuchi14}}\hspace{0pt}
.
The key to forming SCJs in \mbox{
\citet{Kikuchi14} }\hspace{0pt}
was that the orbits
of scattered cores get circularised efficiently via the accretion of high
angular momentum gas in the outer disc\DIFdelbegin \DIFdel{as in \mbox{
\citet{Kikuchi14}}\hspace{0pt}
. Therefore, we could not form SCJs via scenario
(1) efficiently, because our core tends to }\DIFdelend  . Since our disc only stretches
out to 100 au, even when there was a scattered core, the core would
repeatedly encounter the larger planet that scattered the core in
the first place, and an SCJ does not form efficiently. 

\begin{figure*}
\noindent %
\noindent\begin{minipage}[t]{1\columnwidth}%
\begin{center}
\includegraphics[width=0.9\paperwidth]{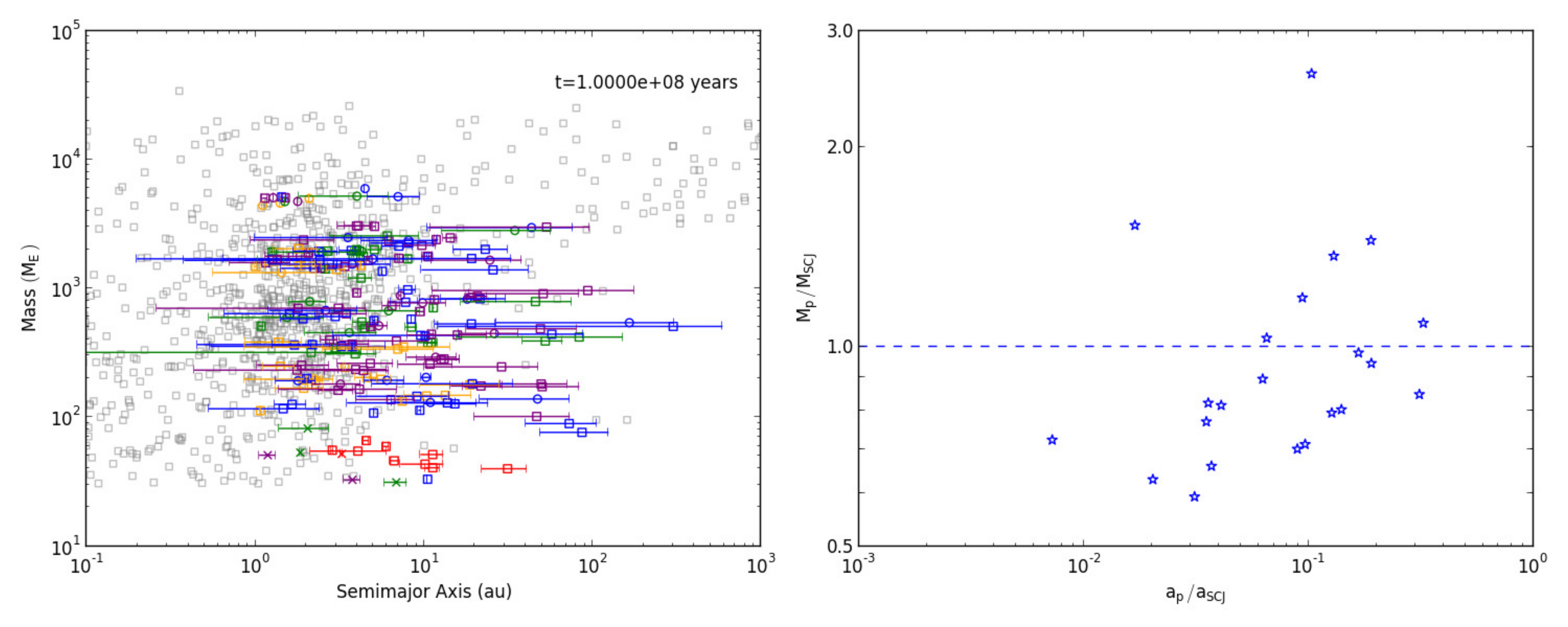} 
\par\end{center}%
\end{minipage}

\caption{Left: Semi-major axis mass distribution of giant planets at 100 Myr.
The error bars represent the locations of pericentre and apocentre
and thus show how eccentric orbits are. As before, red, orange, green,
blue, and purple correspond to stellar metallicities of -0.5,
-0.3, 0.0, 0.3, and 0.5, respectively. The circles, squares, and crosses
correspond to single-planet, multiple-giant-planet, and multiple-planet
but single giant planet systems, respectively. Right: Mass and
semi-major axis ratios for the furthermost SCJ and its companion. \label{fig:SCJs}}
\end{figure*}

When an SCJ is formed via scenario (2), our simulations indicate the
following. First, a stellar metallicity tends to be high $\left[{\rm Fe/H}\right]\gtrsim0.0$
in a system with a SCJ. As seen in the left panel of Figure \ref{fig:SCJs},
all but one giant planet beyond 20 au are from such systems. One giant
planet is at around 30 au in the system with $\left[{\rm Fe/H}\right]=-0.5$,
but its mass is relatively low and $\sim40\,M_{\oplus}$. This trend
is not surprising, because higher metallicity discs tend to form multiple
giant planets without much migration (see Figure \ref{fig:GPformtime})
and thus are prone to dynamical instabilities that can scatter giant
planets outward.

Second, the orbit of an SCJ tends to be eccentric and out to a few
hundred au. It is difficult to scatter a giant planet beyond that
and retain it when it is formed within $\sim20\,$au. The tendency
towards a non-zero eccentricity is a direct outcome of the scattering
event and thus is not surprising. In our simulations, the eccentricity
of SCJs ranges  over $e\sim0.2-0.9$ with the mean value of 0.56.

Third, an SCJ tends to be accompanied by another giant planet\DIFdelbegin \DIFdel{interior
to its orbit}\DIFdelend . Our
simulations show $\sim78.6\,\%$ of SCJs (22/28) are in two- or three-giant-planet
systems, while $\sim21.4\,\%$ (6/28) are in single-planet systems.
The right panel of Figure \ref{fig:SCJs} shows the semi-major axis
ratio of a companion planet to the furthermost SCJ as well as their
mass ratio. In multiple-planet systems, the furthermost SCJ tends
to have a less massive companion interior to its orbit ($15/22\sim68.2\,\%$)
rather than a more massive one ($7/22\sim31.8\,\%$). Thus, the scattering
event tends to place a more massive planet outward and a lower-mass
companion inward in our simulations.

The preference for the inner, lower-mass companion is opposite
\DIFdelbegin \DIFdel{trend }\DIFdelend to what has been observed in previous studies \citep[e.g. ][]{Marzari02,Nagasawa08,Ida13},
where a more massive planet was scattered \DIFdelbegin \textit{\DIFdel{inward}}
\DIFdelend  \textit{inward} with a higher
probability of $>80\,\%$. However, all of these studies started with
the configuration where the inner planet is the most massive one $(2\,M_{J},\,M_{J},\,M_{J})$.
Since the higher mass planets tend to stay closer to the initial location
upon scattering \citep{Chatterjee08}, it may not be surprising that
the more massive planet stayed in the innermost orbit in these studies.
In our simulations, on the other hand, more massive planets tend to
form in the outer part of the disc (see Figure \ref{fig:insituPF},
for example). Therefore, it \DIFdelbegin \DIFdel{may not be }\DIFdelend  is not surprising that there is a preference
\DIFdelbegin \DIFdel{for }\DIFdelend  towards the outermost planets being more massive than the inner ones
because of their inertia and the difficulty of switching orbits.
We discuss their formation and distributions further in Section
\ref{subsec:Form_SCJs}.

\subsection{The effects of other parameters}

In this sub-section, we show the effects of the turbulent viscosity
$\alpha_{{\rm turb}}$ as well as the pebble accretion efficiencies.

\subsubsection{The effects of the turbulent viscosity alpha parameter $\alpha_{{\rm turb}}$\label{subsec:singlePF}}

For the bulk simulations, we assumed $\alpha_{{\rm turb}}=10^{-4}$.
In this sub-section, we investigate the dependence of pebble accretion
simulations on the turbulent alpha parameter by \DIFdelbegin \DIFdel{studying the cases
with }\DIFdelend  using single-planet-core
simulations and by comparing $\alpha_{{\rm turb}}=10^{-3}$ and $10^{-5}$
\DIFdelbegin \DIFdel{besides the default }\DIFdelend  cases to the default ones with $\alpha_{{\rm turb}}=10^{-4}$\DIFdelbegin \DIFdel{in single protoplantary-core simulations}\DIFdelend . We
did \DIFdelbegin \DIFdel{focus on initial conditions of }\DIFdelend  not use all the disc models but focused on representative cases
of \DIFdelbegin \DIFdel{the
bulk simulations here. For the disc models, we choose }\DIFdelend Discs 2, 5, and 8, for which the initial disc mass is the same,
$0.06\,M_{\odot,}$ \DIFdelbegin \DIFdel{while }\DIFdelend  and the diffusion timescales are \DIFdelbegin \DIFdel{different: }\DIFdelend 0.1, 1.0, and
10 Myr, respectively. The initial semi-major axes of protoplanets are
chosen to be $\sim3$, 5, 7, and \DIFdelbegin \DIFdel{$10\,{\rm AU}$}\DIFdelend  $10\,$au, which are all beyond the
snowline initially. This is a rather narrow range compared to $\sim0.5-15\,$au
in the bulk simulations, but they are chosen so that we can test the
likelihood of forming giant planets. The stellar metallicity for all
the simulations in this sub-section is the solar value of ${\rm [Fe/H]}=0.0$.
Other parameters used for the default case are $(b,c)=(8,3)$ for
the Kelvin-Helmholtz timescale, and the pebble accretion efficiency
of $\epsilon_{{\rm IGM16}}$.

\begin{figure*}
\noindent %
\noindent\begin{minipage}[t]{1\columnwidth}%
\begin{center}
\includegraphics[width=0.9\paperwidth]{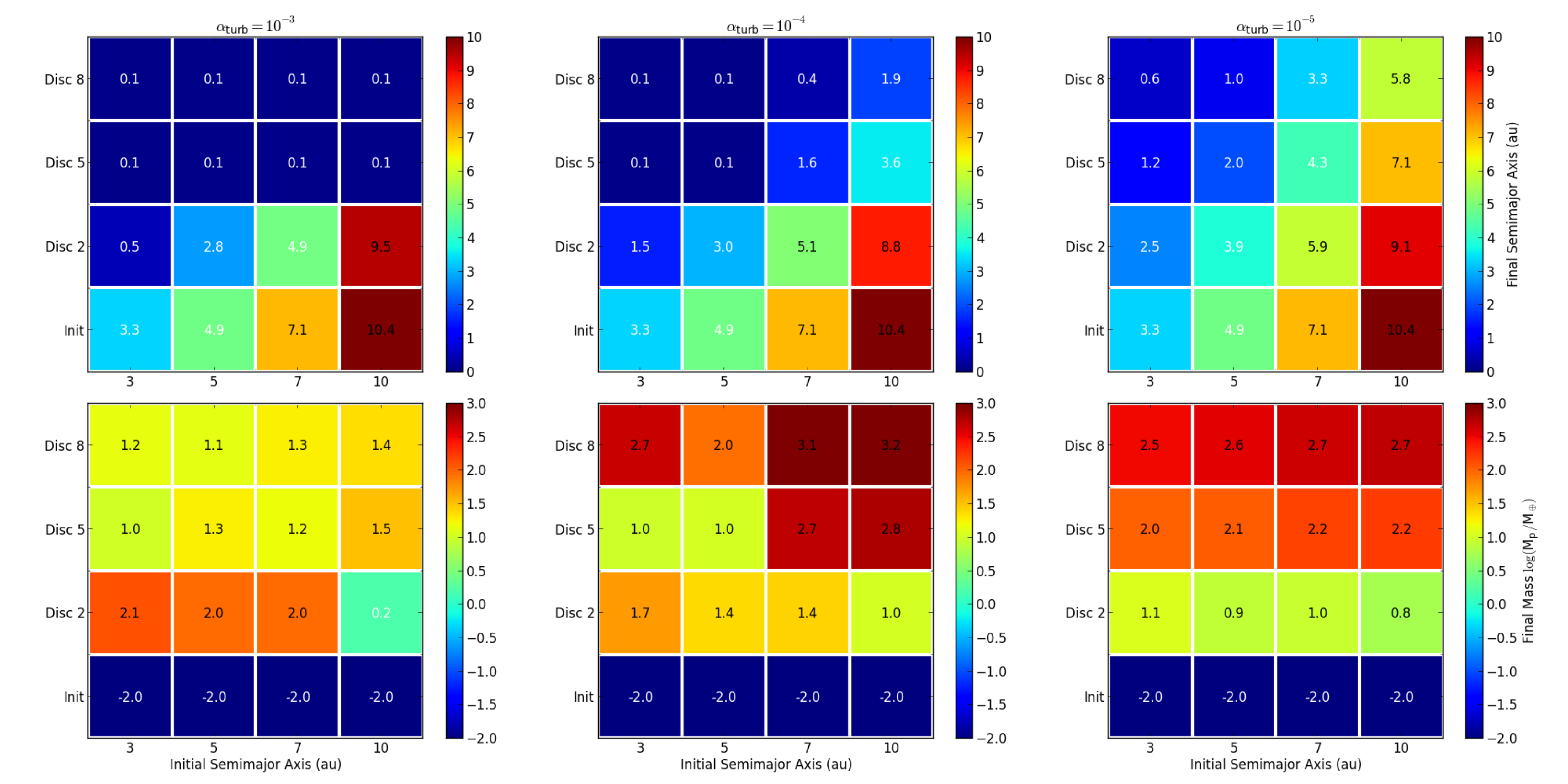} 
\par\end{center}%
\end{minipage}

\caption{Outcome of single-planet formation simulations with the default
setting (i.e. $(b,c)=(8,3)$ for the Kelvin-Helmholtz timescale, ${\rm [Fe/H]}=0.0$,
and the pebble accretion efficiency by IGM16) with $\alpha_{{\rm turb}}=10^{-3}$
(left panels), $10^{-4}$ (middle panels), and $10^{-5}$ (right panels).
The top panels show the final semi-major axis of a single planet for
different initial semi-major axes and disc models. The bottom cells
marked as `Init' correspond to the initial semi-major axes that
are common for all disc models. The bottom panels present the final
planetary masses in log scale. The bottom cells represent the initial
protoplanetary masses of $0.01M_{\oplus}$ for all the cases.\label{fig:aMsingle2}}
\end{figure*}

Figure \ref{fig:aMsingle2} shows the outcomes of single-core simulations
with $\alpha_{{\rm turb}}=10^{-3},\,10^{-4},$ and $10^{-5}$. \DIFdelbegin \DIFdel{The
}\DIFdelend  From
the cases with $\alpha_{{\rm turb}}=10^{-4}$\DIFdelbegin \DIFdel{are shown here as well to
make the comparison easier. We can confirm }\DIFdelend  , we can confirm the
trends described in \DIFdelbegin \DIFdel{the
}\DIFdelend previous sections: Disc 2 forms rather low-mass
CJs, while Discs 5 and 8 form fully grown giant planets across a range
of \DIFdelbegin \DIFdel{disc }\DIFdelend  orbital radii.

For the cases of $\alpha_{{\rm turb}}=10^{-3}$, most planets migrate
to the edge of the disc ($0.1\,$au here), except for Disc 2, where
relatively low-mass giant planets form over a range of orbits. The
loss of the majority of protoplanetary cores to the disc's inner edge
is similar to \citet{Matsumura17}, where we have also adopted $\alpha_{{\rm turb}}=10^{-3}$.
However, differently from \citet{Matsumura17}, the formation of relatively
low-mass \DIFdelbegin \DIFdel{, }\DIFdelend CJs is possible here largely due to the new migration prescription
we adopted (see Section \ref{subsec:OrbEvol}). The results are
consistent with what we would expect from those in Section \ref{subsec:multiplePF}.
For the higher viscosity $\alpha_{{\rm turb}}$, the PIM is higher,
and thus planet formation becomes too slow (compared to type I migration)
to form giant planets in Discs 5 and 8. The giant planets are formed
only in the most rapidly dissipating disc (Disc 2), which provides
a high mass flux for a short period of time, but their masses are
low for giant planets.

For the cases of $\alpha_{{\rm turb}}=10^{-5}$, the formation rate
of giant planets, especially CJs beyond 1 au, is even higher than
the default case of $\alpha_{{\rm turb}}=10^{-4}$ for Discs 5 and
8, while only super-Earths are formed for Disc 2. The trends are again
consistent with those seen in Section \ref{subsec:multiplePF}. For
the lower $\alpha_{{\rm turb}}$ , both the PIM and the migration
transition mass are lower. Therefore, protoplanetary cores start gas
accretion at lower mass and also tend to switch from type I to type
II migration earlier and survive. In Disc 2, the cores are too small
to become giant planets within the disc's lifetime, while in Discs
5 and 8, CJs can be formed over a wide range of orbital radii.

\subsubsection{The effects of the pebble accretion efficiency \label{subsec:epsOL18}}

\begin{figure*}
\noindent %
\noindent\begin{minipage}[t]{1\columnwidth}%
\begin{center}
\includegraphics[width=0.9\paperwidth]{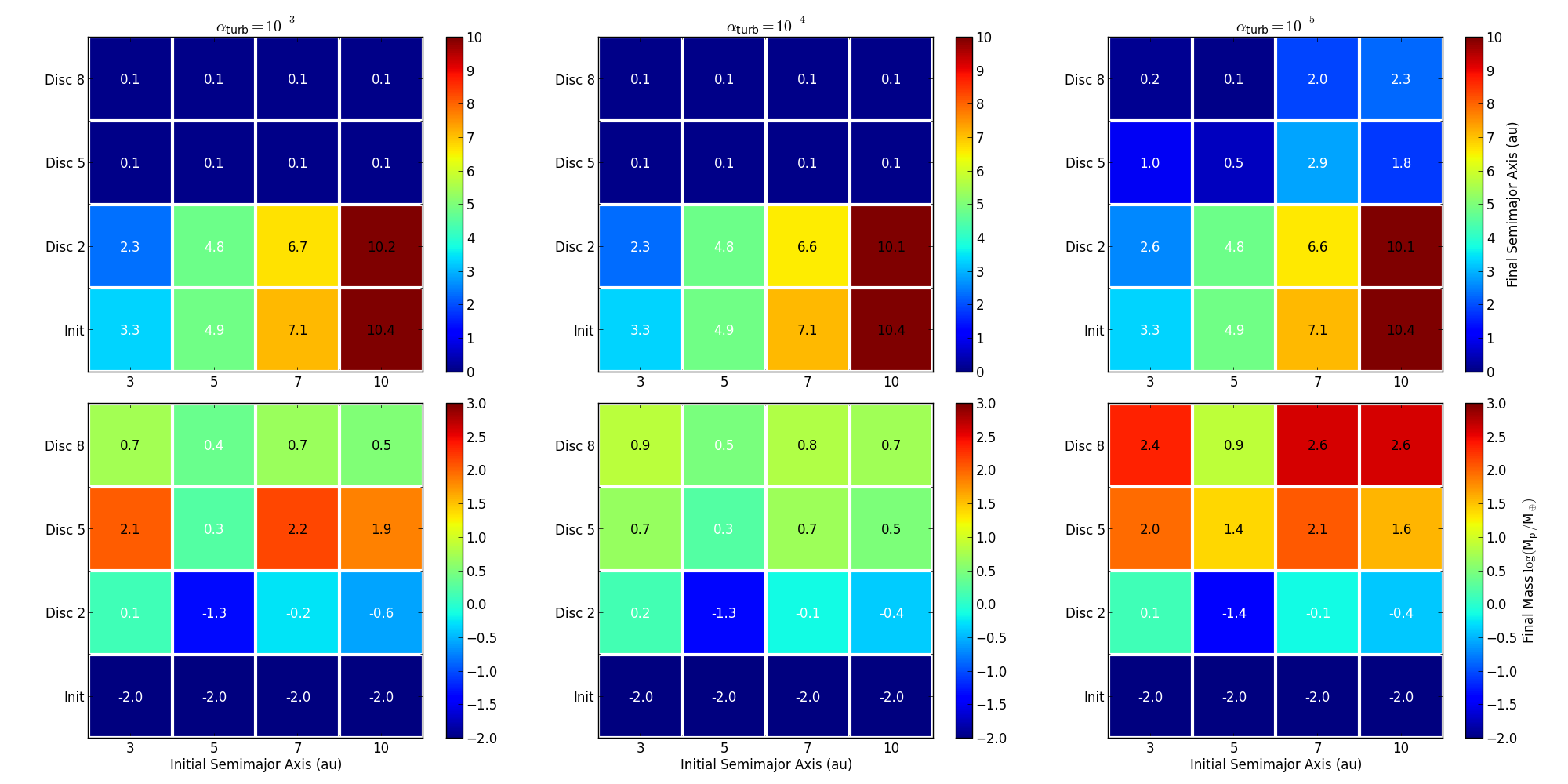} 
\par\end{center}%
\end{minipage}

\caption{Same as Figure \ref{fig:aMsingle2}, but with the pebble accretion
efficiency of \citet{Ormel18} rather than that of \citet{Ida16a}.\label{fig:aMsingle3}}
\end{figure*}

Figure \ref{fig:aMsingle3} shows the same as Figure \ref{fig:aMsingle2},
but with the pebble accretion efficiency by \citet{Ormel18} $\epsilon_{{\rm OL18}}$
(see Equation \ref{eq:epsOL18}) instead of that by \citet{Ida16a}
$\epsilon_{{\rm IGM16}}$ (see Equation \ref{eq:epsIda16}). Despite
the small differences between the two accretion efficiencies seen
in Figure \ref{fig:eps}, these two figures look dramatically different.
There are a couple of reasons here. First, a factor of a few difference
in accretion efficiencies actually leads to a large difference. For
example, a factor of 3 difference corresponds to the metallicity difference
between ${\rm [Fe/H]=0.0}$ and ${\rm [Fe/H]=0.5}$. Second, the accretion
efficiency by \citet{Ormel18} is more sensitive to the inclination.
For example, among cases shown in Figure \ref{fig:aMsingle3}, a protoplanet
at $\sim5\,$au consistently produces a much lower mass planet compared
to the neighbouring ones. This is because the randomly chosen initial
inclination for this core is $i\sim0.26$ degree (i.e. $e\sim0.01$)
as opposed to $i\lesssim0.14$ degrees (i.e. $e\lesssim0.005$) for
the others. Compared to the circular, coplanar case, the accretion
efficiency $\epsilon_{{\rm OL18}}$ decreases by more than an order
of magnitude with $e\sim0.01$ and $i\sim0.26$ degree, while the
difference is about a factor of a few with $e\lesssim0.005$ and $i\lesssim0.14$
degrees.

Besides the slow growth of a protoplanet at $\sim5\,$au, the outcomes
shown in Figure \ref{fig:aMsingle3} are consistent with Figure \ref{fig:aMsingle2}
and the expectations from Section \ref{subsec:multiplePF}. With $\alpha_{{\rm turb}}=10^{-5}$,
giant planets are formed across a range of disc radii for Discs 5
and 8. Compared to the corresponding cases in Figure \ref{fig:aMsingle2},
although the final masses of giant planets are comparable, the final
semi-major axes are smaller in these cases. This is because it takes
longer to form planets with lower efficiencies, and planetary
cores migrate more before becoming giant planets in these cases. In
Disc 2, only very low-mass planets are formed, which is also consistent
with this expectation.

With $\alpha_{{\rm turb}}=10^{-4}$, none of the cases with the solar
metallicity lead to giant planet formation. With $\alpha_{{\rm turb}}=10^{-3}$,
trends are similar to the cases in $\alpha_{{\rm turb}}=10^{-4}$,
but HJs are formed at the disc's inner edges in Disc 5. Since no giant
planets are formed in the corresponding cases in Figure \ref{fig:aMsingle2},
these cases may appear surprising. What happened here is that protoplanets
grew slower with the lower pebble accretion efficiencies, stayed longer
beyond the snow line, and obtained slightly more massive cores before
migrating into the inner disc region, where they become HJs by accreting
gas.

\begin{figure*}

\noindent\begin{minipage}[t]{1\columnwidth}
\begin{center}
\includegraphics[width=0.9\paperwidth]{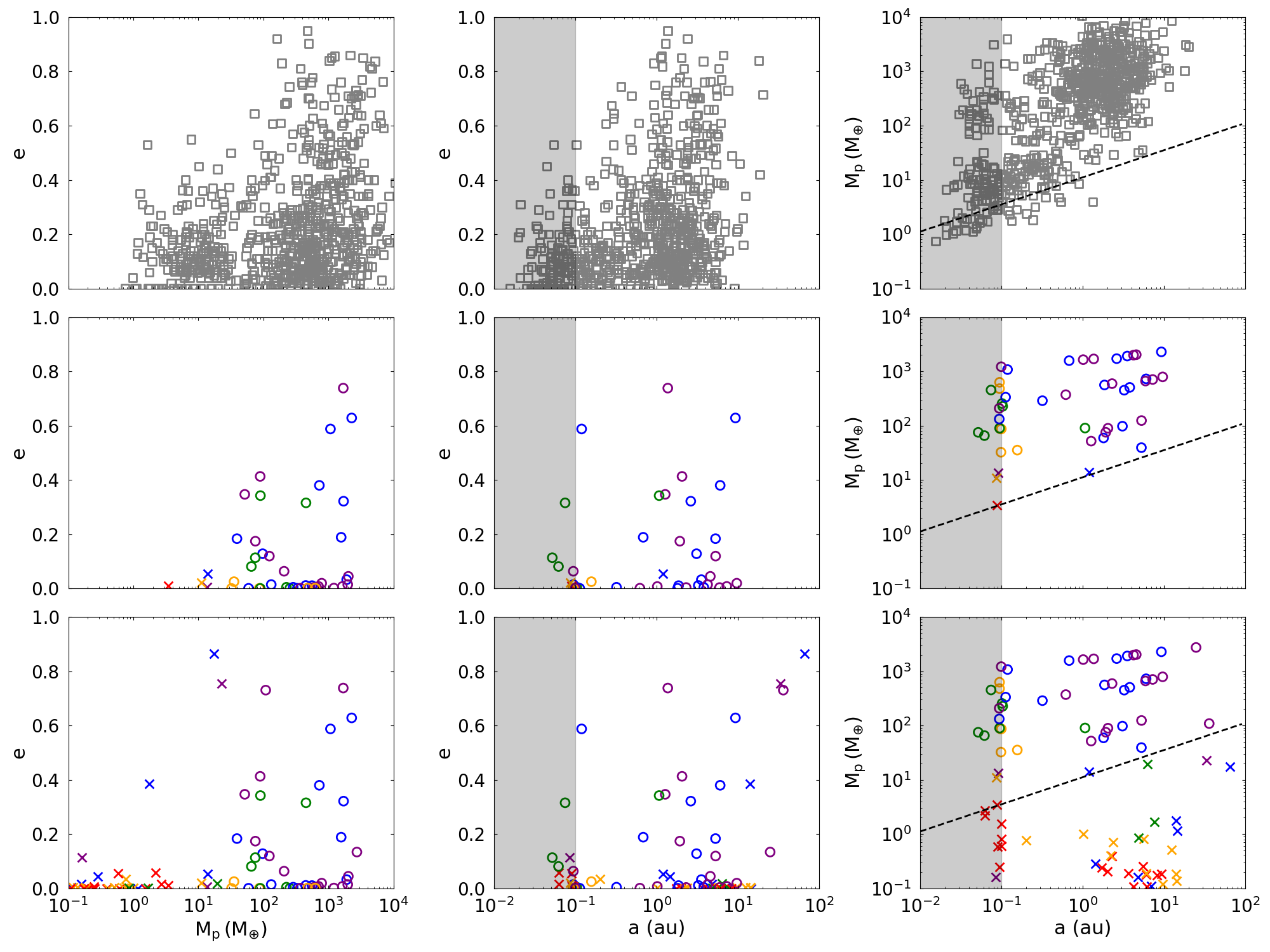}
\par\end{center}
\end{minipage}

\caption{Same as Figure \ref{fig:aM_Me_2panels}, but with the pebble accretion
efficiency $\epsilon_{{\rm OL18}}$ from \mbox{
\citet{Ormel18}}\hspace{0pt}
. Only Discs 2, 5, and 8 are used here.
The overall distributions are similar to Figure \ref{fig:aM_Me_2panels}, but giant planets are
formed mostly with high metallicities ${\rm [Fe/H]}\gtrsim0.3$ (see
text for further discussion). \label{fig:Me-ae-aM_OL18}}

\end{figure*}

For completion, we also ran a set of multiple-core simulations
with $\epsilon_{{\rm OL18}}$ and showed the results in Figure \ref{fig:Me-ae-aM_OL18}.
Here, we only use Discs 2, 5, and 8, but the other parameters are
the same as those shown in Section \ref{subsec:multiplePF} with the
default accretion efficiency $\epsilon_{{\rm IGM16}}$. The overall
distributions are similar to Figure \ref{fig:aM_Me_2panels} for all
the parameters, except that giant planets are mostly formed with high
metallicities ${\rm [Fe/H]}\gtrsim0.3$. This is consistent with the
expectation from single-core simulations shown above. 
\DIFadd{The lack of very massive giant planets ($\gtrsim 2000\,M_E$) compared to Figure \ref{fig:aM_Me_2panels} 
is unlikely to be the real feature.  As seen in Figure \ref{fig:aMgiants}, most massive giant planets are formed 
in most massive discs ($0.2\,M_{\odot}$ in our case), while all the discs used in this figure have $0.06\,M_{\odot}$. }\DIFaddend 
Therefore, to
have distributions similar to Figure \ref{fig:aM_Me_2panels} with
the pebble accretion efficiency $\epsilon_{{\rm OL18}}$ by \citet{Ormel18}, 
we would probably need to assume lower turbulence alpha $\alpha_{{\rm turb}}<10^{-4}$
and very low inclinations for initial protoplanetary cores.

 \section{Discussion \label{sec:Discussion}}

 In this section, we further discuss our results by focusing on several
key topics. First, we compare planet occurrence rates obtained from
our simulations with observations in Section \ref{subsec:occurrence}.
Then, we discuss formation of HJs (Section \ref{subsec:Form_HJs}) and 
SCJs (Section \ref{subsec:Form_SCJs}). 
We also discuss to what orbital radii planets can be formed via pebble accretion 
and how massive planets can become in Section \ref{subsec:Max_Mp_a}, and 
discuss metal and gas mass fractions of planets in Section \ref{subsec:MassFrac_Metals}.
Finally, we compare our results with planetesimal accretion work (Section
\ref{subsec:plsml_vs_pebble}) as well as other pebble accretion work
(Section \ref{subsec:Comp_Bitsch19}).

\subsection{Planet occurrence rates \label{subsec:occurrence}}

\begin{figure*}
\begin{minipage}[t]{1\columnwidth}%
\begin{center}
\DIFdelbeginFL 
\DIFdelendFL \includegraphics[scale=1.0]{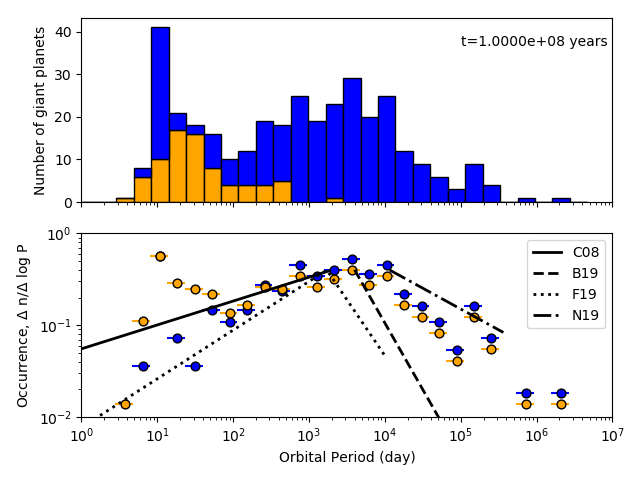} 
\par\end{center}%
\end{minipage}\caption{Top: Distribution of orbital periods of giant planets at the end of
the simulations (blue) and those of giant planets whose orbits are
tidally evolved for 3 Gyr (orange, see Section \ref{subsec:HJs}).
Bottom: Corresponding occurrence rates of giant planets as a function
of orbital period. The blue symbols only include simulated giant planets,
while the orange ones include both simulated and tidally evolved giant
planets. The rates are calculated for the number of planets per star
per $\Delta\log P=0.23$ to be compared with Figure 3 of \citet{Winn18}.
\DIFdelbeginFL \DIFdelFL{Grey area indicate the }\DIFdelendFL The black slopes \DIFdelbeginFL \DIFdelFL{suggested by \mbox{
\citet{Cumming08} }\hspace{0pt}
of }\DIFdelendFL are occurrence rates estimated from observed planets.
The solid, dashed, dotted, and dot-dashed lines correspond to $dn/d\ln P\propto P^{0.26\pm0.1}$
for giant planets with $>0.3\,M_{J}$ within $2000\,$days \DIFdelbeginFL \DIFdelFL{and those
suggested }\DIFdelendFL from \DIFdelbeginFL \DIFdelFL{\mbox{
\citet{Baron19} }\hspace{0pt}
of }\DIFdelendFL 
\citet{Cumming08}, $dn/d\ln P\propto P^{-1.45_{-0.44}^{+0.51}}$ for giant planets $5-5000\,$au
from \mbox{
\citet{Baron19}}\hspace{0pt}
, an asymmetric broken power law of $dn/d\ln P\propto P^{0.53\pm0.09}$
and $dn/d\ln P\propto P^{-1.22\pm0.47}$ with a break period of $P_{{\rm break}}=1717\pm432\,$days
for giant planets with $0.1-20\,M_{J}$ with periods $\sim1-10^{4}\,$days
from \mbox{
\citet{Fernandes19}}\hspace{0pt}
, and $dn/d\ln P\propto P^{-0.453}$ for giant
planets with $5-13\,M_{J}$ with periods $\sim1.16\times10^{4}-3.65\times10^{5}\,$days
($\sim10-100\,$au) from \mbox{
\citet{Nielsen19}}\hspace{0pt}
, respectively. \label{fig:occurrence}}
\end{figure*}

\DIFdelbegin \textbf{\DIFdel{\textcolor{blue}{The top panel of Figure \ref{fig:occurrence}
shows the overall distribution of orbital periods of giant planets.
The blue and orange histograms correspond to the distributions of
simulated planets and tidally evolved planets from Section \ref{subsec:HJs},
respectively. The strong peak seen around $10\,$days comes from the
inner disc edge being set at $0.1\,$au in our simulations, and the
actual distribution around this period should be broader depending
on the locations of the inner disc edges. Otherwise, the overall distribution
shows that CJs near $10^{3}-10^{4}\,$days are most abundant and the
number decreases for planets interior and exterior to these orbits. }}}
\DIFdelend  
The top panel of Figure \ref{fig:occurrence} shows the overall distribution
of orbital periods of giant planets from our work. The blue and orange
histograms correspond to the distributions of simulated planets and
tidally evolved planets from Section \ref{subsec:HJs}, respectively.
The strong peak seen around $10\,$days comes from the inner disc
edge being set at $0.1\,$au in our simulations, and the actual distribution
around this period should be broader depending on the locations of
the inner disc edges. Otherwise, the overall distribution shows that
CJs near $10^{3}-10^{4}\,$days are most abundant, and the number decreases
for planets interior and exterior to these orbits.

\DIFdelbegin \textbf{\DIFdel{\textcolor{blue}{The bottom panel of Figure \ref{fig:occurrence}
shows the corresponding occurrence rates compared with observed slopes
(shaded). Following the approach by \citet{Winn18}, we write the
occurrence rate as: $\frac{\Delta n}{\Delta\log P}$, where $n=N_{{\rm pl}}/N_{*}$
is the ratio of the number of giant planets to that of the host star,
while $\Delta\log P=0.23$ is chosen so that the comparison with Figure
3 of \citet{Winn18} is easier. The blue and orange symbols correspond
to the occurrence rates calculated for simulated planets only and
simulated and tidally evolved planets together, respectively. The
total number of simulated giant planets is $N_{{\rm pl,\,sim}}=274$
and the corresponding number of stars is $N_{{\rm *,\,sim}}=240$,
while we just assume the number of the tidally-evolved planets and
and that of stars are the same: $N_{{\rm pl,\,tide}}=N_{{\rm *,\,tide}}=76$.
Two shaded areas indicate the slopes of $\frac{dn}{d\ln P}\propto P^{-\beta}$
obtained from two observational studies \citep{Cumming08,Baron19}.
\citet{Cumming08} suggested $\beta=0.26\pm0.1$ for observed giant
planets with masses $>0.3\,M_{J}$ and periods $<2000\,$days, while
\citet{Baron19} suggested $\beta\sim-1.45_{-0.44}^{+0.51}$ for directly-imaged
planets with masses $1-20\,M_{J}$ and orbital radii of $5-5000\,$au.
Note that the scaling factors used for these shaded areas are chosen
to make the comparison with our data easier, and the actual observed
occurrence rates of these planets are much lower since $N_{*}$ is
larger and the detection probability is not $100\,\%$ \citep[e.g.][]{Winn18}.
Here, we only attempt to compare the slopes to see the relative occurrence
rates of different types of giant planets. }}}
\DIFdelend  
The bottom panel of Figure \ref{fig:occurrence} shows the corresponding
occurrence rates compared with observed slopes. Following the approach
by \mbox{
\citet{Winn18}}\hspace{0pt}
, we write the occurrence rate as: $\frac{\Delta n}{\Delta\log P}$,
where $n=N_{{\rm pl}}/N_{*}$ is the ratio of the number of giant
planets to that of the host star, while $\Delta\log P=0.23$ is chosen
so that the comparison with Figure 3 of \mbox{
\citet{Winn18} }\hspace{0pt}
is easier.
The blue and orange symbols correspond to the occurrence rates calculated
for simulated planets only and simulated and tidally evolved planets
together, respectively. The total number of simulated giant planets
is $N_{{\rm pl,\,sim}}=274,$ and the corresponding number of stars
is $N_{{\rm *,\,sim}}=240$, while we just assume the number of the
tidally evolved planets and that of stars are the same: $N_{{\rm pl,\,tide}}=N_{{\rm *,\,tide}}=76$.

\DIFdelbegin \textbf{\DIFdel{\textcolor{blue}{The observed slopes agree well with the trends
seen for simulated planets \textemdash{} the occurrence rates increase
with orbital periods out to $\sim750\,{\rm day}\sim1.6\,$au, decrease
with periods slightly beyond $\sim10^{4}\,{\rm day}\sim9\,$au, and
have a broad peak between these periods. As seen in the figure, the
ascending and descending trends on both sides of the peak are reasonably
well-explained by the slopes of the occurrence rates estimated by
\citet{Cumming08} and \citet{Baron19}. The occurrence rates increasing
beyond $\sim1\,$au is also consistent with estimates both by RV surveys
\citep{Mayor11ap} and by Kepler transit survey \citep{Santerne16}.
The overall shape is also consistent with \citet{Bryan16} who suggested
a broad peak in the distribution between 3 and 10 au and positive
and negative $\beta$ values interior and exterior to these regions. }}}
\DIFdelend  
The lines shown in the figure indicate the slopes of $\frac{dn}{d\ln P}\propto P^{-\beta}$
obtained from four studies of observed systems \citep{Cumming08,Baron19,Fernandes19,Nielsen19}.
\citet{Cumming08} used RV-detected planets around FGK stars and suggested
$\beta=0.26\pm0.1$ for giant planets with masses $>0.3\,M_{J}$ and
periods $<2000\,$days, while \mbox{
\citet{Fernandes19} }\hspace{0pt}
investigated both
RV and Kepler studies and found a broken power-law function with $\beta=0.53\pm0.09$
and $\beta=-1.22\pm0.47$ with the break period of $1717\pm\,432$days.
The other two studies focus on outer planets observed by direct imaging
methods. \mbox{
\citet{Baron19} }\hspace{0pt}
suggested $\beta\sim-1.45_{-0.44}^{+0.51}$
for giant planets with masses $1-20\,M_{J}$ and orbital radii of
$5-5000\,$au, while \mbox{
\citet{Nielsen19} }\hspace{0pt}
found $\beta\sim-0.453$ for
planets with $5-13\,M_{J}$ and orbital radii of $10-100\,$au. 
We note that the scaling factors used for these lines are chosen to make the
comparison with our data easier, and the actual observed occurrence
rates of these planets are much lower since $N_{*}$ is larger and
the detection probability is not $100\,\%$ \mbox{
\citep[e.g.][]{Winn18}}\hspace{0pt}
.
Here, we only attempt to compare the slopes to see the relative occurrence
rates of different types of giant planets.

\DIFdelbegin \textbf{\DIFdel{\textcolor{blue}{Within $\sim200\,$day, the occurrence rates
estimated from simulated and tidally-evolved planets (orange) show
a much flatter distribution and do not agree so well with the observed
trends. However, \citet{Cumming08} also reports that the observed
data is equally well-fitted with $\beta=0$ out to $\sim300\,$day
followed by a sharp increase, suggesting that such a flatter distribution
is possible as well. }}}
\DIFdelend  
The overall trends of observed slopes agree well with those seen for
simulated planets. The occurrence rates estimated from our simulations
increase with orbital periods out to $\sim750\,{\rm days}\,(\sim1.6\,{\rm au})$,
decrease with periods slightly beyond $\sim10^{4}\,{\rm days}\,(\sim9\,{\rm au})$,
and have a broad peak between these periods. As seen in the figure,
the inner ascending trend is well explained by the slopes estimated
by \mbox{
\citet{Cumming08} }\hspace{0pt}
and \mbox{
\citet{Fernandes19}}\hspace{0pt}
. The fact that the occurrence rates
increase beyond $\sim1\,$au is also consistent with estimates both
from RV surveys \mbox{
\citep{Mayor11ap} }\hspace{0pt}
and the Kepler transit survey \mbox{
\citep{Santerne16}}\hspace{0pt}
.
The outer descending trend is also 
\DIFadd{broadly consistent with the estimates by 
\citet{Nielsen19}, \citet{Baron19}, and \citet{Fernandes19}, though details are different. 
\citet{Baron19} proposed a much steeper slope compared to our predictions, and }\DIFaddend
\citet{Fernandes19} estimated 
the break point to be at a shorter orbital period. 
The overall shape is also consistent with \mbox{
\citet{Bryan16}, }\hspace{0pt}
who suggested
a broad peak in the distribution between 3 and 10 au and positive
and negative $\beta$ values interior and exterior to these regions.
However, where the occurrence rates start decreasing is still unclear.
The future observations such as Gaia may provide a population of planets
that help to constrain this trend better.

\DIFdelbegin \textbf{\DIFdel{\textcolor{blue}{The total occurrence rates are estimated
to be $\sim10\,\%$ for giant planets with $\gtrsim0.3\,M_{J}$ and
$\lesssim3-4.6\,$au \citep{Cumming08,Mayor11ap}, $\sim52.4\,\%$
for planets with $1-20\,M_{J}$ and $5-20\,$au \citep{Bryan16},
and $2.61_{-1.00}^{+6.97}\,\%$ for giant planets with $1-20\,M_{J}$
and $20-5000\,$au. The trends of these occurrence rates are broadly
consistent with those seen in our simulations. }}}
\DIFdelend  
The total occurrence rates are estimated to be $\sim10\,\%$ for giant
planets with $\gtrsim0.3\,M_{J}$ and $\lesssim3-4.6\,$au \citep{Cumming08,Mayor11ap}, 
$\sim52.4\,\%$ for planets with $1-20\,M_{J}$ and $5-20\,$au \mbox{
\citep{Bryan16}}\hspace{0pt}
,
and $2.61_{-1.00}^{+6.97}\,\%$ for giant planets with $1-20\,M_{J}$
and $20-5000\,$au. The trends of these occurrence rates are broadly
consistent with those seen in our simulations.

\DIFdelbegin \textbf{\DIFdel{\textcolor{blue}{Table \ref{tab:occurrence} summarises the
numbers of HJs, WJs, CJs, and SCJs formed in our simulations. Note
that the fractions shown here are determined with respect to the total
number of giant planets and are different from the occurrence rates.
We find that $>50\,\%$ of all the giant planets are CJs, and thus
more than half the giant planets stay near the formation region. }}}
\DIFdelend  Table \ref{tab:occurrence} summarises the numbers of HJs, WJs, CJs,
and SCJs formed in our simulations. We note that the fractions shown
here are determined with respect to the total number of giant planets
and are different from the occurrence rates. We find that $>50\,\%$
of all the giant planets are CJs, and thus more than half the giant
planets stay near the formation region.

\begin{table*}[h!]
\centering
\begin{tabular}{|c|c|c|c|c|c|}
\hline 
 & HJs  & WJs  & CJs  & SCJs  & Total\tabularnewline
\hline 
 & $\leq0.1\,$au  & $0.1-1\,$au  & $1-20\,$au  & $>20\,$au  & \tabularnewline
\hline 
\hline 
$N_{{\rm pl,\,sim}}$  & $33$  & $43$  & $174$  & $24$  & $274$\tabularnewline
\hline 
$\%$  & $12.0$  & $15.7$  & $63.5$  & $8.8$  & $100$\tabularnewline
\hline 
$N_{{\rm pl,\,sim}}+N_{{\rm pl,\,tide}}$  & $50$  & $96$  & $180$  & $24$  & $350$\tabularnewline
\hline 
$\%$  & $14.3$  & $27.4$  & $51.4$  & $6.9$  & $100$\tabularnewline
\hline 
\end{tabular}

\caption{Numbers and fractions of HJs, WJs, CJs, and SCJs as defined in
the semi-major axis range shown. $N_{{\rm sim}}$ corresponds to the
number of simulated giant planets, while $N_{{\rm tide}}$ corresponds
to the number of giant planets for which tidal evolution is further
calculated for 3 Gyr (see Section \ref{subsec:HJs}). \label{tab:occurrence}}
\end{table*}

\subsection{Formation of hot Jupiters \label{subsec:Form_HJs}}

Any HJ formation model needs to explain that (1) HJs tend not to have nearby companions
\citep[e.g.][]{Wright09,Steffen12,Huang16}, and (2) planets
around higher metallicity stars tend to have a wider range of eccentricities
\citep[e.g.][]{Dawson13,Shabram16,Buchhave18}. Our simulations naturally
explain both of these trends.

When HJs are formed via the tidal circularisation of highly eccentric
orbits, the underlying assumption is that planetary systems have undergone
some sort of dynamical evolution \DIFdelbegin \DIFdel{increasing eccentricities}\DIFdelend  that increased eccentricities, such
as planet-planet scattering \citep[e.g.][]{Ford08,Chatterjee08,Juric08},
the secular chaos \citep[e.g. ][]{Wu11}, or the Kozai-Lidov mechanism
\DIFdelbegin \DIFdel{\mbox{
\citep[e.g.][]{Naoz16}}\hspace{0pt}
}\DIFdelend  \mbox{
\citep[e.g.][]{Fabrycky07,Naoz16}}\hspace{0pt}
. The evolution of such planets often
removes neighbouring planets, and the resulting HJs tend to be alone
and have a range of eccentricities and inclinations. In our simulations,
these HJs \DIFdelbegin \DIFdel{tend
to }\DIFdelend  often belong to high-metallicity and/or massive discs, which
typically produce a few giant planets that are prone to dynamical
instabilities.
On the other hand, when HJs form in situ (i.e. via the migration of cores
followed by gas accretion in our simulations), they often belong to
low-metallicity discs, which typically produce 1-2 giant planets. Therefore,
the resulting HJs tend to be alone and have low eccentricities and
inclinations due to damping by the disc. HJs can also be formed in situ
 or via migration in high-metallicity discs. Such discs usually form
multiple giant planets initially, including HJs, WJs, and CJs, but
HJs and/or WJs are often removed via dynamical instabilities later
on, leaving either (i) only CJs, or (ii) HJs and CJs  \mbox{
\citep[also see][]{Mustill15}}\hspace{0pt}
.
Therefore, even in these cases, HJs usually have no neighbouring giant
planets (but could be accompanied by CJs further out). 

\DIFdelbegin \DIFdel{It is also interesting to compare the standard migration scenario
(i.e. a giant planet forms in the outer part of the disc, and then
migrates to the inner region to become a HJ) with the in-situ HJ formation
scenario. In the standard scenario, if a giant planet forms randomly
at some point of the disc's lifetime, we may expect a uniform distribution
of giant planets from very close to the star out to beyond the snow
line. However, giant planets are known to be clustered within $\sim0.1\,$au
and beyond $\sim1\,$au \mbox{
\citep[e.g.][]{Wright09}}\hspace{0pt}
. In our simulations,
the distribution of giant planets is significantly influenced by stellar
metallicities (as well as disc dissipation timescales and disc masses).
In low-metallicity discs, planet formation is slower and thus protoplanetary
cores in such discs migrate significantly before accreting gas, except
for massive, rapidly dissipating discs. Therefore, the distribution
of giant planets in low-metallicity discs tends to be binary with
HJs and CJs but few WJs. On the other hand, more metal-rich discs
can produce all sorts of giant planets (HJs, WJs, CJs, and SCJs),
leading to a more broader distribution of planets. The right panel
of Figure \ref{fig:ecc-obl-met} shows the orbital distribution of
observed giant planets for different stellar metallicities. For low-metallicity
stars, giant planets are either HJs or CJs, agreeing with our simulations.
It is possible that the trend changes if type I migration in the wind-driven
accretion disc is indeed slowed down as proposed by \mbox{
\citet{Ogihara15}
}\hspace{0pt}
(see Section \ref{subsec:Form_SEs} as well). The future observational
and theoretical/numerical studies can either support or disprove this
trend.
}\DIFdelend  

There may be some difficulties for the scenario in which HJs are formed in situ,
in particular in low-metallicity discs. First, gas accretion may be
stalled near the central stars, because the gas envelope becomes nearly
isothermal and the timescale of replenishing the atmosphere with the
disc gas becomes shorter than the Kelvin-Helmholtz gas accretion timescale
\citep[e.g.][]{Ormel15a,Ormel15b,AliDib20}. However, the gas accretion
may still be possible when the disc is heated by the stellar radiation
alone and temperature and entropy in the inner disc region is low
\citep{AliDib20}, which may be the case towards the end of the disc's
lifetime. Second, the near-infrared disc fraction study by \citet{Yasui10}
suggested that the disc fraction of the low metallicity clusters declines
within 1 Myr as opposed to several Myr for the solar-metallicity clusters.
If the trend were to be confirmed by longer-wavelength studies, this may
place a strong constraint on planet formation timescales in the low-metallicity
environments. In our simulations, giant planets could be formed within
$\sim1\,$Myr even in the low-metallicity discs as long as disc masses
are high enough (see Figure \ref{fig:GPformtime}), but the disc's
lifetimes are much longer than that. Since our current studies do
not take account of the feedback of the planets onto discs, the future
study needs to investigate this further.

\subsection{Upper limits on masses and planet formation radii \label{subsec:Max_Mp_a}}

It is unclear what \DIFdelbegin \DIFdel{would be }\DIFdelend the maximum planetary mass generated by the core
accretion scenario  would be. Recently, \citet{Schlaufman18} pointed
out that giant planets with $\lesssim4\,M_{J}$ tend to be around
metal-rich dwarfs, while the same trend is not observed for $\gtrsim10\,M_{J}$.
He suggested that these two distinct populations may correspond to
core accretion and gravitational instability scenarios, with a transition
mass being $\sim4-10\,M_{J}$. The transition mass is also comparable
to planetary masses generated by the gravitational instability ($\gtrsim3-5\,M_{J}$)
that are reported in the semi-analytic studies \citep[e.g.][]{Kratter10,Forgan11}
and the numerical studies \citep[e.g.][]{Stamatellos08,Hall17}. Most
of our simulations do not generate planets above $\sim10\,M_{J}$,
and planetary masses of $\sim20\,M_{J}$ were only found in massive,
long-lived discs (Disc 7). As discussed in Section \ref{subsec:Initial-Conditions},
it will probably be difficult to go above $\sim50\,M_{J}$ with our
model. Thus, although the highest-mass planets or brown dwarfs 
may be formed by gravitational instability, there is probably a significant overlap
for masses of bodies both scenarios can generate.

How far away a planet can be formed via core accretion is not well
understood either. In this work, we did not place planetary cores
beyond $\sim20\,$au because pebble accretion becomes inefficient
for a Ceres-size planetesimal to grow in the outer disc in our model
(see Section \ref{subsec:Initial-Conditions}). However, as we  also show in Section \ref{subsec:Initial-Conditions}, pebble accretion
may still lead to giant planet formation out to and beyond $100\,$au
if much larger planetesimals form there. If there are massive enough
planetesimals available, pebble accretion (along with planetesimal
accretion) may allow the growth of giant planets far beyond $20\,$au.
Recent disc observations including ALMA regularly find gaps and rings
over a wide range of disc radii, which may be attributed to growing
planets \citep[e.g.][]{Zhang16,Huang18,Long18}. \citet{Lodato19}
estimated planetary masses from 48 observed gaps located around $\sim10-130\,$au
and found planetary masses ranging from $10^{-3}\,M_{J}\sim0.3\,M_{\oplus}$
to over $10\,M_{J}$. If these gaps are indeed signposts of planets,
the fact that a range of planetary masses is observed in the outer
disc may indicate that core accretion is active out to these radii
because gravitational instability is unlikely to form lower end mass
planets. There is, however, a discrepancy between the frequency of
observed SCJs \DIFdelbegin \DIFdel{\mbox{
\citep[$<4.1\%$ for FGK stars,][]{Bowler16} }\hspace{0pt}
}\DIFdelend  \mbox{
\citep[<4.1 for FGK stars,][]{Bowler16} }\hspace{0pt}
and the fact
that the substructures such as rings and gaps are ubiquitous among
observed discs \citep[e.g.][]{Andrews18,Huang18}. Thus, if all of
these rings and gaps correspond to planets, such planets would have
to be removed from the outer disc region efficiently. This could be
achieved if forming planets are redistributed due to migration \citep{Lodato19}
or removed due to episodic accretion to the star \citep{Brittain20}.
The future studies should investigate the connection between forming
planets and evolving protoplanetary discs further.

\subsection{Formation of super cold Jupiters \label{subsec:Form_SCJs}}

In Section \ref{subsec:SCJs}, we discuss two different pathways
of forming SCJs via core accretion: (1) SCJs form in situ, either
directly in the outer disc or 
via scattering of protoplanetary cores to the outer disc followed by gas accretion there \citep[e.g.][]{Kikuchi14}, 
or (2) giant planets first form within $\sim20\,$au and are then scattered into
the outer disc. These two different scenarios are likely to lead to
different distributions of physical and orbital parameters.

First, the in situ formation scenario of SCJs may lead to the \DIFdelbegin \DIFdel{distribution
}\DIFdelend  $a-M_{p}$
correlation where the planetary mass increases with the semi-major
axis\DIFdelbegin \DIFdel{. This is
}\DIFdelend  , because the disc aspect ratio increases with \DIFdelbegin \DIFdel{the semimajor axis }\DIFdelend  $a,$ and thus the
critical gap-opening mass for giant planets also becomes higher \citep{Ida13}.
On the other hand, the formation-then-scattering scenario might lead
to an opposite trend because a lower-mass planet is more easily scattered
further. We checked the Spearman's rank correlation for simulated
giant planets beyond $1\,$au (see the left panel of Figure \ref{fig:SCJs})
and found the correlation coefficient of $r_{s}\sim-0.18$ with the
p-value of $p=0.011$ ($r_{s}\sim-0.19$ with the p-value of $p=0.0015$
for all). Since the critical value for $p<0.05$ is $\mid r_{s}\mid\sim0.2$
for the size of our data, there is little or very weak negative correlation
between the semi-major axis and the mass for giant planets. Due to
the observational bias for high-mass planets, it is difficult to assess
whether there is any trend for observed planets.

Second, the eccentricity distributions may also be different between
these two scenarios. The \DIFdelbegin \DIFdel{scenario (2) (i.e. }\DIFdelend formation-then-scattering \DIFdelbegin \DIFdel{)
}\DIFdelend  scenario is likely
to lead to an isolated eccentric giant planet, while the in situ formation
scenario is more likely to lead to single- or multiple-giant planets
on nearly circular orbits. In our simulations, giant planets scattered
beyond $20\,$au have eccentricities ranging from $e\sim0.2-0.8$
with the mean value of $0.56$ (see Figure \ref{fig:aM_Me_2panels}).
\citet{Kikuchi14}, on the other hand, showed that the eccentricity
of the scattered core is reduced to $e<0.2$ before reaching the Saturn
mass via gas accretion in the outer disc. The eccentricities of multiple
giant planets could increase via dynamical instabilities after formation,
as long as they are not too far from each other \citep[e.g. ][]{Chatterjee08}.

Recently, \citet{Bowler20} studied the eccentricity distributions
of 18 brown dwarfs ($15-75\,M_{J}$) and nine giant planets ($2-15\,M_{J}$)
with orbital radii $5-100\,$au, and they found that the estimated eccentricity
distribution is very broad and ranges over $e=0-1$ for brown dwarfs
and is much narrower with a peak at $\bar{e}=0.13$ for giant planets,
which may support the in situ formation scenario of SCJs more than the
formation-then-scattering one. However, the peak eccentricity becomes
$\bar{e}=0.23$ for five giant planets excluding four HR 8799 planets
on nearly circular orbits. The future observations will  further constrain
the eccentricity distribution \DIFdelbegin \DIFdel{further }\DIFdelend and provide us with an important clue regarding the formation pathways of SCJs.

\subsection{Mass fractions and metal contents \label{subsec:MassFrac_Metals}}

\begin{figure*}
\noindent 
\noindent\begin{minipage}[t]{1\columnwidth}
\begin{center}
\includegraphics[width=0.9\paperwidth]{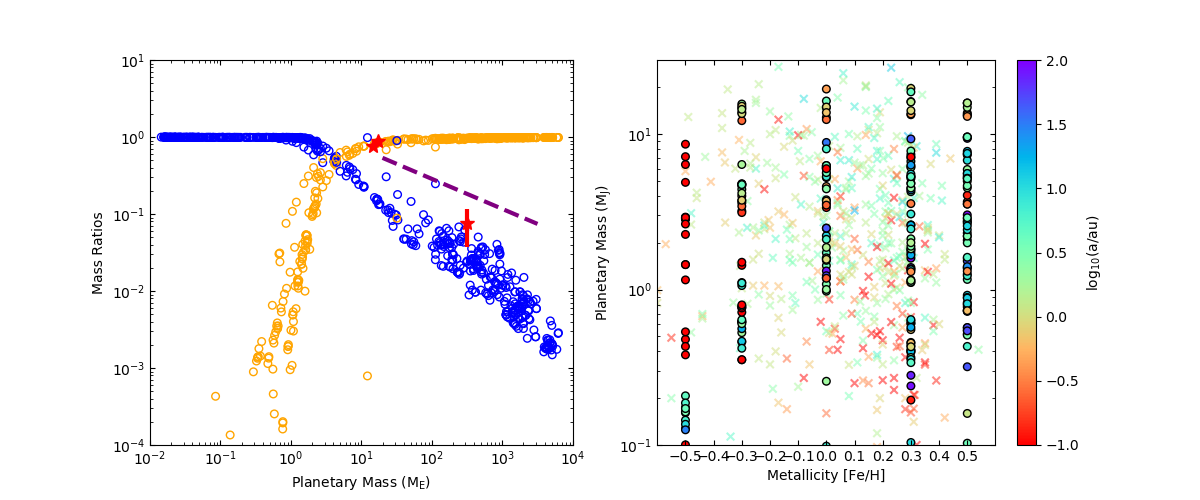} 
\par\end{center}
\end{minipage}\caption{Left: Mass fractions of the rocky core (blue) $\frac{M_{{\rm core}}}{M_{p}}$
and the gaseous envelope (orange) $\frac{M_{{\rm env}}}{M_{p}}$ for
all the planets formed in our simulations. The core and the envelope
have comparable masses $M_{{\rm core}}\sim M_{{\rm env}}$ when a
planetary mass is $\sim4\,M_{\oplus}$. The purple dashed line shows
the metal fractions of observed giant planets estimated by \mbox{
\citet{Thorngren16}.}\hspace{0pt}
The red stars with error bars indicate the metal-mass fractions estimated
for Jupiter, Uranus, and Neptune \mbox{
\citep{Fortney10}}\hspace{0pt}
. Right: Planetary
mass and metallicity for simulated (circles) and observed (crosses)
giant planets. The colours indicate the final and observed semi-major
axes, respectively. Masses are generally independent of the stellar
metallicity. \label{fig:Mass_core_env}}
\end{figure*}

The left panel of Figure \ref{fig:Mass_core_env} shows fractions
of a planetary mass in the core (blue) and in the envelope (orange)
for all the simulated planets. The envelope should usually also
contain some metals, but here we added all the accreted dust particles
to the core mass. Also plotted are the fitted metal-mass fractions
of extrasolar planets from \mbox{
\citet{Thorngren16} }\hspace{0pt}
(dashed line) and
the estimated metal fractions in Jupiter, Uranus, and Neptune 
\citep[stars with error bars,][]{Fortney10,Helled20}.
As we can see, mass fractions of planetary cores of our simulations
are lower than the estimate by \mbox{
\citet{Thorngren16} }\hspace{0pt}
by a factor of
a few or more, though the distribution is consistent with the estimated
metal-mass fraction of Jupiter.

This underestimation of the metal-mass fractions in our simulations
may result from our assumption of gas accretion. For our simulations,
the \DIFdelbegin \DIFdel{orbital distribution of SEs is more
binary with either close-in, hot SEs or cold SEs beyond $\sim1\,$au,
while that of Earth-sized planets is much broader.
This binary distribution
occurs because cold SEs are formed in rapidly dissipating discs (Discs
1-3),
while hot SEs are formed in longer-lived discs where SEs migrated
all the way to the inner disc edge (0.1 au). Such a binary orbital
distribution is clearly different }\DIFdelend  core mass and the envelope mass become comparable to each other:
$M_{{\rm core}}\sim M_{{\rm env}}$ when $M_{p}\sim4\,M_{E}$. As
seen in the figure, this is much lower than the crossover mass expected
from the observed \DIFdelbegin \DIFdel{trend. }\DIFdelend  planets ($\sim15\,M_{E}$). The crossover mass represents
a group of planets that are on the verge of becoming giant planets
but failed to do so because the gas disc dissipated (i.e. $t_{{\rm diff}}\sim\tau_{KH}$):

\DIFdelbegin \DIFdel{This discrepancy could be resolved if a chain of planets could be
trapped in mean motion resonances at the disc inner edge, which becomes
dynamically unstable as the gas disc dissipates and leads to formation
of multiple non-resonant SEs \mbox{
\citep[e.g. ][]{Ogihara10,Lambrechts19,Izidoro19ap}}\hspace{0pt}
.This method, however, requires a sharp enough disc edge \mbox{
\citep{Ogihara10}}\hspace{0pt}
,
and non-saturated or weakly-saturated corotation resonances \mbox{
\citep{Brasser18}}\hspace{0pt}
. }

\DIFdel{Another possibility of saving these type I migrators may be to take
account of a more careful model of the wind-driven accretion. For
example, \mbox{
\citet{Ogihara17} }\hspace{0pt}
showed that }\DIFdelend  
\begin{equation}
\frac{M_{p}}{M_{E}}\sim10^{\frac{b-6}{c}}\left(\frac{t_{{\rm diff}}}{10^{6}\,{\rm yr}}\right)^{-\frac{1}{c}}\sim4.64\left(\frac{t_{{\rm diff}}}{10^{6}\,{\rm yr}}\right)^{-\frac{1}{3}}.
\end{equation}
In \mbox{
\citet{Matsumura17}}\hspace{0pt}
, the crossover mass was $M_{{\rm core}}\sim M_{{\rm env}}\sim10\,M_{\oplus}$
as seen in Figure 11, where we adopted $(b,\,c)=(9,\,3)$
rather than $(8,\,3)$. Thus, it is likely that we need to adopt $(b,\,c)=(9,\,3)$
to form planets with more appropriate rock-gas mass ratios. The reason
why we adopted $(b,\,c)=(8,\,3)$ instead was to speed up gas
accretion and to avoid the loss of planets due to type I migration\DIFdelbegin \DIFdel{could be
slowed in }\DIFdelend  .
As discussed in Section \ref{subsec:OrbEvol}, the pebble isolation
mass is reached before the mass of migration transition from type
I to type II. Therefore, the gas accretion timescale in this phase needs
to be short compared to the migration timescale for a growing planet
to switch from type I to type II regime. This further confirms the
need for a proper mechanism of type I migration.

The right panel of Figure \ref{fig:Mass_core_env} compares the mass-metallicity
relation for simulated (circles) and observed (crosses) giant planets.
The apparent lack of simulated planets with $\sim0.1-1\,M_{J}$ for
the solar metallicity case is most likely due to the \DIFdelbegin \DIFdel{inner disc region due to the flattening of the surface
mass density profile, although whether such a density profile would
actually be achieved is inconclusive from numerical studies \mbox{
\citep{Suzuki16,Bai16}}\hspace{0pt}
. }
\DIFdel{More recently, migration in the wind-driven accretion discs has been
explored via hydrodynamic simulations for both low-mass \mbox{
\citep{McNally20}
}\hspace{0pt}
and high-mass planets \mbox{
\citep{Kimming20}}\hspace{0pt}
. \mbox{
\citet{McNally20} }\hspace{0pt}
performed
3D hydrodynamic simulations by including a thin surface layer mimicking
the wind-driven accretion, and found that the dynamical corotation
torque speeds inward type I migration up rather than slowing it down. This is the
opposite outcome from the expectation of the 2D inviscid
disc models \mbox{
\citep{Paardekooper14}}\hspace{0pt}
. \mbox{
\citet{Kimming20} }\hspace{0pt}
performed
2D hydrodynamic simulations of a gap-opening planet by taking account
of the effects of the torque caused by the disc wind, and showed that
planet could rapidly migrate outward when the mass loss rate is high
enough because the mass loss leads to a strongly asymmetric mass distribution
in the co-orbital region. The overall effect of the wind-driven accretion on migration is yet to be explored, but it may have a significant
effect on formation and survival of planets }\DIFdelend  choice of disc
models rather than the real feature, because planets of this mass
range are undergoing rapid gas accretion and have difficulties in forming unless
the formation timescale is comparable to the disc dissipation timescale. There
is an indication that the maximum masses of giant planets may increase
for higher metallicity environments or even peak near ${\rm [Fe/H]}\sim0$,
which is also observed by planetesimal accretion work by \mbox{
\citet{Mordasini12}}\hspace{0pt}
.
Otherwise, as stated in Section \ref{subsec:HJs}, masses of giant
planets do not strongly depend on stellar metallicities. \mbox{
\citet{Thorngren16}
}\hspace{0pt}
studied RV and transit planets and found a similar overall trend in
the mass-metallicity relation (see their Figure 9). 

 \subsection{Planetesimal accretion and pebble accretion \label{subsec:plsml_vs_pebble}}

Planet formation via planetesimal accretion has been studied and compared
with observed systems by many groups \citep[e.g.][]{Ida04a,Ida13,Mordasini09,Mordasini12,Thommes08,Coleman16b}.
Here, we compare our  pebble accretion work with that of \citet{Mordasini12}
and \citet{Ida13}. Although a direct comparison is difficult, \DIFdelbegin \DIFdel{the basic trends we
observe
for our pebble accretion study is very similar to those reported in
these planetesimal accretion studies. We will point out }\DIFdelend  we discuss a few similarities and differences \DIFdelbegin \DIFdel{below}\DIFdelend  in planet formation
outcomes.

\citet{Mordasini12} used the single-planet formation population synthesis
model and studied the effects of the disc's metallicity, mass, and lifetime;
while \citet{Ida13} used a similar population synthesis model but
took the effects of multiple-planet formation and dynamical
interactions into account. The disc properties adopted in these studies are different,
but they do overlap with ours. We considered the disc radii
over $0.1-100\,$au with disc masses $0.02-0.2\,M_{\odot}$ as well
as stellar metallicities ${\rm [Fe/H]}=[-0.5,\,0.5]$. \citet{Mordasini12}
considered the disc outer radii of $30\,$au with disc masses $0.004-0.09\,M_{\odot}$
and metallicities ${\rm [Fe/H]}=[-0.5,\,0.5]$. Applying their surface
mass density out to $100\,$au, the disc masses become $0.008-0.16\,M_{\odot}$.
\citet{Ida13}, on the other hand, considered the disc outer radii
of $30\,$au with disc masses $1.6\times10^{-4}-0.16\,M_{\odot}$
and metallicities ${\rm [Fe/H]}=[-0.2,\,0.2]$. Again, applying their
surface mass density to the disc out to $100\,$au, the disc masses
correspond to $0.0053-0.53\,M_{\odot}$.

First, we highlight the difference in dependences of planet formation
rates on disc properties such as metallicities, masses, and dissipation
timescales. In terms of these disc parameters, the pebble mass accretion
rate has the following dependency (see Equations \ref{eq:dotMf_final},
\ref{eq:dotM}, and \ref{eq:alpha_acc}):

\begin{equation}
\dot{M}_{{\rm core}}=\epsilon\dot{M}_{F}\propto\left(10^{{\rm [Fe/H]}}\right)^{2}\alpha_{{\rm acc}}^{-1}\dot{M}_{*}\propto\left(10^{{\rm [Fe/H]}}\right)^{2}M_{D},\label{eq:Mcore_peb}
\end{equation}
where $M_{D}$ is the disc mass and the function of both the initial
disc mass $M_{D,0}$ and the dissipation timescale $t_{{\rm diff}}$.
On the other hand, a similar equation for planetesimal accretion is
proportional to the dust surface mass density $\Sigma_{d}=10^{{\rm [Fe/H]}}\Sigma_{g}$
and has the \DIFdelbegin \DIFdel{dependence }\DIFdelend  following dependency \citep{Mordasini12}:

\begin{equation}
\dot{M}_{{\rm core,plsml}}\propto\Sigma_{d}\propto10^{{\rm [Fe/H]}}M_{D}.
\end{equation}
\citet{Ida13} took account of the gas drag effect on planetesimals
\citep{Kokubo02} and thus had a slightly different \DIFdelbegin \DIFdel{dependence }\DIFdelend  dependency of:

\begin{equation}
\dot{M}_{{\rm core,plsml}}\propto10^{{\rm [Fe/H]}}M_{D}^{7/5}.
\end{equation}
Therefore, in both pebble and planetesimal accretion models, increasing
either metallicity or disc mass makes planet formation more efficient.
This complementary effect of the metallicity and the disc mass was also pointed out by \citet{Mordasini12}.

In pebble accretion, the metallicity effect is more significant than
the disc mass effect, as seen in Equation \ref{eq:Mcore_peb}. More
specifically, changing the metallicity effect by an order of magnitude
(e.g. from ${\rm [Fe/H]=-0.5}$ to $0.5$ in $10^{{\rm [Fe/H]}}$)
has a much larger effect on the accretion rate than changing the disc
mass by an order of magnitude (e.g. from $0.02\,M_{\odot}$ to $0.2\,M_{\odot}$).
Indeed, Figure \ref{fig:GPformtime} shows that \DIFdelbegin \DIFdel{the }\DIFdelend the giant planet formation
timescale is different by $\sim2$ orders of magnitude for the same
disc mass with ${\rm [Fe/H]}=-0.5$ and $0.5$, while the timescale
is different by $\sim1$ order of magnitude for the same metallicity
with $M_{D,0}=0.02\,M_{\odot}$ and $0.2\,M_{\odot}$.

The final mass, on the other hand, is less affected by the stellar
metallicity both in planetesimal accretion \citep{Mordasini12} and
in pebble accretion, though the reasons may be slightly different.
For pebble accretion, the metallicity does not greatly affect \DIFdelbegin \DIFdel{on }\DIFdelend final
masses, partly because the PIM is independent of the stellar metallicity
(see Equation \ref{eq:PIM_A18}) and partly because the final planetary
mass is determined by the disc-limited gas accretion and thus is controlled
by the available gas disc mass. For planetesimal accretion, however,
the isolation mass depends on the dust density and thus masses of
low-mass planets or cores of giant planets are likely to be affected
by metallicities. Similarly to us, both \citet{Mordasini12} and \citet{Ida13} assumed that gas accretion is limited by the disc
supply towards the end:

\[
\dot{M}_{{\rm gas}}\propto\dot{M}_{*}\propto M_{D},
\]
though we took the reduced surface mass density effect
in the gap into consideration (Equation \ref{eq:M_TT16}), as shown in Equation \ref{eq:Mgas}.
Thus, the final mass of giant planets in particular is likely determined
by the balance between the disc mass and the disc dissipation timescale,
rather than by the metallicity.

In our work, this can be seen in Figure \ref{fig:aMgiants}, where
planetary masses are higher for the higher disc mass and for the longer
diffusion timescale. We can also see that planetary masses are higher
for the same initial disc mass if the dissipation time is longer (i.e.
when the disc experiences the lower mass accretion rate for a longer
period of time, rather than the higher mass accretion rate for a shorter
period of time). On the other hand, planetary masses are similar for
the same disc model with different metallicities (see Section \ref{subsec:HJs}
for the discussion on the increasing mass trend in $a-M_{p}$).

One major difference between pebble and planetesimal accretion is
that the formation timescale of giant planets in pebble accretion is short
compared to the disc's lifetime over a range of disc radii \citep[e.g.][]{Lambrechts12}.
This makes it easier for giant planets to form even in less-optimal
environments such as lower-mass, lower-metallicity, and/or more rapidly
dissipating discs. Furthermore, as is also discussed in Section \ref{subsec:Max_Mp_a},
pebble accretion may allow giant planet formation far beyond the radii
in which we placed the cores for our simulations.

Finally, our Figure \ref{fig:aM_Me_2panels} can be compared with
Figure 8 of \citet{Ida13}. Although adopted models and initial conditions
are very different between these two studies, both reproduce the observed
$M_{p}-e$, $a-e$, and $a-M_{p}$ distributions well and show similar
trends in simulated distributions. Perhaps the main difference is that,
compared to \citet{Ida13}, our work has produced more eccentric planets.
\citet{Ida13} reported that only $\sim10\,\%$ of their planets have
moderate eccentricities of $e\gtrsim0.2$. In our work, \DIFdelbegin \DIFdel{$\sim40.3\,\%$
of observable planets (middle panels) and }\DIFdelend $\sim26.2\,\%$
of all simulated planets (bottom panels)  and $\sim40.3\,\%$ of observable
planets (middle panels) have $e\geq0.2,$ as opposed to $\sim42\,\%$
of observed planets (top panels). Thus, our work agrees much better
with observations, at least for eccentricities  (also see Figure \ref{fig:hist_aMe}). 

There may be a few factors leading to this difference. The dynamical
instabilities generally require a few giant planets that are
close enough to each other \citep[e.g.][]{Chatterjee08}. \citet{Ida13} adopted
planetesimal accretion, which is slower than pebble accretion, and
therefore it may have been more difficult to form multiple giant planets
nearby, because the inner one may have migrated inward and away from
outer-growing protoplanets. \citet{Ida13} did not take account of
secular perturbations between giant planets in their Monte Carlo method
for planet-planet scattering either, which can also trigger orbital
instabilities and is of course automatically included in our N-body
simulations. Furthermore, \citet{Ida13} explored a relatively narrow
range of metallicities of ${\rm [Fe/H]}=[-0.2,\,0.2]$ as opposed
to our ${\rm [Fe/H]}=[-0.5,\,0.5]$. Since the number of giant planets
per system is more sensitive to disc masses for lower metallicities
(e.g. see cases with ${\rm [Fe/H]}\lesssim0.0$ in Figure \ref{fig:evol_num}),
they may have generated a lower number of giant planets per system
on the average. Although, this effect may have been mitigated to an
extent since they covered a wide range of disc masses (spanning two
orders of magnitude), and the low metallicities can be compensated
for by more massive discs.  Also, it is possible that the eccentricity
damping effect was too strong in their simulations, which prevented
the occurrence of dynamical instabilities \mbox{
\citep[see next sub-section as well as][]{Bitsch20}}\hspace{0pt}
.

\subsection{Comparison with \citet{Bitsch19} \label{subsec:Comp_Bitsch19}}

Here, we briefly compare our work with a similar N-body work
on giant planet formation by \citet{Bitsch19}. A direct comparison
is not possible since the details of their models are different from
ours (e.g. equation of motion, pebble and gas accretion models, pebble
isolation mass). However, we list differences in disc parameters below
and compare their results with ours for similar disc parameters. We
find that there are both similarities and differences.

Their gas accretion rate follows \citet{Hartmann98} and \citet{Bitsch15},
and it is the same as the formula adopted by \citet{Matsumura17}:

\begin{equation}
\log\left(\frac{\dot{M}_{*,\,B19}}{M_{\odot}\,{\rm yr}^{-1}}\right)=-8-\frac{7}{5}\log\left(\frac{t+10^{5}\,{\rm yr}}{10^{6}\,{\rm yr}}\right).\label{eq:dotMs_B19}
\end{equation}
Although we did not use the formula for this paper, their accretion
rate very closely follows Disc 2 of our model (which corresponds
to $M_{d}=0.06\,M_{\odot}$ with $t_{{\rm diff}}=0.1\,$Myr and thus
\DIFdelbegin \DIFdel{$\alpha\sim7.4\times10^{-2}$}\DIFdelend  $\alpha_{{\rm acc}}\sim7.4\times10^{-2}$). Since they evolve the
disc from 2 Myr to 5 Myr, their mass accretion changes from $\dot{M}_{*}\sim3.8\times10^{-9}\,M_{\odot}$
to $\sim1\times10^{-9}\,M_{\odot}$. They also adopted $\alpha_{{\rm acc}}=5.4\times10^{-3}$
for disc accretion and $\alpha_{{\rm turb}}=5.4\times10^{-4}$ and
$10^{-4}$ for disc turbulence. Although the mass evolution is similar
to our Disc 2, their disc could be close to our Disc 6 (i.e. $M_{d}=0.02\,M_{\odot}$
with $t_{{\rm diff}}=1\,$Myr and thus \DIFdelbegin \DIFdel{$\alpha\sim7.4\times10^{-3}$}\DIFdelend  $\alpha_{{\rm acc}}\sim7.4\times10^{-3}$)
since they started with an evolved disc and assumed $\alpha_{{\rm acc}}=5.4\times10^{-3}$.
In our simulations, both produce similar types of planets (see Figure
\ref{fig:aMgiants} and a discussion below).

On the other hand, their pebble mass flux is

\begin{equation}
\dot{M}_{F,\,B19}\propto Z^{7/3}\alpha^{-1}\dot{M}_{*,\,B19},
\end{equation}
where $Z=1\%$ is the metallicity. Although they assumed the solar
metallicity for their simulations, they scaled the pebble mass flux
by the factor of $S_{{\rm peb}}=1-10$, which corresponds to exploring
the metallicities of ${\rm [Fe/H]}=0.0$ to $0.43$. Since our pebble
mass flux is $\dot{M}_{F}\propto Z^{2}\alpha^{-1}\dot{M}_{*}$, the
dependence on each parameter is similar.

One of the conclusions by \citet{Bitsch19} was that formation of
CJs was possible for cores originating from 10 au with $\alpha_{{\rm turb}}=5.4\times10^{-4}$
and those from 5 au with $\alpha_{{\rm turb}}=10^{-4}$. From their
Figures $4-6$, the formation of giant planets up to $\sim300\,M_{\oplus}$
was possible for both $\alpha_{{\rm turb}}$ cases above $S_{{\rm peb}}=2.5$
(which corresponds to ${\rm [Fe/H]}\gtrsim0.17$), while no giant planets
formed for the solar metallicity case. Our simulations show similar
trends. In Disc 6 with $\alpha_{{\rm turb}}=10^{-4}$, giant planets
primarily become low-mass CJs with $\sim100-400\,M_{\oplus}$ for
metallicities ${\rm [Fe/H]}=0.3$ and $0.5,$ while no giant planets
form for ${\rm [Fe/H]}\leq0.0$ (see Figure \ref{fig:aMgiants}).
In Disc 2, the outcomes are similar except that planetary masses are
lower $\sim100-300\,M_{\oplus,}$ and very low-mass giant planets can
be formed with ${\rm [Fe/H]}=0.0$ (but not for lower metallicities).
Out of formed giant planets, those starting from $3-5\,$au typically
become the closest-in CJs with orbital radius beyond $\sim1\,$au.
For higher (lower) $\alpha_{{\rm turb}}$, planets tend to migrate further (less) 
(see Section \ref{subsec:singlePF} and Figure \ref{fig:aMsingle2})
as indicated by \citet{Bitsch19}.

One major difference between \citet{Bitsch19} and our work is that
their simulations are left with a number of giant planets with low
eccentricities, while we have successfully reproduced the eccentricity
distribution of giant planets (see Figure \ref{fig:aM_Me_2panels}).
The result may be surprising at a glance because the number of giant
planets that formed and survived in simulations by \citet{Bitsch19} is
higher than in our simulations, and typically $\sim5$ or more from their Figures
6 and 7. We note that the higher number of surviving planets is not likely
due to the difference in the initial number of cores (60 for their
simulations as opposed to 10 for ours). \citet{Juric08} showed that,
even when there were 50 giant planets, the final number of planets
would be 2-3, as long as the dynamical instabilities occur. The eccentricity
distribution from such simulations  with an initially high number of
giant planets agrees well with observations, and also with dynamical
instability simulations with lower number of giant planets \citep[e.g.][]{Juric08,Chatterjee08}.
Thus, the large number of surviving giant planets in \citet{Bitsch19}
indicates that the dynamical instability among giant planets was rare
in their simulations. Indeed, even their long-term evolution of 100
Myr did not lead to a dramatic increase in orbital eccentricities
\citep{Bitsch19}.

\DIFdelbegin \DIFdel{We suspect that this lack of dynamical instabilities may be caused
because their choice of $\alpha_{{\rm acc}}=5.4\times10^{-3}$ is
inconsistent with the time evolution of the stellar mass accretion
rate they have adopted (Equation \ref{eq:dotMs_B19}). The fit to
this observed mass accretion rate requires a rather high viscosity
$\alpha_{{\rm acc}}\sim0.01$ for the disc size of $10-100\,$au \mbox{
\citep{Hartmann98}
}\hspace{0pt}
and even higher $\alpha_{{\rm acc}}$ for a larger disc. By assuming
the lower $\alpha_{{\rm acc}}$ for the same accretion rate,
the estimated
surface mass density and thus the disc mass becomes higher, which
leads to more efficient eccentricity and inclination damping . In fact, }\DIFdelend 
 While this manuscript was under revision, we became aware of the work
by \mbox{
\citet{Bitsch20},}\hspace{0pt}
in which they studied the effects of eccentricity
and inclination damping efficiencies on the eccentricity distribution
of giant planets. They parameterised the eccentricity and inclination
damping timescales as $\tau_{a}=K\,\tau_{e}=K\,\tau_{i}$ with $K=5,\,50,\,500$,
and $5000$, and found that the observed eccentricity distribution
of giant planets can be recovered for slow damping with $K\sim5-50$.
Since \mbox{
\citet{Bitsch19} }\hspace{0pt}
adopted a faster damping with $K=100$, it
may be the reason why they did not obtain eccentric giants.

As seen in Figure \ref{fig_typeItypeII}, we also have $K\sim100,$ 
though we managed to reproduce the eccentricity distribution of giants.
Since our eccentricity and inclination damping prescriptions are similar to 
those by \citet{Bitsch19}, it is possible that subtle differences in disc conditions 
changed the dynamical outcomes of simulations.
For example, the choice
of $\alpha_{{\rm acc}}=5.4\times10^{-3}$ in \citet{Bitsch19} 
may be inconsistent with the time evolution of the stellar mass accretion rate 
they adopted (Equation \ref{eq:dotMs_B19}). The fit to this observed mass accretion
rate requires a rather high viscosity $\alpha_{{\rm acc}}\sim0.01$
for the disc size of $10-100\,$au \citep{Hartmann98} and even higher
$\alpha_{{\rm acc}}$ for a larger disc. By assuming the lower $\alpha_{{\rm acc}}$
for the same accretion rate, the estimated surface mass density and
thus the disc mass becomes higher, which leads to more efficient eccentricity
and inclination damping. In fact, we observed a similar lack
of dynamical instabilities in the no-migration simulations of \citet{Matsumura17} 
--- we adopted the same stellar mass accretion equation as Equation
\ref{eq:dotMs_B19} and used $\alpha\leq5\times10^{-3}$ when $3-5$
giant planets were formed (except for one case with 3 giant planets
with $\alpha=0.01$).
For completeness, \citet{Matsumura17} had $K\sim100$ in the type I regime and $K\sim10$ in
the type II regime, which should favour more eccentric systems. 
Moreover, we had twice as many cores in a much narrower range of disc radii ($0.3-5\,$au) 
compared to the current runs.
It is possible, however, since these simulations did not include migration, that
planets were separated too far from one another to invoke dynamical instabilities 
within simulation times. 
In that case, convergent migration may also play an important role in determining 
the eccentricity and inclination distributions of planetary systems.

\section{Conclusion \label{sec:Summary}}

For this paper, we studied the formation of planetary systems via
pebble accretion by using N-body simulations, and we investigated the
effects of disc parameters such as masses, dissipation timescales,
and metallicities. This is a continuation of \citet{Matsumura17},
in which we modified the N-body code SyMBA \citep{Duncan98} to incorporate
the pebble accretion model by \citet{Ida16a}, gas accretion, type
I and type II migration, and eccentricity and inclination damping.
In this work, we updated the code as detailed in Section \ref{sec:Methods}
to take account of the recent development of the field, and we also adopted
a two-$\alpha$ disc model, where mass accretion and disc turbulence
have different drivers.

\DIFdelbegin \DIFdel{The }\DIFdelend  We find that the disc masses, dissipation timescales, and stellar
metallicities all affect the efficiency of planet formation. The effects
of each parameter can be summarised as follows (see Section \ref{subsec:plsml_vs_pebble}): 
\begin{itemize}
\item Disc metallicities ${\rm [Fe/H]}$ affect the formation timescales of
protoplanetary cores, but they do not strongly affect the final planetary
masses. 
\item Initial disc masses $M_{D,0}$ affect both core formation and gas
accretion timescales, and thus the final planetary masses. 
\item Disc diffusion timescales $t_{{\rm diff}}$ set time limits \DIFdelbegin \DIFdel{for }\DIFdelend  on planet
formation in the disc, and thus affect the final planetary masses.
\DIFdelbegin 

\DIFdelend \end{itemize}
We identified two types of giant planet formation trends, depending
on whether planet formation is fast compared to the disc's dispersal
or not. When a disc's dissipation timescales are long in comparison to typical
planet formation timescales (Discs 4, 5, 7, and 8 in our simulations,
formation-dominated case), giant planets are massive ($\sim M_{J}$
or higher) and distributed over a wide range of orbital radii (from
$\sim0.1\,$au to over $100\,$au). On the other hand, when a disc's
dissipation timescales are comparable to planet formation timescales
(Discs 1, 2, 3, and 6, disc-dissipation-limited case), giant planets
tend to be low-mass ($\sim M_{J}$ or lower) and CJs (with $a\gtrsim1\,$au).
The formation timescale depends both on stellar metallicities and
disc masses \textemdash{} the timescale is shorter for more massive,
more metal-rich discs. Therefore, protoplanetary cores tend to migrate
significantly before accreting gas to become giant planets in low
metallicity discs, while giant planets can form in situ in the outer
part of high-metallicity, massive discs. For low-mass, low-metallicity
discs, giant planet formation is difficult.

Our main \DIFdelbegin \DIFdel{results }\DIFdelend  findings are the following: 
\begin{itemize}
\item Differently from \citet{Matsumura17}, we successfully reproduced
\DIFadd{the overall distribution trends} of semi-major axes $a$, eccentricities $e$,
and planetary masses $M_{p}$ \DIFadd{of extrasolar giant planets (see Section \ref{subsec:overall} and
Figure \ref{fig:aM_Me_2panels}), though we tend to
overproduce CJs compared to HJs and WJs.} The success of reproducing
the $a-M_{p}$ distribution, especially for CJs, is largely due to
the new type II migration formula and the two-$\alpha$ disc
model, as proposed by \citet{Ida18}. The success in reproducing the
$e$ distribution is likely due to a more self-consistent disc model\DIFdelbegin \DIFdel{(see Section \ref{subsec:Comp_Bitsch19}) and }\DIFdelend  ,
a higher number of giant planets formed per system compared to \citet{Matsumura17},
and not too efficient eccentricity and inclination damping in the
disc (see Section \ref{subsec:Comp_Bitsch19}). 
\item \DIFdelbegin \textbf{\DIFdel{\textcolor{blue}{The overall occurrence rates of giant planets
as a function of orbital periods agree well with observed trends (see
Section \ref{subsec:occurrence}). The occurrence rates increase with periods 
in the inner region, decrease in the outer region, and peak at $\sim1-10\,$au. The most
abundant giant planets are CJs ($>50\%$), and thus more than half
the giant planets stay near their formation region. }}}
\DIFdelend  The overall occurrence rates of giant planets as a function of orbital
periods agree well with observed trends (see Section \ref{subsec:occurrence}).
The occurrence rates increase with periods in the inner region, decrease
in the outer region, and peak at $\sim1-10\,$au. The most abundant
giant planets are CJs ($>50\%$), and thus more than half the giant
planets in our simulations stay near their formation region.  
\item  As discussed in Section \ref{subsec:Form_HJs}, 
our simulations naturally explain why HJs tend to be alone \citep[e.g.][]{Wright09,Steffen12,Huang16}, 
and also why the eccentricities of HJs are low around low-metallicity
stars and vary more widely around high-metallicity ones \citep[e.g.][]{Dawson13,Shabram16,Buchhave18}.
The same trend is expected for stellar obliquities of their host stars,
and the current observations support that (see Section \ref{subsec:HJs}).
\begin{itemize}
\item  In low-metallicity discs, HJs tend to form in situ: protoplanetary
cores migrate to the inner disc and accrete gas there. This is because
planet formation is slower in the low-metallicity discs, which leads
to greater migration of a protoplanetary core before it reaches the
PIM and starts accreting a significant gas to become a giant planet.
Since the low-metallicity discs tend to form just 1-2 giant planets,
HJs tend to be alone and on nearly circular and coplanar orbits. 
\item In high-metallicity discs, HJs can be formed either via tidal circularisation
of highly eccentric orbits or via \DIFdelbegin \DIFdel{``}\DIFdelend  a migration scenario (including in situ
\DIFdelbegin \DIFdel{'' formation(}\DIFdelend  formation; see Section \ref{subsec:HJs}). The higher metallicity
discs tend to produce a number of giant planets that are prone to
dynamical instabilities. A HJ could be formed from a WJ/CJ as its
eccentric orbit is circularised. Alternatively, HJs could be first
formed \DIFdelbegin \DIFdel{``}\DIFdelend in situ \DIFdelbegin \DIFdel{'' }\DIFdelend (i.e. via core migration followed by gas accretion)
 or via migration, along with WJs and CJs. The dynamical instabilities
in such systems often remove either HJs and/or WJs, leaving either
(i) only CJs, or (ii) HJs with CJs. HJs formed in high-metallicity discs \DIFdelbegin \DIFdel{tend to }\DIFdelend have
a wider variety of eccentricities and inclinations and also tend
to be alone.
\end{itemize}
 \item If an SCJ is formed, as a giant planet grows within $\sim20\,$au and
then gets scattered outward, we expect that such an SCJ (1) was born
in a high-metallicity disc ($\left[{\rm Fe/H}\right]\gtrsim0.0$),
(2) has an eccentric orbit, and (3) tends to be accompanied \DIFdelbegin \DIFdel{($\sim80\,\%$)
by an inner }\DIFdelend  by another
giant planet ( $\sim80\,\%$) (see Section \ref{subsec:SCJs}). 
\item Most warm Jupiters ($0.1\,{\rm au}\lesssim a\lesssim1\,$au) are formed
in the formation-dominated discs (i.e. Discs 4, 5, 7, and 8 in our
simulations). In other words, in our simulations, it is difficult
to form WJs in rapidly dissipating, low-mass and/or low-metallicity
discs. 
\item CJs \DIFdelbegin \DIFdel{can }\DIFdelend  tend to be formed in high-mass and/or high-metallicity discs,
where the planet formation timescale is \DIFdelbegin \DIFdel{short compared to }\DIFdelend  comparable to or shorter than
the disc dissipation timescale. 
\end{itemize}
 Finally, there are still several issues that need to be resolved/explored
in our work. Most importantly, type I migration is still too fast
and we tend to lose SEs. For example, type I migration can be slowed
in the inner disc region if we fully adopt the wind-driven disc, as in \citet{Ogihara18}.
Resolving the migration issue
is also important when choosing a more appropriate gas accretion formula,
which would provide more accurate planetary compositions (see Section
\ref{subsec:MassFrac_Metals}). Furthermore, when $\alpha_{{\rm turb}}\ll\alpha_{{\rm acc}}$
as we assumed, the gap depth may also be affected by the wind-driven
accretion.

\begin{acknowledgements}
\DIFdelbegin \DIFdel{SM would }\DIFdelend  We thank Man Hoi Lee and Eduard Vorobyov for useful discussions and
an anonymous referee for detailed comments. SM is grateful to Aurora
Sicilia-Aguilar for valuable discussions and also for kindly sharing
her data from \mbox{
\citet{SiciliaAguilar10}}\hspace{0pt}
. SM would also like to thank
the Earth-Life Science Institute at Tokyo Institute of Technology
for its hospitality, where part of this work has been done. SM is
\DIFdelbegin \DIFdel{also grateful to Aurora Sicilia-Aguilar for valuable discussions
and also for kindly sharing her data from \mbox{
\citet{SiciliaAguilar10}}\hspace{0pt}
.
SM is }\DIFdelend partly supported by the STFC grant number ST/S000399/1 (The Planet-Disk
Connection: Accretion, Disk Structure, and Planet Formation). RB acknowledges
financial assistance from the Japan Society for the Promotion of Science
(JSPS) Shingakujutsu Kobo (JP19H05071). SI is supported by MEXT Kakenhi
18H05438.

\end{acknowledgements}

\bibliographystyle{aa}
\bibliography{REF}

\end{document}